\documentclass[sn-mathphys-num]{sn-jnl}% Math and Physical Sciences Numbered Reference Style 
\usepackage{epstopdf}
\usepackage{amssymb}
\usepackage{graphicx}%
\usepackage{multirow}%
\usepackage{amsmath,amssymb,amsfonts}%
\usepackage{amsthm}%
\usepackage{mathrsfs}%
\usepackage[title]{appendix}%
\usepackage{xcolor}%
\usepackage{textcomp}%
\usepackage{manyfoot}%
\usepackage{booktabs}%
\usepackage{algorithm}%
\usepackage{algorithmicx}%
\usepackage{algpseudocode}%
\usepackage{listings}%
\usepackage{subfig}
\usepackage{graphicx}
\usepackage{ytableau}
\usepackage[all,cmtip]{xy}
\usepackage{enumitem}
\usepackage[right]{lineno}
\newcommand{\bfa}{{\mathbf a}}
\newcommand{\bfb}{{\mathbf b}}
\newcommand{\bfe}{{\mathbf e}}
\newcommand{\bfu}{{\mathbf u}}
\newcommand{\bfv}{{\mathbf v}}

\newcommand{\bfx}{{\mathbf x}}

\newtheorem{theorem}{Theorem}
\newtheorem{proposition}{Proposition}% 
\newtheorem{example}{Example}%
\newtheorem{remark}{Remark}%
\newtheorem{lemma}{Lemma}%
\newtheorem{definition}{Definition}%

\newtheorem{theoremA}[subsubsection]{Theorem}
\newtheorem{propositionA}[subsubsection]{Proposition}% 
\newtheorem{exampleA}[subsubsection]{Example}%
\newtheorem{remarkA}[subsubsection]{Remark}%
\newtheorem{lemmaA}[subsubsection]{Lemma}%
\newtheorem{corollaryA}[subsubsection]{Corollary}%
\newtheorem{definitionA}[subsubsection]{Definition}%

\begin{document}

\title[Conditions for Morphology-Based Topological Filtrations and Applications to Firn Data Analysis]{Conditions for Morphology-Based Topological Filtrations and Applications to Firn Data Analysis}


%\author*[1]{\fnm{Chuan-Shen} \sur{Hu}}\email{iauthor@gmail.com}

%\author[2]{\fnm{Yu-Min} \sur{Chung}}\email{iiauthor@gmail.com}

%\author[3]{\fnm{Kaitlin} \sur{Keegan}}\email{iiiauthor@gmail.com}

%\affil*[1]{\orgdiv{Department}, \orgname{Organization}, \orgaddress{\street{Street}, \city{City}, \postcode{100190}, \state{State}, \country{Country}}}

%\affil[2]{\orgdiv{Department}, \orgname{Organization}, \orgaddress{\street{Street}, \city{City}, \postcode{10587}, \state{State}, \country{Country}}}

%\affil[3]{\orgdiv{Department}, \orgname{Organization}, \orgaddress{\street{Street}, \city{City}, \postcode{610101}, \state{State}, \country{Country}}}

\author*[1,2]{\fnm{Chuan-Shen} \sur{Hu}}\email{chuanshenhu1@nuk.edu.tw}\email{chuanshenhu.official@gmail.com}

\author[3]{\fnm{Yu-Min} \sur{Chung}}%\email{yumchung@alumni.iu.edu}

\author[4]{\fnm{Kaitlin} \sur{Keegan}}%\email{kkeegan@unr.edu}

\author[1]{\fnm{Kun-Siang} \sur{Yu}}%\email{chuanshenhu.official@gmail.com}

\affil*[1]{\orgdiv{Department of Applied Mathematics}, \orgname{National University of Kaohsiung}, \orgaddress{\street{700, Kaohsiung University Rd., Nanzih District, Kaohsiung City}, \postcode{81148 }, \country{Taiwan}}}

%\affil*[2]{\orgdiv{Department of Mathematics}, \orgname{National Central University}, \orgaddress{\street{No. 300, Zhongda Rd., Zhongli District, Taoyuan City}, \postcode{320317}, \country{Taiwan}}} 

\affil*[2]{\orgdiv{Department of Mathematics}, \orgname{National Taiwan Normal University}, \orgaddress{\street{162, Section 1, Heping E. Rd., Taipei City}, \postcode{106}, \country{Taiwan}}}

\affil[3]{\orgname{Eli Lilly and Company}, \city{Indianapolis}, \postcode{46285}, \state{IN}, \country{USA}}

\affil[4]{\orgdiv{Department of Geological Sciences and Engineering}, \orgname{University of Nevada Reno}, \orgaddress{\street{1664 N. Virginia St.}, \city{Reno}, \postcode{89557}, \state{NV}, \country{USA}}}


\abstract{Persistent homology (PH), a key tool in topological data analysis (TDA), captures global topological features of digital images through \emph{topological filtrations}. Alternatively, mathematical morphology (MM), rooted in set theory and lattice theory, provides operations such as opening and closing to modify local geometric structures in digital images. This motivates incorporating local geometric information into a PH framework via \emph{morphological filtrations}, yielding an MM-based PH framework. However, the validity of such filtrations depends on the \emph{absorption property} of MM operations, which may fail for arbitrary \emph{structuring elements}, the components defining MM operators. To address this issue, we introduce \emph{shift inclusion} as a sufficient condition for ensuring absorption, provide a formal proof, and demonstrate its utility in pore-structure analysis, highlighting the synergy between MM and PH for image and scientific data analysis.}



\maketitle
%\linenumbers

%%%%%%%%%%%%%%%%%%%%%%%%%%%%%%%%%%%%%%%%%%%
%%%%%%%%%%%%%%%%%%%%%%%%%%%%%%%%%%%%%%%%%%%
%%%%%%%%%%%%%%%%%%%%%%%%%%%%%%%%%%%%%%%%%%%
%%%%%%%%%%%%%%%%%%%%%%%%%%%%%%%%%%%%%%%%%%%
\section{Introduction}\label{Section: Introduction}
%%%%%%%%%%%%%%%%%%%%%%%%%%%%%%%%%%%%%%%%%%%
%%%%%%%%%%%%%%%%%%%%%%%%%%%%%%%%%%%%%%%%%%%
%%%%%%%%%%%%%%%%%%%%%%%%%%%%%%%%%%%%%%%%%%%
%%%%%%%%%%%%%%%%%%%%%%%%%%%%%%%%%%%%%%%%%%%

Persistent homology (PH) is a fundamental and widely used tool in topological data analysis (TDA), offering a multiscale framework for capturing intrinsic shape and structural features in data. By tracking topological features such as connected components, loops, and voids across varying scales, PH, together with its representation via the persistence barcode (PB) or persistence diagram (PD), enables robust and interpretable analysis, particularly in high-dimensional settings. Owing to these strengths, PB or PD, which encodes the birth--death lifecycles of topological features, has been successfully applied across a broad range of domains, including image processing~\cite{edelsbrunnerTDAImage,chenChao_NeurlPS2023,chung2018topological, Chung-Day-Hu-2022}, machine learning~\cite{JMLR:v16:bubenik15a,chen2021sars,songdechakraiwut2023topological}, biomedical and biological data science~\cite{songdechakraiwut2023topological,Bhaskar2023}, chemistry and materials science~\cite{nakamura2015persistent,hiraoka2016hierarchical,murakami2019ultrahigh,HuQC2025}, and molecular biology~\cite{CangWei2020PcH,chen2020topology,chen2021sars,zhu2023tidal}.

From the perspective of TDA, a binary image can be modeled as a cubical complex, while a grayscale image defines a weighted version, inducing a sublevel set filtration via pixel intensities~\cite{kaczynski2004computational}. 
This yields a multiscale representation of image structure, where the PB captures the evolution of topological features such as connected components and holes across thresholds; in particular, holes associated with high-intensity regions tend to have larger death values. Such representations have been widely applied in image analysis, including texture classification~\cite{YMC_Austin_2020_CVPRW}, medical imaging~\cite{MPF_mitochondria_2024_Chung}, and microstructure characterization~\cite{chung2018topological,MPF_mitochondria_2024_Chung}. 
Recent work integrates PH with geometric features, such as distance transforms and Minkowski functionals, to enhance its expressiveness in digital imaging~\cite{Hu2021ICCV,armstrongPorous2018, korkmaz2025cumperlay}.

\begin{figure}
\centering
\includegraphics[width=0.9\linewidth]{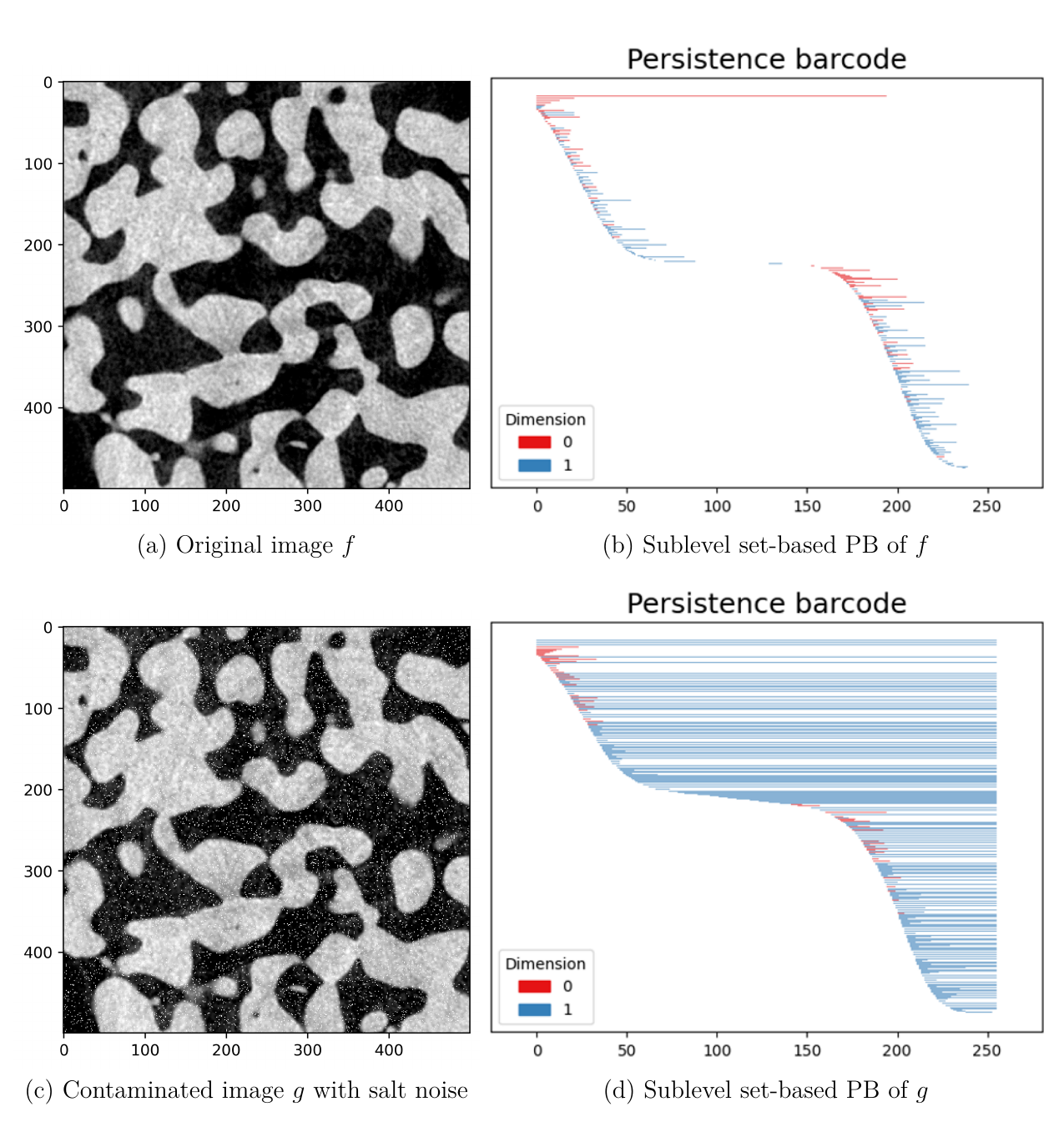}
\caption{Illustration of a grayscale image $f$, corresponding to one micro-CT-derived cross-sectional image of firn, and its salt-noise-contaminated counterpart $g$, generated by introducing salt noise into $f$. Panels (b) and (d) show the corresponding persistence barcodes derived from sublevel set filtrations.}
\label{Fig. Weakness of conventional PB}
\end{figure}

Although sublevel set-based PH effectively captures topological structures associated with pixel intensities in grayscale images, it primarily reflects the intensity values at which features appear and disappear, and thus provides only limited information about their geometric shape and spatial organization. 
This limitation is illustrated in Fig.~\ref{Fig. Weakness of conventional PB}. 
The grayscale image \( f \) in Fig.~\ref{Fig. Weakness of conventional PB}(a) and its salt-noise-contaminated counterpart \( g \) in Fig.~\ref{Fig. Weakness of conventional PB}(b) exhibit similar visual appearances, yet their corresponding PBs differ significantly in the number of \(\beta_1\) intervals. 
This discrepancy arises because the sublevel set filtration is highly sensitive to extreme pixel intensities, which can introduce spurious one-dimensional hole structures with large death values.

To capture additional spatial information of topological structures in digital images, we leverage \textit{mathematical morphology} (MM) to construct topological filtrations~\cite{Chung-Day-Hu-2022,Hu2021ICCV}. MM is a well-established framework in image processing, grounded in set theory, lattice algebra, and discrete geometry, among others, that enables the analysis of image structures via spatially localized operations~\cite{lucVincent1994,soille2013,Najman-Mathematical-Morphology,huijianLi2016,haralick1987}.
In particular, morphological operations, such as erosion, dilation, opening, closing, and distance transformation, form a fundamental toolkit for structural image analysis and have been widely applied in (bio)medical imaging, computer vision, and texture analysis~\cite{naranjo2020,kour2014real,angulo2015,klein1972texture,soille2013,jonesSoille1996}.

However, constructing a valid topological filtration through morphological opening and closing operations requires an additional condition, known as the \emph{absorption property}, which does not hold in general (see, e.g., Example~\ref{ExampleA: B1 subseteq B2, but not opening decreasing}). Specifically, this property depends on the choice of \emph{structuring elements} (SEs), which encode the neighborhood information of pixels. The absorption property holds only for sequences of SEs that satisfy certain constraints.

In this paper, we present a systematic and practically implementable framework for constructing valid topological filtrations induced by morphological opening and closing operations, centered around the notion of \textit{shift inclusion}. The theoretical framework underlying this work is rooted in the first author's doctoral dissertation~\cite{Hu_PhD_Dissertation}, where a more general algebraic formulation of morphological filtrations was developed. In the present paper, we focus on a concrete, practically implementable setting for digital images on $\mathbb{Z}^2$ and provide a streamlined, systematically organized presentation of the theory, including refined, simplified proofs adapted to this setting. We further establish a coherent framework for constructing valid topological filtrations and demonstrate its practical utility through applications to firn images. We further investigate the geometric significance of the resulting birth--death information and demonstrate the practical utility of the proposed framework through applications to firn images, which exhibit complex topological structures relevant to glaciological studies.

%%%%%%%%%%%%%%%%%%%%%%%%%%%%%%%%%%%%%%%%%%%
%%%%%%%%%%%%%%%%%%%%%%%%%%%%%%%%%%%%%%%%%%% 
\subsection{Related Work and Contributions}
%%%%%%%%%%%%%%%%%%%%%%%%%%%%%%%%%%%%%%%%%%%
%%%%%%%%%%%%%%%%%%%%%%%%%%%%%%%%%%%%%%%%%%%

The structuring element (SE) defines the neighborhood structure in MM. Originally introduced as a \emph{morphological filter} on a topological space~\cite{morphFilter-1,morphFilter-2,morphFilter-3,serra1984image}, it has been extended to weighted and spatially variant forms and integrated with tools from measure theory and vector field theory~\cite{legazAparicio2018}. However, in this work, we focus on fixed SEs to analyze local image structures. Different choices of SEs induce distinct erosion, dilation, opening, and closing operations, leading to varying degrees of geometric modification and structural alteration in images (see, e.g., Section~\ref{Section: Erosion, Dilation, Opening, and Closing}).

In many real-world applications of MM, SEs are often assumed to satisfy the absorption property. However, this property does not hold in general for arbitrarily chosen SEs (see, e.g., Example~\ref{ExampleA: B1 subseteq B2, but not opening decreasing}). Several approaches have been proposed to enforce this condition. For instance, for images on $\mathbb{Z}^2$, sequences of SEs constructed via Minkowski sums of \textit{periodic line segments} satisfy the absorption property~\cite[Fig.~11.8]{soille2013}. Alternatively, the openness property of SE sequences provides a criterion for verifying this condition~\cite{heijmans1994MorImgOperators,heijmans1995morphological}. Nevertheless, a systematic and practically implementable construction of such SEs remains lacking, and the geometric interpretation of birth--death values in opening- and closing-based persistent homology is still not fully understood.

From this perspective, we establish a rigorous foundation for selecting increasing SE sequences to construct valid topological filtrations via opening and closing operations. This amounts to identifying SE sequences for which the induced operations satisfy the absorption property (see Section~\ref{Section: Absorption Property}). We introduce \emph{shift inclusion} as a relation on SEs that guarantees this property (see Section~\ref{Section: Shift Inclusion for Topological Filtration Construction}), and further propose its generalization, \emph{weak shift inclusion} (see Appendix~\ref{Section: Weak Shift Inclusion}), proving its equivalence to the absorption property even on non-rectangular domains. This yields a unified framework for analyzing opening and closing operations in irregular settings and facilitates the integration of MM and TDA.

Based on topological filtrations induced by opening and closing operations, the resulting MM-based PH framework complements conventional sublevel set-based PH by capturing the shape and size of locally connected components and holes in digital images. This yields richer geometric information beyond standard sublevel set filtrations (see Fig.~\ref{Fig. closing filtrations and curves}). Notably, building on the shift inclusion framework, the MM-based spatial characterization of white and black clusters in binary images~\cite{Chung-Day-Hu-2022} shows strong potential for analyzing and mitigating salt-and-pepper noise.

Furthermore, the theoretical results of this paper extend naturally to a broader algebraic setting. Specifically, images can be modeled as functions defined on a subset $P$ of an abelian group $M$, with morphological operations induced by SEs in $M$~\cite{Hu_PhD_Dissertation}. 
Under this setting, the notion of shift inclusion and the associated theoretical results extend beyond the lattice $\mathbb{Z}^2$ considered here (e.g., to $M=\mathbb{Z}^m$), providing a unified perspective for analyzing morphological filtrations in higher-dimensional or more general discrete settings. In this paper, however, we restrict to $M=\mathbb{Z}^2$ for concreteness and to emphasize applications to digital image analysis.

To illustrate the geometric significance of the MM-based PB, we analyze firn, a porous intermediate material between snow and glacial ice that consists of an interconnected network of ice grains and air-filled pore spaces. 
The topology and geometry of these pore networks govern processes such as densification, meltwater percolation, and gas transport, making their characterization essential for understanding firn evolution and climate records derived from ice cores. 
Using MM-based PH, we extract birth--death information from persistence intervals to characterize the spatial organization of pore and ice structures across layers. 

%%%%%%%%%%%%%%%%%%%%%%%%%%%%%%%%%%%%%%%%%%%
%%%%%%%%%%%%%%%%%%%%%%%%%%%%%%%%%%%%%%%%%%%
\subsection{Paper Organization}
%%%%%%%%%%%%%%%%%%%%%%%%%%%%%%%%%%%%%%%%%%%
%%%%%%%%%%%%%%%%%%%%%%%%%%%%%%%%%%%%%%%%%%%

Section~\ref{Section: Mathematical Background} reviews the necessary background on digital images, persistent homology, and morphological operations (erosion, dilation, opening, and closing). 
Section~\ref{Section: Main Results} presents the main theoretical contributions, introducing \emph{shift inclusion} as a sufficient condition for constructing topological filtrations via opening and closing operations, along with its application to firn image analysis. Section~\ref{Section: Discussion and Conclusion} discusses broader implications and concludes the paper. Additional technical details and theoretical results are provided in the appendices (Appendices~\ref{Appendix: Digital Images and Cubical Complexes}--\ref{Appendix: Further Discussion on Shift Inclusion}), including the generalized notion of \emph{weak shift inclusion} in Appendix~\ref{Section: Weak Shift Inclusion}.

%%%%%%%%%%%%%%%%%%%%%%%%%%%%%%%%%%%%%%%%%%%
%%%%%%%%%%%%%%%%%%%%%%%%%%%%%%%%%%%%%%%%%%%
%%%%%%%%%%%%%%%%%%%%%%%%%%%%%%%%%%%%%%%%%%%
%%%%%%%%%%%%%%%%%%%%%%%%%%%%%%%%%%%%%%%%%%%
\section{Mathematical Background}\label{Section: Mathematical Background}
%%%%%%%%%%%%%%%%%%%%%%%%%%%%%%%%%%%%%%%%%%%
%%%%%%%%%%%%%%%%%%%%%%%%%%%%%%%%%%%%%%%%%%%
%%%%%%%%%%%%%%%%%%%%%%%%%%%%%%%%%%%%%%%%%%%
%%%%%%%%%%%%%%%%%%%%%%%%%%%%%%%%%%%%%%%%%%%

Section~\ref{Section: Mathematical Background} introduces the essential mathematical framework of this work, including digital images (Section~\ref{Section: Digital Images}), persistent homology (Section~\ref{Section: Persistent Homology}), and mathematical morphology (Section~\ref{Section: Mathematical Morphology}), highlighting their connections in image analysis. Rather than relying on traditional MM terminology or lattice-theoretic notation, we formulate the theory within elementary set theory and Euclidean topology, providing a unified perspective that bridges the MM and TDA communities.

%%%%%%%%%%%%%%%%%%%%%%%%%%%%%%%%%%%%%%%%%%%
%%%%%%%%%%%%%%%%%%%%%%%%%%%%%%%%%%%%%%%%%%%
\subsection{Digital Images}\label{Section: Digital Images}
%%%%%%%%%%%%%%%%%%%%%%%%%%%%%%%%%%%%%%%%%%%
%%%%%%%%%%%%%%%%%%%%%%%%%%%%%%%%%%%%%%%%%%%

The standard notations \( \mathbb{Z} \), \( \mathbb{N} \), and \( \mathbb{R} \) denote the sets of integers, positive integers, and real numbers, respectively, serving as the fundamental number systems considered in this paper. The set of non-negative integers is denoted by \( \mathbb{Z}_{\geq 0} \), which is equivalent to \( \mathbb{N} \cup \{ 0\} \). Similarly, \( \mathbb{R}_{\geq 0} \) represents the set of all non-negative real numbers. The extended real number system, denoted by $\overline{\mathbb{R}} = \mathbb{R} \cup \{ \pm \infty \}$, is considered in certain circumstances. A topological space, such as a subspace of \( \mathbb{R}^2 \) (e.g., the geometric realization of a 2D cubical complex), is typically denoted using calligraphic font, such as \( \mathcal{X} \), \( \mathcal{Y} \), or \( \mathcal{Z} \).

A \textit{\(2\)-dimensional lattice} is the set $\mathbb{Z}^2$ of integer pairs $(x,y)$, which serves as the ambient space for digital image pixels. 
We denote its elements by boldface letters, e.g., $\mathbf{u}=(x,y)\in\mathbb{Z}^2$, to distinguish them from scalars. 
The standard basis of $\mathbb{R}^2$ is $\{\mathbf{e}_1=(1,0), \mathbf{e}_2=(0,1)\}$, which generates $\mathbb{Z}^2$ over $\mathbb{Z}$, and we use the canonical projections $\pi_i:\mathbb{Z}^2\to\mathbb{Z}$ defined by $\pi_1(x,y)=x$ and $\pi_2(x,y)=y$. For $A,B \subseteq \mathbb{Z}^2$, the \textit{Minkowski sum} and \textit{difference} are defined by
\begin{equation*}
A + B := \{ \bfa + \bfb \mid \bfa \in A, \bfb \in B \}, \qquad
A - B := \{ \bfa - \bfb \mid \bfa \in A, \bfb \in B \}.
\end{equation*}
For simplicity, for $\bfx \in \mathbb{Z}^2$ and $B \subseteq \mathbb{Z}^2$, we write $\bfx + B$ and $\bfx - B$ in place of $\{\bfx\}+B$ and $\{\bfx\}-B$. Geometrically, the sets $\bfx + B$ and $\bfx - B$ can be viewed as discrete neighborhoods of $\bfx$ in $\mathbb{Z}^2$ and play a central role in defining erosion, dilation, opening, and closing.

An \textit{image domain} is defined as a nonempty subset $P$ of $\mathbb{Z}^2$. In particular, a \textit{discrete rectangle} is a subset of the form $(I \times J) \cap \mathbb{Z}^2$, where $I$ and $J$ are closed intervals in $\mathbb{R}$. In this paper, apart from certain discussions and generalizations, we will mainly focus on rectangular image domains. Elements of $P$ are called \textit{pixels}, and $P$ is typically represented as a \textit{grid diagram}, i.e., a collection of unit squares corresponding to the pixels. For example, if $P = ([0,2] \times [0,2]) \cap \mathbb{Z}^2 = \{ 0,1,2 \} \times \{ 0, 1, 2\}$, then the grid diagram of $P$ is represented by
\begin{equation}
\label{Eq. block diagram}
\ytableausetup{centertableaux}
P = \begin{ytableau}
 \ &  \ & \  \\
  &   &  \\
\mathbf{0}  &   & \\  
\end{ytableau} \quad \text{or} \quad 
P = \begin{ytableau}
 \ &  \ & \  \\
  &   &  \\
 \ &   & \\  
\end{ytableau}\quad,
\end{equation}
where $\mathbf{0} = (0,0)$ denotes the origin in $\mathbb{Z}^2$. In practice, since the shape of $P$ is the main concern, the origin $\mathbf{0}$ is often omitted from the grid diagram representation.

A \textit{digital image} on a domain $P \subseteq \mathbb{Z}^2$ is a function $g: P \to \mathbb{R}_{\geq 0}$ that assigns to each pixel a non-negative real value, called the \textit{pixel value}. 
Under the grid diagram representation, a digital image is visualized by assigning values to the corresponding unit squares. 
For example, for $P$ defined in \eqref{Eq. block diagram}, a digital image $g: P \to \mathbb{R}_{\geq 0}$ can be represented as follows:
\begin{equation}
\label{Eq. block diagram-image representation}
\ytableausetup{centertableaux}
g = \begin{ytableau}
g_{11} & g_{12} & g_{13} \\
g_{21} & g_{22} & g_{23} \\
g_{31} & g_{32} & g_{33} \\
\end{ytableau} \quad.
\end{equation}
Here, the entries $g_{ij}$ represent pixel values in $\mathbb{R}_{\geq 0}$. If $g$ takes values in $\{0,1\}$, it is called a \textit{binary image}; otherwise, it is referred to as a \textit{grayscale image}. We denote by $\mathcal{I}_P = \{g: P \to \mathbb{R}_{\geq 0}\}$ the set of all images on $P$, and by $\mathcal{BI}_P = \{g: P \to \{0,1\}\}$ the set of all binary images.

In this work, we explore the geometry and shape of the black pixel set of a given image. Typically, pixels in an image are considered \textit{black pixels} if they have a pixel value of $0$. On the other hand, in a binary image, pixels with a value of $1$ are called \textit{white pixels}. Mathematically, the sets of black and white pixels of a binary image $g \in \mathcal{BI}_P$ are the preimages given by
\begin{equation}
\begin{split}
\{ g = 0 \} := g^{-1}(0) &= \{ \mathbf{x} \in P \ | \ g(\mathbf{x}) = 0 \}, \\
\{ g = 1 \} := g^{-1}(1) &= \{ \mathbf{x} \in P \ | \ g(\mathbf{x}) = 1 \}.
\end{split}
\end{equation}
Furthermore, by considering the partial order $\leq$ on $\mathcal{I}_P$ defined by 
\begin{equation}
\label{Eq. partial order on images}
f \leq g \text{ \ if and only if \ } f(\mathbf{x}) \leq g(\mathbf{x})  \text{ \ whenever \ } \mathbf{x} \in P,
\end{equation}
it follows that $f^{-1}(0) \subseteq g^{-1}(0)$ whenever $g \leq f$. Moreover, these two conditions are equivalent if $f$ and $g$ are both binary images.

Under the grid diagram representation, the collection of black pixels in an image can be identified with a union of closed unit squares in $\mathbb{R}^2$, where any two squares intersect only along shared edges or vertices. 
This forms a \textit{cubical complex} (see, e.g.,~\cite{kaczynski2004computational}). 
Below is an example of an image on a $3\times 3$ domain and the corresponding geometric realization of the cubical complex of its black pixels:
\begin{equation}\label{Eq. Cubical complex representation}
g = \begin{ytableau}
 0 & 0 & 0 \\
 1 & 0 & 0 \\
 0 & 1 & 0 \\
\end{ytableau} \quad \text{and} \quad \mathcal{X}_g = \begin{ytableau}
 *(gray) & *(gray) & *(gray) \\
 \none[] & *(gray) & *(gray) \\
 *(gray) & \none[] & *(gray) \\
\end{ytableau}~.  
\end{equation}
The set $\mathcal{X}_g$ is referred to as the topological representation of the image $g$, and is a compact subset of $\mathbb{R}^2$. For a detailed formulation of the correspondence between a binary image $g$ and its topological representation $\mathcal{X}_g$, see Appendix~\ref{Appendix: Digital Images and Cubical Complexes}.

An \textit{image operation} on an image domain $P$ is a function $\xi: \mathcal{I}_P \to \mathcal{I}_P$ that maps each image on $P$ to another. Analogous to \eqref{Eq. partial order on images}, we define a partial order on image operations by
\begin{equation}
\label{Eq. the partial order on image operations}
\xi \leq \zeta \quad \text{if} \quad \xi(f) \leq \zeta(f) \quad \text{for all} \quad f \in \mathcal{I}_P.
\end{equation}
The \textit{identity operation} on $\mathcal{I}_P$ is the identity map $\operatorname{id}_{\mathcal{I}_P}$, which maps each element to itself. In this paper, image operations are essential for generating topological filtrations of digital images and form the foundation for computing persistent homology. In particular, fundamental operations such as \emph{morphological erosion}, \emph{dilation}, \emph{opening}, and \emph{closing}, introduced in Section~\ref{Section: Mathematical Morphology}, are employed in this context.

\bigskip
\begin{remark}
More generally, an image operation in mathematical morphology can be extended to a function \( f: \mathcal{I}_P \to \mathcal{I}_Q \) between different image domains \( P \) and \( Q \). However, in this paper, we restrict our focus to the case where \( P = Q \).
\end{remark}
\bigskip

A fundamental example of an image operation is \textit{thresholding}. For $t \in \mathbb{R}_{\geq 0}$, the thresholding operation $\tau_t:\mathcal{I}_P \to \mathcal{I}_P$ is defined by
\begin{equation}
\label{Eq. Thresholding operation}
\tau_t(g)(\mathbf{x}) :=
\begin{cases}
0, & \text{if } g(\mathbf{x}) \leq t, \\
1, & \text{otherwise.}
\end{cases}
\end{equation}
We often write $g_{[t]} := \tau_t(g)$. Thresholding converts a grayscale image into a binary image and is widely used in image processing, while in TDA it provides an efficient way to construct filtrations of cubical complexes~\cite{soille2013,Najman-Mathematical-Morphology,sonka2013image,chung2018topological,kaczynski2004computational,Kovalevsky2006}. In particular, conventional PH computation is based on filtrations induced by thresholding operations, such as $\tau_0 \geq \tau_1 \geq \cdots \geq \tau_{255}$ (see Fig.~\ref{Fig. Weakness of conventional PB}). Relevant properties are summarized in Appendix~\ref{Section: Thresholding Operations}. Fig.~\ref{Fig. Definition of images} summarizes the key notions of this section, including the image domain, grayscale and binary images, the geometric realization of a cubical complex, and the thresholding operation.

\begin{figure}
    \centering
    \includegraphics[width=0.9\linewidth]{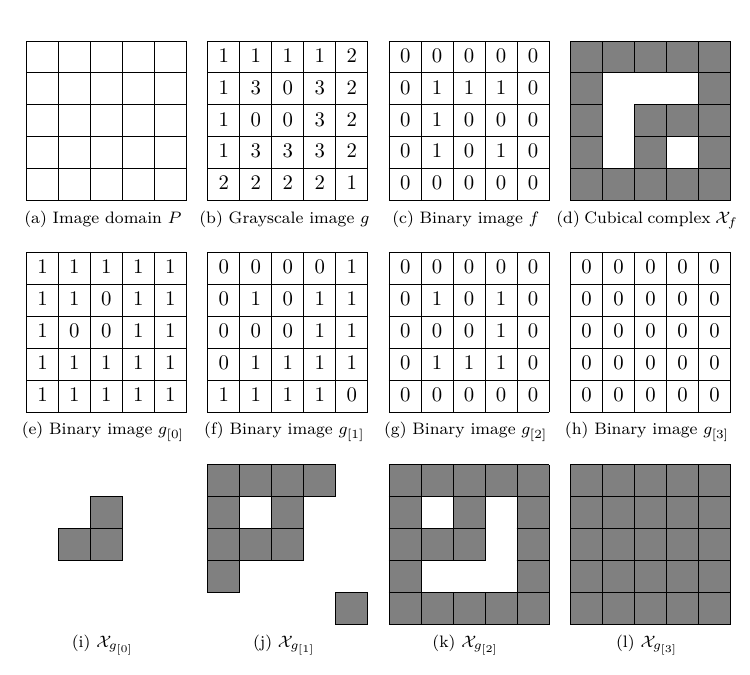}
    \caption{Top row: (a) a $5\times5$ image domain $P \subseteq \mathbb{Z}^2$, (b) a grayscale image $g \in \mathcal{I}_P$, (c) a binary image $f \in \mathcal{BI}_P$, and (d) the topological representation $\mathcal{X}_f \subseteq \mathbb{R}^2$. In (d), pixels with value $0$ correspond to black squares in $\mathbb{R}^2$. Middle row: images (e)--(h) form the sequence $g_{[0]} \geq g_{[1]} \geq g_{[2]} \geq g_{[3]}$ obtained via thresholding at levels $0,1,2,3$. Bottom row: images (i)--(l) form a topological filtration $\mathcal{X}_{g_{[0]}} \subseteq \mathcal{X}_{g_{[1]}} \subseteq \mathcal{X}_{g_{[2]}} \subseteq \mathcal{X}_{g_{[3]}}$, corresponding to the sequence in (e)--(h).}
    \label{Fig. Definition of images}
\end{figure}

%%%%%%%%%%%%%%%%%%%%%%%%%%%%%%%%%%%%%%%%%%%
%%%%%%%%%%%%%%%%%%%%%%%%%%%%%%%%%%%%%%%%%%%
\subsection{Persistent Homology}
\label{Section: Persistent Homology}
%%%%%%%%%%%%%%%%%%%%%%%%%%%%%%%%%%%%%%%%%%%
%%%%%%%%%%%%%%%%%%%%%%%%%%%%%%%%%%%%%%%%%%%

This paper focuses on homology and persistent homology (PH) based on cubical complexes and their filtrations derived from digital images. This section briefly introduces the key concepts that form the foundation of the paper. Specifically, Section~\ref{Section: The homology of topological spaces} presents the homology of a cubical complex along with the geometric intuition behind the associated Betti numbers. Section~\ref{Section: Persistent homology and persistence barcodes} introduces the PH of a filtration of cubical complexes and explains the geometric interpretation of the resulting persistence barcodes.

%%%%%%%%%%%%%%%%%%%%%%%%%%%%%%%%%%%%%%%%%%%
\subsubsection{The homology of topological spaces}
\label{Section: The homology of topological spaces}
%%%%%%%%%%%%%%%%%%%%%%%%%%%%%%%%%%%%%%%%%%%

Algebraic topology provides a framework for studying intrinsic properties of geometric objects, such as connected components, loops, and higher-dimensional holes. Homology theory captures these features via Betti numbers, defined as the dimensions of homology groups over a field~\cite{Greenberg,Vick,munkres2018elements}. This perspective underlies topological data analysis (TDA), which studies the shape of data. In this paper, we focus on $2$D digital images and their $0$th and $1$st Betti numbers ($q=0,1$), corresponding to connected components and holes.

Formally, the $q$-th homology of a topological space $\mathcal{X}$ over a field $\mathbb{F}$ is denoted by $H_q(\mathcal{X};\mathbb{F})$, and its dimension $\beta_q(\mathcal{X}) := \dim_{\mathbb{F}} H_q(\mathcal{X};\mathbb{F})$ is called the $q$-th Betti number. 
In practice, we take $\mathbb{F}=\mathbb{F}_2=\mathbb{Z}/2\mathbb{Z}$ and write $H_\bullet(\mathcal{X})$ for $H_\bullet(\mathcal{X};\mathbb{F}_2)$. In Fig.~\ref{Fig. Illustration of PB}(i)--(l), the topological spaces $\mathcal{X}_{g_{[0]}}, \mathcal{X}_{g_{[1]}}, \mathcal{X}_{g_{[2]}}, \mathcal{X}_{g_{[3]}}$ have Betti pairs $(\beta_0,\beta_1) = (1,0), (2,1), (1,2), (1,0)$, respectively, corresponding to the numbers of connected components and $1$-dimensional holes. In particular, $\mathcal{X}_{g_{[1]}}$ has two connected components, one of which contains a hole, while $\mathcal{X}_{g_{[2]}}$ has one connected component containing two holes.

%%%%%%%%%%%%%%%%%%%%%%%%%%%%%%%%%%%%%%%%%%%
\subsubsection{Persistent homology and persistence barcodes}
\label{Section: Persistent homology and persistence barcodes}
%%%%%%%%%%%%%%%%%%%%%%%%%%%%%%%%%%%%%%%%%%%

Beyond a single space, persistent homology (PH) studies a \textit{filtration} of spaces and tracks the evolution of topological features. Formally, a filtration is a nested sequence
\begin{equation*}
\mathcal{X}_\bullet: \quad \emptyset = \mathcal{X}_0 \subseteq \mathcal{X}_1 \subseteq \mathcal{X}_2 \subseteq \cdots \subseteq \mathcal{X}_n.    
\end{equation*}
By the functoriality of homology~\cite{munkres2018elements,Greenberg,Vick}, this induces the sequence 
\begin{equation}
\label{Eq. PH}
\operatorname{PH}_q(\mathcal{X}_\bullet): \quad 0 \xrightarrow{ \ \ f_0 \ \ } H_q(\mathcal{X}_{1}) \xrightarrow{ \ \ f_1 \ \ } H_q(\mathcal{X}_{2}) \xrightarrow{ \ \ f_2 \ \ } \cdots \xrightarrow{ \ \ f_{n-1} \ \ } H_q(\mathcal{X}_{n}) 
\end{equation}
of homology groups, called the persistent homology of the filtration. Here, each \( H_\bullet(\mathcal{X}_i) \) denotes the homology of \( \mathcal{X}_i \), each homomorphism \( f_i \) is induced by the inclusion map \( \mathcal{X}_i \hookrightarrow \mathcal{X}_{i+1} \), and \( q \in \mathbb{Z}_{\geq 0} \) indicates the homological dimension of interest. 

Using the algebraic structure of the homology groups, PH tracks the lifespan of topological features across a filtration, where each feature is associated with a \textit{persistence interval} (or simply an \textit{interval}) $(b,d)$ determined by its \textit{birth} and \textit{death} times. These intervals are summarized by PB~\cite{barannikov1994framed,carlsson2004persistence,ghrist2008barcodes}, or equivalently by PD, which represents each interval as a point $(b,d)$ in $\overline{\mathbb{R}} \times \overline{\mathbb{R}}$~\cite{cohen2005stability}. Several equivalent algebraic approaches can be used to extract birth and death information \cite{zomorodian2004computing,crawley2015decomposition}. Rather than giving formal definitions, we use Fig.~3 to illustrate their geometric meaning.

\begin{figure}
    \centering
    \includegraphics[width=0.9\linewidth]{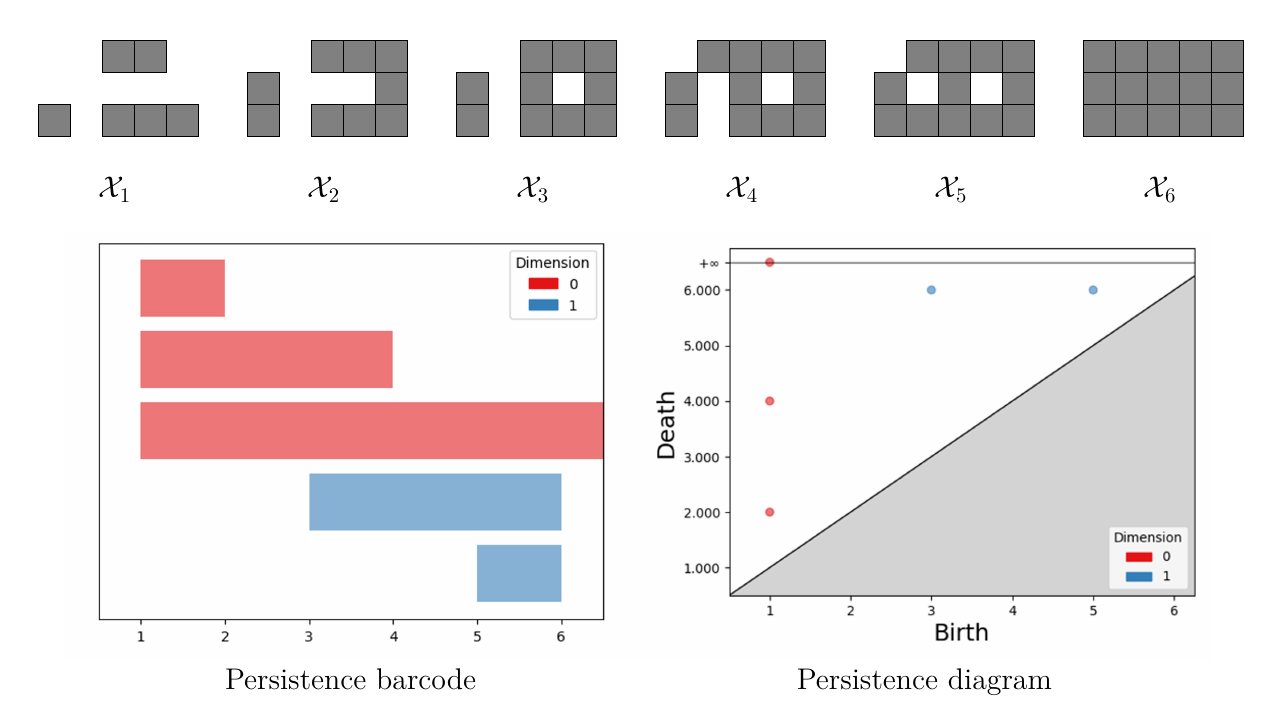}
    \caption{Top row: A topological filtration $\mathcal{X}_{1} \subseteq \mathcal{X}_{2} \subseteq \cdots \subseteq \mathcal{X}_{6}$ of cubical complexes. Bottom row: The corresponding persistence barcode and persistence diagram.}
    \label{Fig. Illustration of PB}
\end{figure}

Fig.~\ref{Fig. Illustration of PB} illustrates an example of a filtration $\mathcal{X}_{1} \subseteq \mathcal{X}_{2} \subseteq \cdots \subseteq \mathcal{X}_{6}$ of topological spaces in \( \mathbb{R}^2 \), along with the corresponding \( 0 \)-th and \( 1 \)-st persistence barcodes and their associated persistence diagrams. Specifically, the \( 0 \)-th and \( 1 \)-st persistence barcodes of the filtration are the collections
\begin{equation*}
\operatorname{PB}_0 = \{ (1,\infty), (1,2), (1,4) \} \quad \text{and} \quad \operatorname{PB}_1 = \{ (3,6), (5,6) \}.   
\end{equation*}
In this example, three connected components are born at $1$, one of which merges with another at $2$, giving rise to the interval $(1,2)$ in $\operatorname{PB}_0$. 
Another interval $(1,4)$ corresponds to a merge occurring at $4$. 
For $\operatorname{PB}_1$, the intervals $(3,6)$ and $(5,6)$ correspond to holes born at $3$ and $5$, respectively, and dying at $6$.

%%%%%%%%%%%%%%%%%%%%%%%%%%%%%%%%%%%%%%%%%%%
%%%%%%%%%%%%%%%%%%%%%%%%%%%%%%%%%%%%%%%%%%%
\subsection{Mathematical Morphology}
\label{Section: Mathematical Morphology}
%%%%%%%%%%%%%%%%%%%%%%%%%%%%%%%%%%%%%%%%%%%
%%%%%%%%%%%%%%%%%%%%%%%%%%%%%%%%%%%%%%%%%%%

In this section, we review the standard image operations in mathematical morphology, namely erosion, dilation, opening, and closing. Mathematical morphology (MM) admits different formalisms: Serra~\cite{serra1984image} models images as subsets using $\oplus$ and $\ominus$, whereas Soille~\cite{soille2013} treats images as real-valued functions on lattices. In computational topology, these symbols often have different meanings (e.g., direct sums), leading to ambiguity. To avoid this, under the notation and settings adopted in this paper, we adopt a set-theoretical formulation that provides a unified framework for integrating MM with topological data analysis.

%%%%%%%%%%%%%%%%%%%%%%%%%%%%%%%%%%%%%%%%%%%
\subsubsection{Structuring Elements}
\label{Section: Structuring Elements}
%%%%%%%%%%%%%%%%%%%%%%%%%%%%%%%%%%%%%%%%%%%

\begin{figure}%
\centering
\includegraphics[width=0.8\linewidth]{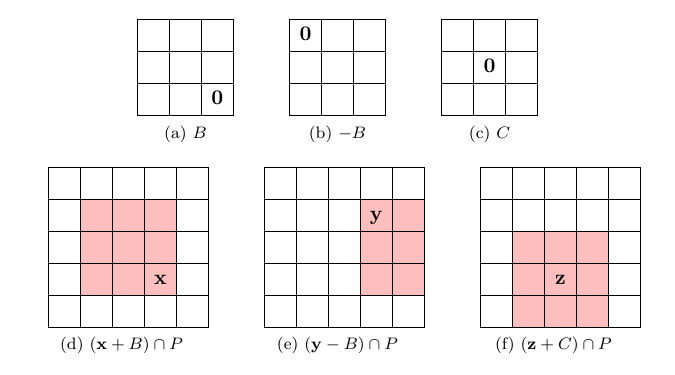}
\caption{Structuring elements as discrete neighborhoods in the image domain. Panels (a)--(c) illustrate three SEs $B$, $-B$, and $C$, all with the same $3\times3$ shape but with different placements of the origin. Panels (d)--(f) show the corresponding neighborhoods $(\mathbf{x}+B)\cap P$, $(\mathbf{y}-B)\cap P$, and $(\mathbf{z}+C)\cap P$ for points $\mathbf{x}, \mathbf{y}, \mathbf{z}\in P$, where $P$ is as in Fig.~\ref{Fig. Definition of images}(a). Shaded regions indicate the included pixels.}
\label{Fig. SEs and nbds}
\end{figure}

Erosion, dilation, opening, and closing are fundamental image operations in mathematical morphology (MM). These operations analyze local neighborhoods in a discrete image domain $P \subseteq \mathbb{Z}^2$. To describe such neighborhoods, \textit{structuring elements} (SEs) are used, modeled as finite subsets $B \subseteq \mathbb{Z}^2$ containing the origin $\mathbf{0}$~\cite{soille2013}. 
Their translations $\mathbf{x}+B$ and $\mathbf{x}-B$ for $\mathbf{x} \in \mathbb{Z}^2$ define discrete neighborhoods of $\mathbf{x}$. 
In this work, for $\mathbf{x} \in P$, we focus on the restricted neighborhoods $(\mathbf{x}+B)\cap P$ and $(\mathbf{x}-B)\cap P$, which capture the local structure within the image domain $P$. A SE $B \subseteq \mathbb{Z}^2$ is said to be \textit{symmetric} if $B = -B := \{ -\bfx \mid \bfx \in B \}$. Fig.~\ref{Fig. SEs and nbds} illustrates examples of SEs and the corresponding neighborhoods using grid diagrams, highlighting their role in morphological image analysis. A sequence of SEs $B_1 \subseteq B_2 \subseteq \cdots \subseteq B_n$ is called an \textit{increasing SE sequence}, or simply an \textit{SE sequence}. 

%%%%%%%%%%%%%%%%%%%%%%%%%%%%%%%%%%%%%%%%%%%
\subsubsection{Erosion, Dilation, Opening, and Closing}
\label{Section: Erosion, Dilation, Opening, and Closing}
%%%%%%%%%%%%%%%%%%%%%%%%%%%%%%%%%%%%%%%%%%%

\begin{figure}%
\centering
\includegraphics[width=0.9\linewidth]{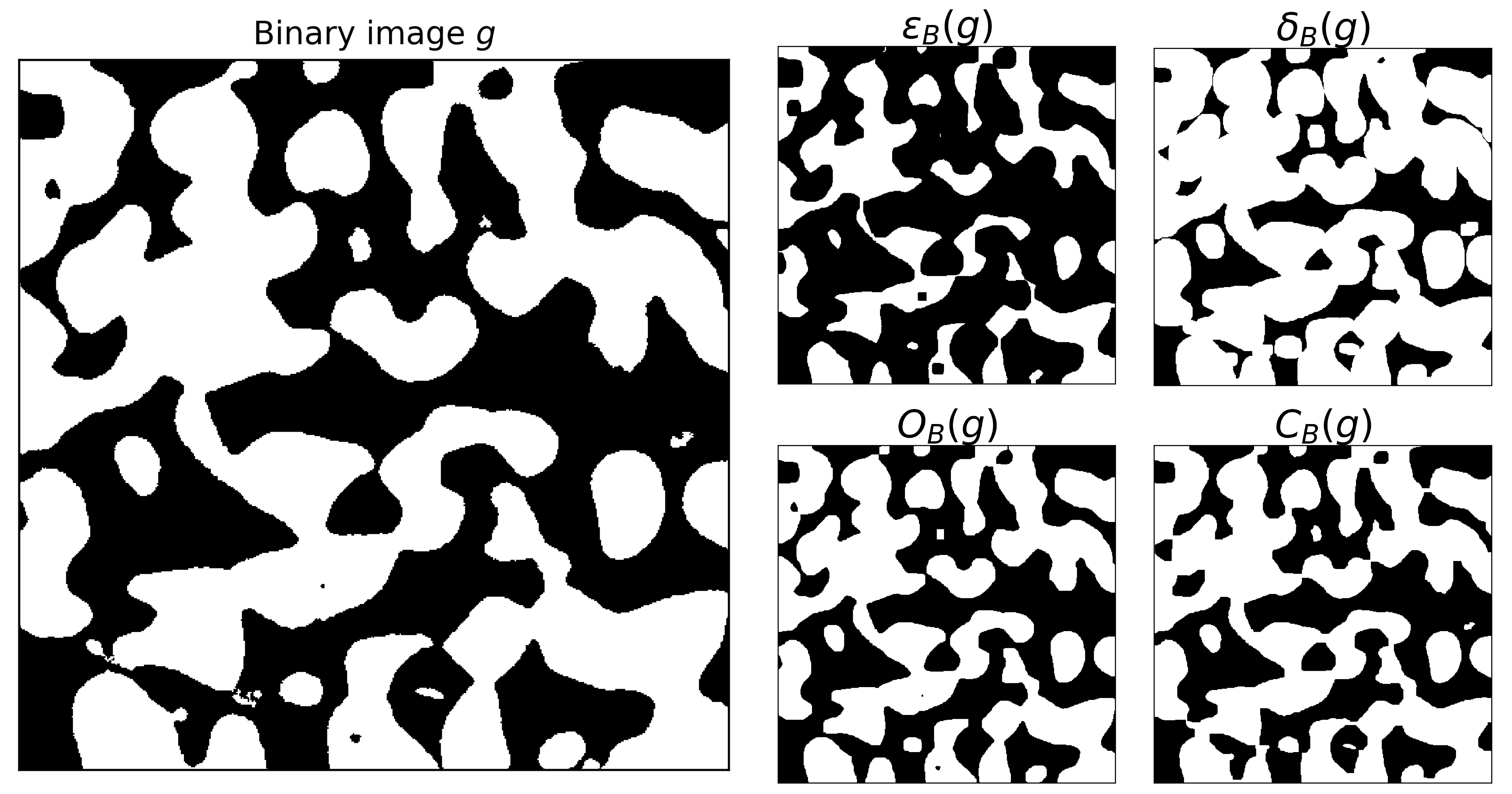}
\caption{Illustration of a binary image $g$ and transformed images $\epsilon_B(g)$, $\delta_B(g)$, $O_B(g)$, and $C_B(g)$. The SE $B$ is an $11\times11$ square centered at the origin $\mathbf{0}$. The image $g$ is a $500\times500$ pixel, binarized micro-CT cross-sectional image of firn from 7-m depth at NEEM, Greenland. The binarization threshold was obtained using the average pixel value as the threshold.}
\label{Fig. Demonstration of erosion, dilation, opening, and closing}
\end{figure}

Given SEs and their induced neighborhoods on $P \subseteq \mathbb{Z}^2$, the morphological operations are defined. For $g \in \mathcal{I}_P$ and an SE $B$, the \textit{erosion} and \textit{dilation} are image operations $\epsilon_B, \delta_B: \mathcal{I}_P \to \mathcal{I}_P$, defined for $\mathbf{x} \in P$ by
\begin{equation}
\label{Equation: Erosion and dilation}
\begin{split}
\epsilon_B(g)(\mathbf{x}) &= \min \{ g(\mathbf{x} + \bfb) \ | \ \bfb \in B, \mathbf{x} + \bfb \in P \}, \\
\delta_B(g)(\mathbf{x}) &= \max \{ g(\mathbf{x}-\bfb) \ | \ \bfb \in B, \mathbf{x}-\bfb \in P \}.
\end{split}    
\end{equation}

\bigskip
\begin{remark}
The dilation formulas in \eqref{Equation: Erosion and dilation} and \eqref{Equation: Opposite dilation} are not equivalent for a general SE $B$. Both definitions appear in different applications and software packages; see Appendix~\ref{Appendix Section: Alternative Definition of Morphological Dilation}. In this paper, we adopt $\delta_B$ in \eqref{Equation: Erosion and dilation} as the definition of dilation, and explain the rationale in the same appendix.    
\end{remark}
\bigskip

Since erosion $\epsilon_B(g)(\mathbf{x})$ and dilation $\delta_B(g)(\mathbf{x})$ compute the minimum and maximum over $(\mathbf{x}+B)\cap P$ and $(\mathbf{x}-B)\cap P$, respectively, it is necessary to verify that $\mathbf{x}\pm\mathbf{b}\in P$ for all $\mathbf{b}\in B$. For this purpose, for $\mathbf{x}\in P$, we define
\begin{equation}
\label{Eq. Sets B(x;P,+) and B(x;P,-)}
B(\bfx; P, +) = \{ \bfb \in B \mid \bfx + \bfb \in P \} \ \text{ and } \ B(\bfx;P, -) = \{ \bfb \in B \mid \bfx - \bfb \in P \}.
\end{equation}
In the case shown in Fig.~\ref{Fig. SEs and nbds}(d), we have \( B(\bfx; P, +) = B \), since the entire highlighted region lies within \( P \). On the other hand, in the case of Fig.~\ref{Fig. SEs and nbds}(e), \( B(\bfx; P, -) \) is a strict subset of \( B \). Using this notation \eqref{Eq. Sets B(x;P,+) and B(x;P,-)}, Equation~\eqref{Equation: Erosion and dilation} can be written as
\begin{equation}
\label{Equation: Erosion and dilation-2}
\begin{split}
\epsilon_B(g)(\bfx) &= \min \{ g(\bfx + \bfb) \ | \ \bfb \in B(\mathbf{x};P,+) \}, \\
\delta_B(g)(\bfx) & = \max \{ g(\bfx-\bfb) \ | \ \bfb \in B(\mathbf{x};P,-)  \}.
\end{split}    
\end{equation}

Through function composition, they give rise to different operations, among which opening and closing are commonly used. The \textit{morphological opening} and \textit{morphological closing} via $B$, denoted by $O_B$ and $C_B$, are image operations on $\mathcal{I}_P$ defined as
\begin{equation}
\label{Equation: definitions of open and close}
    O_B = \delta_B \circ \epsilon_B \ \ {\rm and} \ \ C_B = \epsilon_B \circ \delta_B.
\end{equation}
Notably, $\epsilon_B=\delta_B=\operatorname{id}_{\mathcal{I}_P}$ and $O_B=C_B=\operatorname{id}_{\mathcal{I}_P}$ if $B=\{\mathbf{0}\}$.

Fig. \ref{Fig. Demonstration of erosion, dilation, opening, and closing} illustrates an example of how the operations $\epsilon_B$, $\delta_B$, $O_B$, and $C_B$ operate on images. As presented in Fig. \ref{Fig. Demonstration of erosion, dilation, opening, and closing}, for a given binary image, erosion tends to thin the white regions, while dilation tends to thicken them. As a result, the opening operation, which applies erosion followed by dilation, effectively removes small white noise while preserving the overall shape of larger white regions. Conversely, the closing operation, which applies dilation followed by erosion, fills small holes within white regions and connects nearby white structures. Due to these effects, opening and closing are commonly used for filtering and smoothing digital images.

\begin{figure}%
\centering
\includegraphics[width=0.9\linewidth]{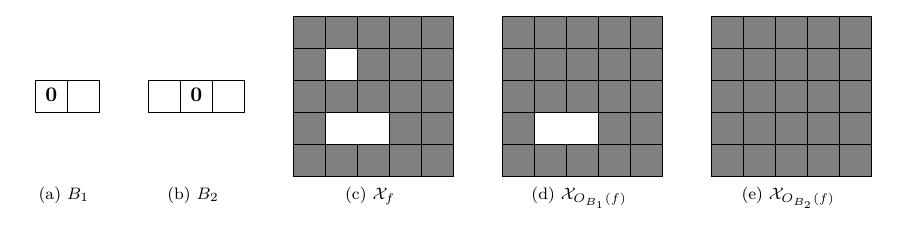}
\caption{A toy example of the topological representation $\mathcal{X}_f$ of a binary image $f$, together with the spaces induced by $O_{B_1}(f)$ and $O_{B_2}(f)$ for SEs $B_1 \subseteq B_2$.}
\label{Fig. SEs and Sizes}
\end{figure}

%%%%%%%%%%%%%%%%%%%%%%%%%%%%%%%%%%%%%%%%%%%
\subsubsection{Absorption Property}
\label{Section: Absorption Property}
%%%%%%%%%%%%%%%%%%%%%%%%%%%%%%%%%%%%%%%%%%%

To construct topological filtrations based on morphological opening and closing, we explore the \textit{absorption property} of these operations~\cite{soille2013}. Formally, two SEs \( B_1 \subseteq B_2 \) are said to satisfy the absorption property if
\begin{equation}
\label{Equation: Absorption property}
O_{B_2}(f) \leq O_{B_1}(f) \quad \text{and} \quad C_{B_1}(f) \leq C_{B_2}(f) \quad \text{whenever} \quad f \in \mathcal{I}_P.
\end{equation}
Under the topological representation adopted in this paper (see Fig. \ref{Fig. Definition of images}), the property in \eqref{Equation: Absorption property} can be expressed as follows:
\begin{equation}
\label{Equation: Absorption property-X-form}
\mathcal{X}_{O_{B_1}(f)} \subseteq \mathcal{X}_{O_{B_2}(f)} \quad \text{and} \quad \mathcal{X}_{C_{B_2}(f)} \subseteq \mathcal{X}_{C_{B_1}(f)} \quad \text{whenever} \quad f \in \mathcal{I}_P,
\end{equation}
where the order of the induced cubical complexes is opposite to the order of opening and closing operations shown in  \eqref{Equation: Absorption property}. This is because an image \( f \) being less than another image \( g \) implies that \( f \) contains more pixels with value \( 0 \).

%Let $B_1 \subseteq B_2 \subseteq \cdots \subseteq B_n$ be SEs and let $g \in \mathcal{I}_P$. If $\mathcal{X}_{O_{B_i}(g)} \subseteq \mathcal{X}_{O_{B_j}(g)}$ for all $i \leq j$, then the family $\{\mathcal{X}_{O_{B_i}(g)}\}_{i=1}^n$ is called an \textit{opening filtration}. Dually, if $\mathcal{X}_{C_{B_j}(g)} \subseteq \mathcal{X}_{C_{B_i}(g)}$ for all $i \leq j$, then $\{\mathcal{X}_{C_{B_i}(g)}\}_{i=1}^n$ is called a \textit{closing filtration}.

The shapes and sizes of SEs influence how erosion, dilation, opening, and closing operations modify local regions in a digital image. For example, in Fig. \ref{Fig. SEs and Sizes}, the topological space $\mathcal{X}_f$ induced from a binary image $f$ contains two $1$-dimensional holes enclosed by the black pixels. In particular, these two holes are of different sizes: one occupies one white pixel, and the other occupies two. When applying the opening operations with different SEs, $B_1$ and $B_2$, this size information can be estimated by the SEs. More precisely, $O_{B_1}$ can only erase the smaller hole, while $O_{B_2}$ can remove both holes. In particular, from a TDA perspective, the changes in hole structures within the filtration $\mathcal{X}_{f} \subseteq \mathcal{X}_{O_{B_1}(f)} \subseteq \mathcal{X}_{O_{B_2}(f)}$ can be reflected by the persistence intervals in the associated PH of the filtration.

For a given digital image $f \in \mathcal{I}_P$, suppose there exists a sequence of SEs $B_1, B_2, \ldots, B_n$ such that $O_{B_j}(f) \leq O_{B_i}(f)$ and $C_{B_i}(f) \leq C_{B_j}(f)$ whenever $i < j$. Then this sequence induces a filtration of cubical complexes in $\mathbb{R}^2$:
\begin{equation*} 
\mathcal{X}_{C_{B_n}(f)} \subseteq \cdots \subseteq \mathcal{X}_{C_{B_2}(f)} \subseteq \mathcal{X}_{C_{B_1}(f)} \subseteq \mathcal{X}_{f} \subseteq \mathcal{X}_{O_{B_1}(f)} \subseteq \mathcal{X}_{O_{B_2}(f)} \subseteq \cdots \subseteq \mathcal{X}_{O_{B_n}(f)}.
\end{equation*}
In particular, this MM-based filtration of cubical complexes provides a natural framework for persistent homology computation, capturing additional spatial and size information beyond conventional PH based on sublevel set filtrations~\cite{Chung-Day-Hu-2022}. Unfortunately, the property \eqref{Equation: Absorption property} or \eqref{Equation: Absorption property-X-form} does not generally hold, even when \( B_1 \) is a subset of \( B_2 \) (see Section \ref{Appendix Section: Counterexamples to the Absorption Property}). 

%%%%%%%%%%%%%%%%%%%%%%%%%%%%%%%%%%%%%%%%%%% 
%%%%%%%%%%%%%%%%%%%%%%%%%%%%%%%%%%%%%%%%%%%
\subsection{Sublevel Set-Based PH vs. MM-Based PH}
\label{Section: Sublevel Set-Based PH vs. MM-Based PH}
%%%%%%%%%%%%%%%%%%%%%%%%%%%%%%%%%%%%%%%%%%%
%%%%%%%%%%%%%%%%%%%%%%%%%%%%%%%%%%%%%%%%%%%

At the end of Section~\ref{Section: Mathematical Background}, we highlight the key geometric differences between sublevel set-based and MM-based filtrations. Sublevel set filtrations are typically applied to grayscale images, whereas MM-based filtrations are originally designed for binary images. As a result, they capture topological features tailored to different data types and applications, providing an intuitive motivation for using morphological operations, such as opening and closing, to construct meaningful filtrations.

Conventional PH for a grayscale image $f: P \to \{0,1,\dots,k\}$ on $P \subseteq \mathbb{Z}^2$ is based on the sublevel set filtration of cubical complexes induced by thresholding, $\mathcal{X}_{f_{[0]}} \subseteq \cdots \subseteq \mathcal{X}_{f_{[k]}}$, where $\mathcal{X}_{f_{[i]}}$ corresponds to threshold $i$ (see \eqref{Eq. Thresholding operation}; cf. Fig.~\ref{Fig. Definition of images}). 
In contrast, MM-based PH analyzes the topology within a fixed sublevel set (i.e., a binary image) via filtrations induced by morphological operations (see Fig.~\ref{Fig. SEs and Sizes}). This captures geometric information, such as the size and shape of connected components and holes, so that persistence intervals carry richer geometric meaning. 

In particular, by combining MM-based filtrations with thresholding over pixel intensities, one obtains a bifiltration (or more generally, a multiparameter filtration) of digital images. Such constructions have shown promising potential in the analysis of general and biomedical images~\cite{Chung-Day-Hu-2022,MPF_mitochondria_2024_Chung}, and provide a natural foundation for multiparameter persistence in topological data analysis \cite{Hu_PhD_Dissertation,Chung-Day-Hu-2022,Hu2021ICCV}. Further discussion is presented in Section~\ref{Section: Main Results}.

%%%%%%%%%%%%%%%%%%%%%%%%%%%%%%%%%%%%%%%%%%%
%%%%%%%%%%%%%%%%%%%%%%%%%%%%%%%%%%%%%%%%%%%
%%%%%%%%%%%%%%%%%%%%%%%%%%%%%%%%%%%%%%%%%%%
%%%%%%%%%%%%%%%%%%%%%%%%%%%%%%%%%%%%%%%%%%%
\section{Main Results}
\label{Section: Main Results}
%%%%%%%%%%%%%%%%%%%%%%%%%%%%%%%%%%%%%%%%%%%
%%%%%%%%%%%%%%%%%%%%%%%%%%%%%%%%%%%%%%%%%%%
%%%%%%%%%%%%%%%%%%%%%%%%%%%%%%%%%%%%%%%%%%%
%%%%%%%%%%%%%%%%%%%%%%%%%%%%%%%%%%%%%%%%%%%
Section~\ref{Section: Shift Inclusion for Topological Filtration Construction} introduces the \textit{shift inclusion} relation, ensuring valid topological filtrations with general SEs, addressing the failure shown in Example~\ref{ExampleA: B1 subseteq B2, but not opening decreasing}. Additionally, the more general \textit{weak shift inclusion} is proposed as an equivalent condition to guarantee the absorption property, with details provided in Appendix~\ref{Section: Weak Shift Inclusion}. Section~\ref{Section: Firn Data Analysis} presents the analysis of the firn data described in Section~\ref{Section: Dataset}, based on the proposed morphological opening and closing filtrations and their corresponding PHs. The main theoretical results in this section are adapted from the first author's doctoral dissertation~\cite{Hu_PhD_Dissertation}, with modifications and simplified proofs tailored to the present setting.

%%%%%%%%%%%%%%%%%%%%%%%%%%%%%%%%%%%%%%%%%%%%%%
%%%%%%%%%%%%%%%%%%%%%%%%%%%%%%%%%%%%%%%%%%%%%%
\subsection{Shift Inclusion Condition for Opening and Closing Filtration Construction}
\label{Section: Shift Inclusion for Topological Filtration Construction}
%%%%%%%%%%%%%%%%%%%%%%%%%%%%%%%%%%%%%%%%%%%%%%
%%%%%%%%%%%%%%%%%%%%%%%%%%%%%%%%%%%%%%%%%%%%%%

As introduced in Sections~\ref{Section: Persistent Homology} and~\ref{Section: Mathematical Morphology}, a major goal of this article is to investigate under what conditions the absorption properties $O_{B_2}(g) \leq O_{B_1}(g)$ and $C_{B_1}(g) \leq C_{B_2}(g)$ hold for every image $g$. As demonstrated in Example~\ref{ExampleA: B1 subseteq B2, but not opening decreasing}, the inequality $O_{B_2}(g) \leq O_{B_1}(g)$ does not hold in general. This suggests that additional constraints on the SEs $B_1 \subseteq B_2$ are necessary. To explore how such conditions might be formulated to ensure the desired absorption property, we begin with a computational observation presented as the following lemma. 
\bigskip
\begin{lemma}\label{Lemma: crucial observation for shift inclusion}
Let $P \subseteq \mathbb{Z}^2$ be an image domain, $\mathbf{x} \in P$, and let $B \subseteq \mathbb{Z}^2$ be an SE. Then, for any image $g \in \mathcal{I}_P$, the following statements hold: 
\begin{itemize}
\item[\rm (a)] $O_{B}(g)(\bfx) = 0$ if and only if for every $\bfb \in B(\bfx;P,-)$, there is a $\bfb' \in B(\bfx-\bfb;P,+)$ such that $g(\bfx - \bfb + \bfb') = 0$;
\item[\rm (b)] $C_{B}(g)(\bfx) = 0$ if and only if there is a $\bfb \in B(\bfx;P,+)$ such that $g(\bfx + \bfb - \bfb') = 0$ for every $\bfb' \in B(\bfx+\bfb;P,-)$.
\end{itemize} 
\end{lemma}
\begin{proof}
We prove statement (a) concerning the opening operation, as the proof of statement (b), regarding the closing operation, follows in a similar argument. By the definition of the opening operation $O_B$, we have
\begin{equation*}
O_{B}(g)(\bfx) = \max_{\bfb \in B(\mathbf{x}; P, -)} \epsilon_{B}(g)(\bfx - \bfb).    
\end{equation*}
Therefore, $O_{B}(g)(\mathbf{x}) = 0$ if and only if $\epsilon_{B}(g)(\bfx - \bfb) = 0$ for all $\mathbf{b} \in B(\mathbf{x}; P, -)$.  By the definition of erosion, this is equivalent to the following condition: every $\mathbf{b} \in B(\mathbf{x}; P, -)$ admits a $\mathbf{b}' \in B(\mathbf{x} - \mathbf{b}; P, +)$ such that $g(\mathbf{x} - \mathbf{b} + \mathbf{b}') = 0$.
\end{proof}
\bigskip
To motivate the conditions under which the absorption property holds, we consider two SEs $B_1$ and $B_2$ where $B_1 \subseteq B_2 \subseteq \mathbb{Z}^2$ and $O_{B_1}(g)(\mathbf{x}) = 0$.  The goal is to identify the conditions under which $O_{B_2}(g)(\mathbf{x}) = 0$. By Lemma \ref{Lemma: crucial observation for shift inclusion}, $O_{B_2}(g)(\mathbf{x}) = 0$ if and only if every $\mathbf{b}_2 \in B_2(\mathbf{x}; P, -)$ admits a $\mathbf{b}_2' \in B_2(\mathbf{x}-\mathbf{b}_2; P, +)$ such that $g(\mathbf{x} - \mathbf{b}_2 + \mathbf{b}_2') = 0$. 
 Since $O_{B_1}(g)(\mathbf{x}) = 0$, every $\mathbf{b}_1 \in B_1(\mathbf{x}; P, -)$ admits a $\mathbf{b}_1' \in B_1(\mathbf{x} - \mathbf{b}_1; P, +)$ such that
\begin{equation*}
0 = g(\mathbf{x} - \mathbf{b}_1 + \mathbf{b}_1') = g(\mathbf{x} - \mathbf{b}_2 + (\mathbf{b}_2 - \mathbf{b}_1 + \mathbf{b}_1')).     
\end{equation*}
In particular, if every $\mathbf{b}_2 \in B_2(\mathbf{x}; P, -)$ admits a corresponding $\mathbf{b}_1 \in B_1(\mathbf{x}; P, -)$ such that $B_1 + (\mathbf{b}_2 - \mathbf{b}_1) \subseteq B_2$, then $\mathbf{b}_2 - \mathbf{b}_1 + \mathbf{b}_1' \in B_2(\mathbf{x}-\mathbf{b}_2;P,+)$. Therefore, $O_{B_2}(g)(\mathbf{x}) = 0$ by the property stated above. This observation, together with its dual counterpart, motivates the following definition.
\bigskip
\begin{definition}
\label{Definition: Shift Inclusion w.r.t P}
Let $P \subseteq \mathbb{Z}^2$ be an image domain, and let $B_1 \subseteq B_2$ be SEs in $\mathbb{Z}^2$. We say that $B_1$ and $B_2$ satisfies the \textbf{shift inclusion with respect to $P$}, denoted by $B_1 \subseteq_{S,P} B_2$, if $B_1$ and $B_2$ satisfy the following properties:
\begin{itemize}
    \item[$({\rm a})$] (\textbf{Positive property}, denoted by $B_1 \subseteq_{S,P,+} B_2$) For all $\bfx \in P$ and $\bfb_2 \in B_2(\mathbf{x};P,+)$, there exists a $\bfb_1 \in B_1(\mathbf{x};P,+)$ such that $B_1 + (\bfb_2 - \bfb_1) \subseteq B_2$;
    \item[$({\rm b})$] (\textbf{Negative property}, denoted by $B_1 \subseteq_{S,P,-} B_2$) For all $\bfx \in P$ and $\bfb_2 \in B_2(\mathbf{x};P,-)$, there exists a $\bfb_1 \in B_1(\mathbf{x};P,-)$ such that $B_1 + (\bfb_2 - \bfb_1) \subseteq B_2$.
\end{itemize}
A sequence $B_1 \subseteq B_2 \subseteq \cdots \subseteq B_n$ of included SEs satisfying the shift inclusion relations is called a \textbf{shift inclusion sequence}.
\end{definition}

\bigskip
\begin{remark}\label{Remark: Positive/negative relationship}
By the definitions of the positive and negative properties, $B_1 \subseteq_{S,P,+} B_2$ if and only if $-B_1 \subseteq_{S,P,-} -B_2$. In particular, if $B_1 \subseteq B_2$ are symmetric SEs, then the positive and negative properties are equivalent.  That is, $B_1 \subseteq_{S,P,+} B_2$ if and only if $B_1 \subseteq_{S,P,-} B_2$. Note that the positive and negative properties are not equivalent in general.
\end{remark}
\bigskip

\begin{figure}%
\centering
\includegraphics[width=0.9\linewidth]{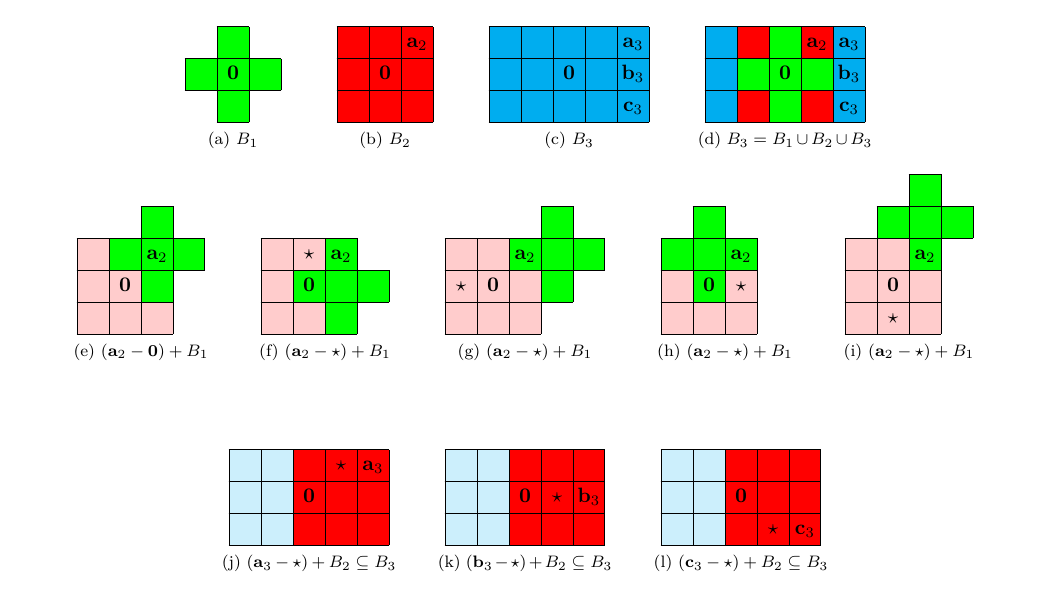}
\caption{Two examples of pairs $B_1 \subseteq B_2$ and $B_2 \subseteq B_3$ of SEs, where one satisfies the shift inclusion relation and the other does not. Panels (a), (b), and (c) depict three SEs $B_1, B_2, B_3$ with $B_1 \subseteq B_2 \subseteq B_3$. Panel (d) illustrates the overlapping relations of the three SEs. Certain points (e.g., $\mathbf{0}$, $\mathbf{a}_2$, $\mathbf{b}_3$, etc.) are specified in the SEs. The second row ((e)--(i)) and the third row ((j)--(l)) explain the reasons for $B_1 \nsubseteq_{S,P} B_2$ and $B_2 \subseteq_{S,P} B_3$ by illustrating the shifted regions of $B_1$ and $B_2$, assuming $P = \mathbb{Z}^2$. The light-colored squares in the second and third rows depict the regions in $B_2$ and $B_3$ that do not belong to the shifted regions, respectively.}
\label{Fig. Counter-example of non-shift included}
\end{figure} 

Fig. \ref{Fig. Counter-example of non-shift included} presents two examples of symmetric SEs $B_1 \subseteq B_2 \subseteq B_3$ with $B_1 \nsubseteq_{S,P} B_2$ and $B_2 \subseteq_{S,P} B_3$, where $P$ is assumed to be $\mathbb{Z}^2$. Specifically, to conclude that $B_1 \nsubseteq_{S,P} B_2$, we observe that $B_1 = \{ \mathbf{0}, \pm \mathbf{e}_1, \pm \mathbf{e}_2 \}$ and $B_2 = B_1 \cup \{ \mathbf{e}_1 \pm \mathbf{e}_2, -\mathbf{e}_1 \pm \mathbf{e}_2 \}$. As depicted in the second row ((e)--(i)), for $\mathbf{a}_2 = \mathbf{e}_1 + \mathbf{e}_2$, $(\mathbf{a}_2 - \mathbf{x}) + B_1 \nsubseteq B_2$ whenever $\mathbf{x} \in B_1$, demonstrating that $B_1 \nsubseteq_{S,P} B_2$. On the other hand, as depicted in the row (j)--(l), for $\mathbf{a}_3, \mathbf{b}_3, \mathbf{c}_3 \in B_3 \setminus B_2$, $\mathbf{a}_3 - \bfe_1, \mathbf{b}_3 - \bfe_1, \mathbf{c}_3 - \bfe_1 \in B_2$ with $B_2 + \mathbf{e}_1 \subseteq B_3$. This shows that $B_2 \subseteq_{S,P,+} B_3$. Because $B_2$ and $B_3$ are symmetric, $B_2 \subseteq_{S,P,-} B_3$, and this shows that $B_2 \subseteq_{S,P} B_3$.

\paragraph{Idempotence and the absorption property}
The absorption property for SEs $B_1$ and $B_2$ can be characterized via the idempotence of the induced opening and closing operators~\cite{heijmans1995morphological}. Namely, under our setting, $O_B \circ O_B = O_B$ and $C_B \circ C_B = C_B$ for any SE $B$. 
It follows that $O_{B_2} \leq O_{B_1}$ (resp. $C_{B_1} \leq C_{B_2}$) if and only if
\begin{equation*}
O_{B_2} \circ O_{B_1} = O_{B_2} = O_{B_1} \circ O_{B_2}
\quad
(\text{resp. } 
C_{B_2} \circ C_{B_1} = C_{B_2} = C_{B_1} \circ C_{B_2}).    
\end{equation*}
However, verifying these identities directly is nontrivial, as the composition of two openings (or closings) involves multiple erosion and dilation operations. In contrast, the notion of positive and negative properties in Definition~\ref{Definition: Shift Inclusion w.r.t P} provides a more tractable and more practical criterion for verifying the absorption property. 

\paragraph{Shift inclusion as a partial order}
At the end of Section~\ref{Section: Shift Inclusion for Topological Filtration Construction}, we present a key observation, stated as Proposition~\ref{prop. rule star has transitivity}, concerning the relations $\subseteq_{S,P,+}$, $\subseteq_{S,P,-}$, and $\subseteq_{S,P}$. Specifically, Proposition~\ref{prop. rule star has transitivity} establishes that these relations, when defined over the set of all SEs in $\mathbb{Z}^2$, are transitive. This property serves as a powerful tool for constructing shift inclusion sequences and underpins the results presented in Theorem~\ref{Theorem: Rectangular SE sequences are shift inclusion} and Theorem~\ref{Theorem: Star structure SE sequences are shift inclusion}.
\bigskip
\begin{proposition}
\label{prop. rule star has transitivity}
Let $P \subseteq \mathbb{Z}^2$ be an image domain, and let $B_1, B_2, B_3$ be SEs in $\mathbb{Z}^2$. Then the following assertions hold:
\begin{enumerate}[label={\rm (\alph*)}]
\item If $B_1 \subseteq_{S,P,+} B_2$ and $B_2 \subseteq_{S,P,+} B_3$, then $B_1 \subseteq_{S,P,+} B_3$
\item If $B_1 \subseteq_{S,P,-} B_2$ and $B_2 \subseteq_{S,P,-} B_3$, then $B_1 \subseteq_{S,P,-} B_3$
\end{enumerate}
Therefore, the relation $\subseteq_{S,P}$ of SEs is transitive.      
\end{proposition}
\begin{proof}
Because $B \subseteq_{S,P,+} B'$ if and only if $-B \subseteq_{S,P,-} -B'$ for SEs $B \subseteq B'$, it is sufficient to prove assertion (a).  Assume $\bfx \in P$ and $\bfb_3 \in B_3(\bfx;P,+)$. Since $B_2 \subseteq_{S,P,+} B_3$, there exists a $\bfb_2 \in B_2(\bfx;P,+)$ such that $B_2 + (\bfb_3 - \bfb_2) \subseteq B_3$. We define $\bfv = \bfb_3 - \bfb_2$. Also, the assumption of $B_1 \subseteq_{S, P, +} B_2$ implies that there is a $\bfb_1 \in B_1(\bfx;P,+)$ such that $B_1 + (\bfb_2 - \bfb_1) \subseteq B_2$. Similarly, we define $\bfu = \bfb_2 - \bfb_1$. Then, $B_1 + (\bfb_3 - \bfb_1) = B_1 + (\bfb_2 - \bfb_1) + (\bfb_3 - \bfb_2) \subseteq B_2 + (\bfb_3 - \bfb_2) \subseteq B_3$, showing that $B_1 \subseteq_{S,P,+} B_3$.
\end{proof}

Since the standard subset relation $\subseteq$ is a prerequisite for the relations $\subseteq_{S,P,+}$ and $\subseteq_{S,P,-}$, both of these relations inherit reflexivity and antisymmetry from $\subseteq$. As a result, for the set $\mathfrak{B}$ of all SEs in $\mathbb{Z}^2$, the relations $\subseteq_{S,P,+}$ and $\subseteq_{S,P,-}$ qualify as partial orders on $\mathfrak{B}$. Consequently, the pair $(\mathfrak{B}, \subseteq_{S,P})$ forms a partially ordered set.

%%%%%%%%%%%%%%%%%%%%%%%%%%%%%%%%%%%%%%%%%%%%%%
\subsubsection{Feasible Examples}
\label{Section: Feasible Examples}
%%%%%%%%%%%%%%%%%%%%%%%%%%%%%%%%%%%%%%%%%%%%%%

\begin{figure}%
\centering
\includegraphics[width=1\linewidth]{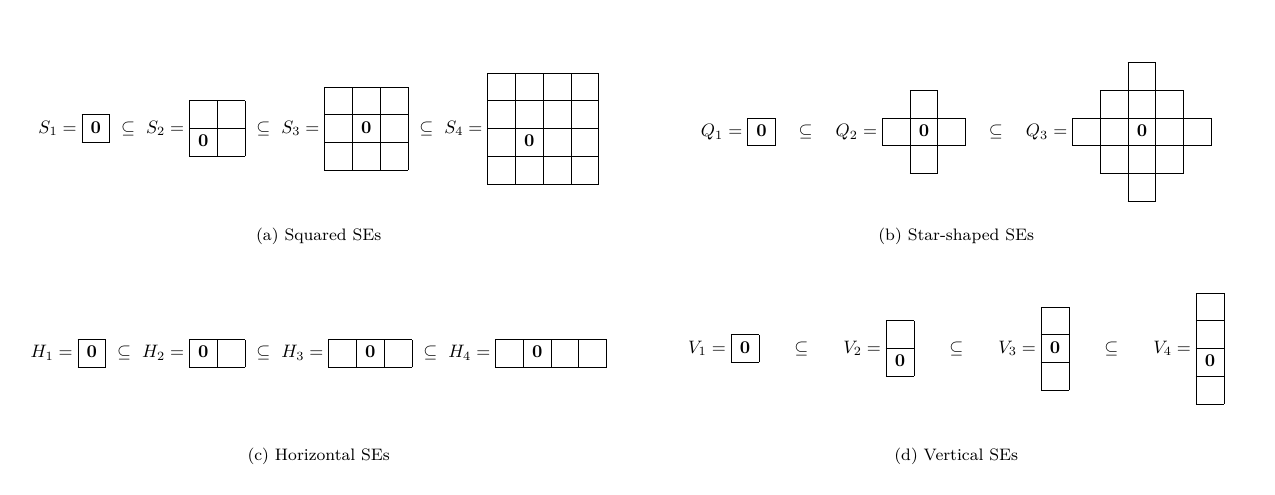}
\caption{Examples of feasible SE sequences satisfying the shift inclusion relation, formed by (a) squares, (b) star-shaped structures, and (c) horizontal and (d) vertical line segments.}
\label{Fig. Examples SE sequences}
\end{figure} 

In Section~\ref{Section: Feasible Examples}, we introduce sequences of SEs that satisfy the shift inclusion relation, referred to as \textit{shift inclusion sequences}.  When the image domain is rectangular, we illustrate several typical examples in $\mathbb{Z}^2$, including sequences of rectangles, squares, and star-shaped structures. Fig.~\ref{Fig. Examples SE sequences} presents their graphical representations via grid diagrams, including sequences formed by squares, star-shaped structures, and horizontal and vertical line segments.   

\paragraph{Rectangular SEs}
Rectangular pixel-based shapes, such as squares and horizontal and vertical segments, are fundamental in image processing. We focus on rectangular SEs in $\mathbb{Z}^2$ containing the origin $\mathbf{0}$ and their nested sequences under inclusion, which form the basis for constructing morphological filtrations and topological descriptors. In particular, Theorem~\ref{Theorem: Rectangular SE sequences are shift inclusion} shows that for a rectangular image domain $P$, any sequence $R_1 \subseteq \cdots \subseteq R_n$ of rectangular SEs forms a shift inclusion sequence. The following Lemma~\ref{Lemma: Shift inclusion lemma} and Theorem~\ref{Theorem: Rectangular SE sequences are shift inclusion} are refined versions of \cite[Lemma~6.2.1 and Theorem~6.2.2]{Hu_PhD_Dissertation}.
\bigskip
\begin{lemma}\label{Lemma: Shift inclusion lemma}
Let $P \subseteq \mathbb{Z}^2$ be a rectangular image domain, and let $B_1, B_2$ be SEs in $\mathbb{Z}^2$. Suppose $B_2 = B_1 \cup (B_1 + \mathbf{v})$ for some $\mathbf{v} \in \mathbb{Z}^2$. If for every $\mathbf{b}_2 \in B_2 \setminus B_1$, the $i$-th coordinate $\pi_i(\mathbf{b}_2 - \mathbf{v})$ of $\mathbf{b}_2 - \mathbf{v}$ lies in the closed interval with endpoints $0 = \pi_i(\mathbf{0})$ and $\pi_i(\mathbf{b}_2)$ for each $i \in \{1, 2\}$, then $B_1 \subseteq_{S,P} B_2$. 
\end{lemma}
\begin{proof}
We first observe the following fact: for any $\mathbf{x}, \mathbf{y} \in P$, where $P \subseteq \mathbb{Z}^2$ is a rectangle, every point $\mathbf{z} \in \mathbb{Z}^2$ such that $\pi_i(\mathbf{z})$ lies in the closed interval between $\pi_i(\mathbf{x})$ and $\pi_i(\mathbf{y})$ for each $i \in \{1, 2\}$ also belongs to $P$.

Suppose $\mathbf{x} = (x_1, x_2) \in P$ and $\mathbf{b}_2 \in B_2(\mathbf{x}; P, +) \setminus B_1$. Let $\mathbf{b}_1 = \mathbf{b}_2 - \mathbf{v}$, then $B_1 + (\mathbf{b}_2 - \mathbf{b}_1) = B_1 + \mathbf{v} \subseteq B_2$. To show that $B_1 \subseteq_{S,P,+} B_2$, it is sufficient to prove that $\mathbf{b}_1 \in B_1(\mathbf{x}; P, +)$. For every $i \in \{ 1, 2 \}$, the $i$-th coordinate of $\mathbf{x} + \mathbf{b}_1$ is
\begin{equation*}
\pi_i(\mathbf{x} + \mathbf{b}_1) = \pi_i(\mathbf{x} + \mathbf{b}_2 - \mathbf{v}) = \pi_i(\mathbf{x}) + \pi_i(\mathbf{b}_2) - \pi_i(\mathbf{v}).  
\end{equation*}
By assumption, the value \( \pi_i(\mathbf{x} + \mathbf{b}_1) \) lies within the closed interval with endpoints \( x_i = \pi_i(\mathbf{x}) \) and \( \pi_i(\mathbf{x}) + \pi_i(\mathbf{b}_2) = \pi_i(\mathbf{x} + \mathbf{b}_2) \). Since \( P \) is a rectangle, \( \mathbf{x} + \mathbf{b}_1 \in P \). Therefore, \( B_1 \subseteq_{S, P, +} B_2 \). On the other hand, we  have \( B_1 \subseteq_{S, P, -} B_2 \) by a symmetric argument. Hence, we conclude that \( B_1 \subseteq_{S, P} B_2 \).
\end{proof}
 
\begin{figure}
    \centering
    \includegraphics[width=0.9\linewidth]{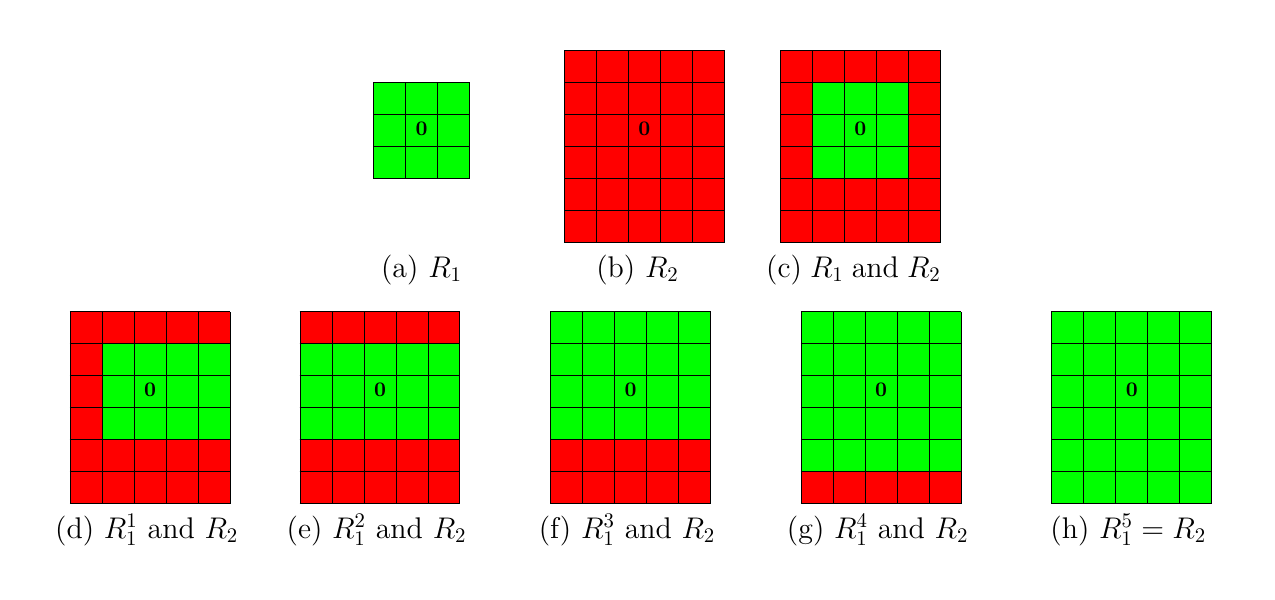}
    \caption{Illustration of rectangular SEs $R_1 = \{ 0, \pm 1 \} \times \{ 0, \pm 1 \}$ and $R_2 = \{ 0, \pm 1, \pm 2 \} \times \{ 0, \pm 1, \pm 2, -3 \}$. As shown in the second row (panels (d)--(h)), the SE $R_2$ can be constructed as a union of translated copies of $R_1$. Explicitly, the construction of $R_2$ can be decomposed step by step into a filtration of SEs: $R_1 \subseteq R_1^1 \subseteq R_1^2 \subseteq R_1^3 \subseteq R_1^4 \subseteq R_1^5 = R_2$, where $R_1^1 = R_1 \cup (R_1 + \mathbf{e}_1)$, $R_1^2 = R_1^1 \cup (R_1^1 - \mathbf{e}_1)$, $R_1^3 = R_1^2 \cup (R_1^2 + \mathbf{e}_2)$, $R_1^4 = R_1^3 \cup (R_1^3 - \mathbf{e}_2)$, and $R_1^5 = R_1^4 \cup (R_1^4 - \mathbf{e}_2)$.}
    \label{fig: Demo_Rectangles_Shifted}
\end{figure}

\begin{theorem}
\label{Theorem: Rectangular SE sequences are shift inclusion}
Let $P \subseteq \mathbb{Z}^2$ be a rectangular image domain, and let $R_1, R_2$ be SEs in $\mathbb{Z}^2$. If $R_1$ and $R_2$ are rectangles and $R_1 \subseteq R_2$, then $R_1 \subseteq_{S,P} R_2$.    
\end{theorem}
\begin{proof}
Assume that \( R_1 = \mathbb{Z}^2 \cap ([a_1, b_1] \times [a_2, b_2]) \) and \( R_2 = \mathbb{Z}^2 \cap ([c_1, d_1] \times [c_2, d_2]) \), where \( a_i, b_i, c_i, d_i \in \mathbb{Z} \) satisfy \( c_i \leq a_i \leq b_i \leq d_i \) for each \( i \in \{1, 2\} \). Let $\{ \mathbf{e}_1, \mathbf{e}_2 \}$ be the standard $\mathbb{Z}$-basis of $\mathbb{Z}^2$, then $R_1 \pm \mathbf{e}_1 = \mathbb{Z}^2 \cap ([a_1 \pm 1, b_1 \pm 1] \times [a_2, b_2])$ and $R_1 \pm \mathbf{e}_2 = \mathbb{Z}^2 \cap ([a_1, b_1] \times [a_2 \pm 1, b_2  \pm 1])$. Therefore, based on the rectangular structures of \( R_1 \) and \( R_2 \), there exists a nested sequence of rectangles \( R_1^0 \subseteq R_1^1 \subseteq R_1^2 \subseteq \cdots \subseteq R_1^n \) such that $R_1 = R_1^0 \subseteq R_1^1 \subseteq R_1^2 \subseteq \cdots \subseteq R_1^n = R_2$, where each \( R_1^{k+1} \) is formed from \( R_1^k \) by expanding in one of the coordinate directions:
\begin{equation*}
R_1^{k+1} = R_1^k \cup (R_1^k + \mathbf{e}_j) \quad \text{or} \quad R_1^{k+1} = R_1^k \cup (R_1^k - \mathbf{e}_j),    
\end{equation*}
for some \( j \in \{1, 2\} \) and for every \( k \in \{0, 1, \ldots, n-1\} \). According to Proposition~\ref{prop. rule star has transitivity}, the proof of the theorem is complete if we can show that \( R_1^k \subseteq_{S, P} R_1^{k+1} \) for every \( k \in \{0, 1, \ldots, n-1\} \). Since \( P \), \( R_1^k \), and \( R_2 \) are rectangles and \( \mathbf{0} \in R_1 \), the conditions of Lemma~\ref{Lemma: Shift inclusion lemma} are satisfied. Therefore, we have \( R_1^k \subseteq_{S, P} R_1^{k+1} \) for each \( k \in \{0, 1, \ldots, n-1\} \). Consequently, the theorem follows.   
\end{proof}
Fig.~\ref{fig: Demo_Rectangles_Shifted} illustrates the SE sequence $R_1^0 \subseteq R_1^1 \subseteq \cdots \subseteq R_1^n$, where $R_1^0 = R_1$ and $R_1^n = R_2$, as constructed in the proof of Theorem~\ref{Theorem: Rectangular SE sequences are shift inclusion}. 
Each inclusion step satisfies the conditions of Lemma~\ref{Lemma: Shift inclusion lemma}, ensuring that the sequence forms a valid shift inclusion sequence with respect to the rectangular domain $P$.  Building on this result, we present examples of shift inclusion sequences formed by representative rectangular SEs, such as \textit{squares}, \textit{horizontal lines}, and \textit{vertical lines}, which serve as fundamental building blocks for constructing topological filtrations.

\begin{example}[Squares]
\label{Example: Square SE sequences are shift inclusion}
For every \( n \in \mathbb{N} \), the \textbf{squared structuring elements} in \( \mathbb{Z}^2 \) are defined by the recursive formula
\begin{equation*}
\begin{split}
    S_1 &= \{ (0,0) \},\\ 
    S_n &= \left\{ \begin{array}{ll}
S_{n-1} \cup (S_{n-1} + \bfe_1) \cup (S_{n-1} + \bfe_2) \cup (S_{n-1} + \bfe_1 + \bfe_2) & \mbox{if $n$ is even}\\
S_{n-1} \cup (S_{n-1} - \bfe_1) \cup (S_{n-1} - \bfe_2) \cup (S_{n-1} - \bfe_1 - \bfe_2) & \mbox{if $n$ is odd}
\end{array}\right.,    
    \end{split}
\end{equation*}
In particular, every $S_n$ is an $n \times n$ rectangle that contains the origin $\mathbf{0} = (0,0)$.  
\end{example}
\bigskip
Fig.~\ref{Fig. Examples SE sequences}(a) illustrates the square SEs $S_1, S_2, S_3, S_4$ using grid diagrams. Since each $S_n$ is a rectangle in $\mathbb{Z}^2$, the sequence $S_1 \subseteq S_2 \subseteq S_3 \subseteq S_4$ forms a shift inclusion sequence by Theorem~\ref{Theorem: Rectangular SE sequences are shift inclusion}. These square SEs are widely used in standard MM libraries, including the \texttt{strel} function in MATLAB and the \texttt{cv2} module in OpenCV. Furthermore, additional rectangular SEs are employed in the firn data analysis in Section~\ref{Section: Firn Data Analysis}, as their simple structures reduce computational complexity while preserving geometric significance. In particular, we focus on \textit{horizontal} and \textit{vertical lines} (see Fig.~\ref{Fig. Examples SE sequences}(c) and (d)), defined as follows.
\bigskip
\begin{example}[Horizontal and vertical lines]
\label{Example: Horizontal and vertical lines}
For every positive integer \( n \in \mathbb{N} \), the \textbf{n-th horizontal line} in \( \mathbb{Z}^2 \) is defined by the recursive formula
\begin{equation*}
\begin{split}
H_1 &= \{ (0,0) \},\\ 
H_n &= \left\{ \begin{array}{ll}
H_{n-1} \cup (H_{n-1} + \bfe_1) & \mbox{if $n$ is even}\\
H_{n-1} \cup (H_{n-1} - \bfe_1) & \mbox{if $n$ is odd}
\end{array}\right.,    
\end{split}
\end{equation*}
Alternatively, the \textbf{n-th vertical line} in \( \mathbb{Z}^2 \) is defined by the recursive formula
\begin{equation*}
\begin{split}
V_1 &= \{ (0,0) \},\\ 
V_n &= \left\{ \begin{array}{ll}
V_{n-1} \cup (V_{n-1} + \bfe_2) & \mbox{if $n$ is even}\\
V_{n-1} \cup (V_{n-1} - \bfe_2) & \mbox{if $n$ is odd}
\end{array}\right. .    
\end{split}
\end{equation*}
Each horizontal (resp. vertical) line corresponds to a rectangle consisting of a single row (resp. column). Consequently, any nested sequence of such SEs forms a shift inclusion sequence. 
\end{example}
\bigskip
Geometrically, opening (resp. closing) operations with horizontal and vertical SEs detect horizontal and vertical distances between connected black (resp. white) components via the induced PB. Moreover, the size of holes enclosed by black pixels can be inferred from the birth--death information in the PB arising from the opening filtration. Further geometric interpretations are discussed in Section~\ref{Subsubsec: On Absorption Property}.

\paragraph{Star-shaped SEs}
Beyond rectangular shapes, to construct more diverse types of SEs, we first introduce the following lemma, which presents an important observation concerning the union property of shift-included SEs.
\bigskip
\begin{lemma}\label{Lemma: Shift inclusion and union lemma}
Let $P \subseteq \mathbb{Z}^2$ an image domain, and let $B, B_1, B_2$ be SEs in $\mathbb{Z}^2$. If $B \subseteq_{S,P} B_1$ and $B \subseteq_{S,P} B_2$, then $B \subseteq_{S,P} B_1 \cup B_2$.   
\end{lemma}
\begin{proof}
Let $C = B_1 \cup B_2$. Suppose $\mathbf{x} \in P$ and $\mathbf{c} \in C(\mathbf{x}; P, +) \subseteq C$, then either $\mathbf{c} \in B_1$ or $\mathbf{c} \in B_2$. Assume $\mathbf{c} \in B_1$. Then $\mathbf{c} \in B_1(\mathbf{x}; P, +)$. Since \( B \subseteq_{S, P, +} B_1 \), there is a $\mathbf{b} \in B(\mathbf{x}; P, +)$ such that $B + (\mathbf{c} - \mathbf{b}) \subseteq B_1 \subseteq C$. This proves that $B \subseteq_{S, P, +} C$. A symmetric argument yields $B \subseteq_{S, P, -} C$, and hence $B \subseteq_{S, P} C$.
\end{proof}
Viewing $\mathbb{Z}^2$ as a metric space equipped with the $\ell^1$-norm $\|\cdot\|_1$, where $\|(a,b)\|_1 = |a| + |b|$ for $(a,b) \in \mathbb{Z}^2$, a \textit{star-shaped structuring element} is defined as the closed ball centered at the origin $\mathbf{0}$ with a fixed radius, i.e.,
\bigskip
\begin{definition}
Let $\omega \in \mathbb{R}$ with $\omega \geq 1$. 
The \textbf{star-shaped structuring element} of radius $\omega$, denoted by $Q_{\omega}$, is defined by
\begin{equation*}
Q_{\omega} = \{ \mathbf{x} \in \mathbb{Z}^2 \mid \|\mathbf{x}\|_1 \leq \omega - 1 \}
= \{ (a,b) \in \mathbb{Z}^2 \mid |a| + |b| \leq \omega - 1 \}.    
\end{equation*}
Each $Q_{\omega}$ is a finite subset of $\mathbb{Z}^2$ containing $\mathbf{0}$, and hence defines a symmetric SE.
\end{definition}
\bigskip
By definition, \( Q_{\omega_1} \subseteq Q_{\omega_2} \) whenever \( \omega_1 \leq \omega_2 \), which induces an SE sequence derived from an increasing sequence of radii. Furthermore, similar to the recursive formulations for square, horizontal, and vertical SEs, the star-shaped SE \( Q_\omega \) can also be represented by the following recursive equation:
\begin{equation}\label{Eq. star structures SEs, recursive formula}
\begin{split}
Q_1 &= \{ (0,0) \},\\ 
Q_{\omega+1} &= Q_{\omega} \cup (Q_{\omega} + \mathbf{e}_1) \cup (Q_{\omega} - \mathbf{e}_1) \cup (Q_{\omega} + \mathbf{e}_2) \cup (Q_{\omega} - \mathbf{e}_2).   
\end{split}
\end{equation}
By leveraging the observations in Lemmas~\ref{Lemma: Shift inclusion lemma} and~\ref{Lemma: Shift inclusion and union lemma}, the validity of the relation \( Q_\omega \subseteq_{S,P} Q_{\omega+1} \), where \( P \) is a rectangular image domain, can be established through the following lemma and theorem.
\bigskip
\begin{lemma}\label{Lamma: Q S,P Q cup (Q+e1)}
Let \( P \subseteq \mathbb{Z}^2 \) be a rectangular image domain, and let \( \omega \in \mathbb{Z}_{\geq 0} \) be a non-negative integer. Then, for every \( i \in \{1, 2\} \), the SEs \( B_1 = Q_\omega \) and \( B_2 = Q_\omega \cup (Q_\omega \pm \mathbf{e}_i) \) satisfy the assumptions of Lemma~\ref{Lemma: Shift inclusion lemma}. In particular, \( B_1 \subseteq_{S, P} B_2 \).    
\end{lemma}
\begin{proof}
By a symmetric argument,  it is sufficient to consider the case of $i = 1$, $B_1 = Q_\omega$, and $B_2 = Q_\omega \cup (Q_\omega + \mathbf{e}_1)$. Let $\mathbf{b}_2 \in B_2 \setminus B_1$, then $\mathbf{b}_2 = \mathbf{b}_1 + \mathbf{e}_1$ for some $\mathbf{b}_1 \in B_1$. Representing $\mathbf{b}_1 = (x_1, y_1)$, we have $\mathbf{b}_2 = (x_1 + 1, y_1)$. Because $\mathbf{b}_2 \in B_2 \setminus B_1$, $\Vert \mathbf{b}_2 \Vert_1 = |x_1 + 1| + |x_2| > \omega$, and it forces that $x_1 \geq 0$ as $\Vert \mathbf{b}_1 \Vert_1 = |x_1| + |y_1| \leq \omega$. In particular,  $\pi_1(\mathbf{b}_1) = x_1 \in [0, x_1+1] = [\pi_1(0), \pi_1(\mathbf{b}_2)]$ and $\pi_2(\mathbf{b}_1) = \pi_2(\mathbf{b}_2)$. Therefore,  the assumptions of Lemma~\ref{Lemma: Shift inclusion lemma} is satisfied, and we conclude that \( B_1 \subseteq_{S, P} B_2 \).      
\end{proof}
\bigskip
\begin{theorem}\label{Theorem: Star structure SE sequences are shift inclusion}
Let \( P \subseteq \mathbb{Z}^2 \) be a rectangle. Then \( Q_\omega \subseteq_{S,P} Q_{\omega+1} \) for all \( \omega \in \mathbb{Z}_{\geq 0} \).     
\end{theorem}
\begin{proof}
By  \eqref{Eq. star structures SEs, recursive formula}, Lemma \ref{Lamma: Q S,P Q cup (Q+e1)}, and Lemma \ref{Lemma: Shift inclusion and union lemma}, the theorem follows.    
\end{proof}

%%%%%%%%%%%%%%%%%%%%%%%%%%%%%%%%%%%%%%%%%%%%%%  
\subsubsection{On Absorption Property}
\label{Subsubsec: On Absorption Property}
%%%%%%%%%%%%%%%%%%%%%%%%%%%%%%%%%%%%%%%%%%%%%%
This section presents the main theoretical results, given in Theorems~\ref{Theorem: Shift Inclusion as a sufficient condition} and~\ref{Theorem: Shift Inclusion as a sufficient condition, strong form}. Building on the notion of shift inclusion introduced in Section~\ref{Section: Shift Inclusion for Topological Filtration Construction}, we show that any shift inclusion sequence $B_1 \subseteq_{S,P} \cdots \subseteq_{S,P} B_n$ induces an increasing opening filtration $\mathcal{X}_{O_{B_1}(g)} \subseteq \cdots \subseteq \mathcal{X}_{O_{B_n}(g)}$ and a decreasing closing filtration $\mathcal{X}_{C_{B_n}(g)} \subseteq \cdots \subseteq \mathcal{X}_{C_{B_1}(g)}$, thereby providing MM-based filtrations for persistent homology. These results are refined versions of \cite[Theorem~6.3.2 and Corollary~6.3.3]{Hu_PhD_Dissertation}.
\bigskip
\begin{theorem}[Main theorem, first form]
\label{Theorem: Shift Inclusion as a sufficient condition}
Let $P \subseteq \mathbb{Z}^2$ be an image domain, and let $B_1, B_2$ be SEs in $P \subseteq \mathbb{Z}^2$. Then the following statements hold:
\begin{itemize}
\item[$({\rm a})$] If $B_1 \subseteq_{S,P,-} B_2$, then $\{ O_{B_1}(g) = 0 \} \subseteq \{ O_{B_2}(g) = 0 \}$ whenever $g \in \mathcal{I}_P$.
\item[$({\rm b})$] If $B_1 \subseteq_{S,P,+} B_2$, then $\{ C_{B_2}(g) = 0 \} \subseteq \{ C_{B_1}(g) = 0 \}$ whenever $g \in \mathcal{I}_P$.
\end{itemize}
In particular, if $B_1 \subseteq_{S, P} B_2$, then
\begin{equation*}
O_{B_1}(g)^{-1}(0) \subseteq O_{B_2}(g)^{-1}(0) \quad \text{and} \quad C_{B_2}(g)^{-1}(0) \subseteq C_{B_1}(g)^{-1}(0)
\end{equation*}
for every image $g \in \mathcal{I}_P$, which is exactly the desired absorption property.
\end{theorem}
\begin{proof}
As shown in the proof of Lemma~\ref{Lemma: crucial observation for shift inclusion} and in the discussion preceding Definition~\ref{Definition: Shift Inclusion w.r.t P}, assertion~(a) holds, while assertion~(b) follows by a dual argument.
\end{proof} 
Theorem~\ref{Theorem: Shift Inclusion as a sufficient condition} establishes that $B_1 \subseteq_{S,P} B_2$ implies $O_{B_1}(g)^{-1}(0) \subseteq O_{B_2}(g)^{-1}(0)$ and $C_{B_2}(g)^{-1}(0) \subseteq C_{B_1}(g)^{-1}(0)$ for every image $g$ on $P$. Furthermore, by applying Corollary~\ref{Corollary: Opening and Closing ordering relations}, Theorem~\ref{Theorem: Shift Inclusion as a sufficient condition} can be reformulated as Theorem~\ref{Theorem: Shift Inclusion as a sufficient condition, strong form}, which corresponds to the stronger property Equation~\eqref{Equation: Absorption property}.
\bigskip
\begin{theorem}[Main theorem, second form]
\label{Theorem: Shift Inclusion as a sufficient condition, strong form}
Let $P \subseteq \mathbb{Z}^2$ be an image domain, and let $B_1, B_2$ be SEs in $P \subseteq \mathbb{Z}^2$. Then the following statements hold:
\begin{itemize}
\item[$({\rm a})$] If $B_1 \subseteq_{S,P,-} B_2$, then $O_{B_2} \leq O_{B_1}$.
\item[$({\rm b})$] If $B_1 \subseteq_{S,P,+} B_2$, then $C_{B_1} \leq C_{B_2}$.
\end{itemize}
\end{theorem}
\begin{proof}
For (a), by Theorem~\ref{Theorem: Shift Inclusion as a sufficient condition}, if $B_1 \subseteq_{S,P,-} B_2$, then $O_{B_1}(g)^{-1}(0) \subseteq O_{B_2}(g)^{-1}(0)$ for every $g \in \mathcal{I}_P$. In particular, $O_{B_1}(g)^{-1}(0) \subseteq O_{B_2}(g)^{-1}(0)$ whenever $g \in \mathcal{BI}_P$, i.e., $O_{B_2}(g) \leq O_{B_1}(g)$ whenever $g \in \mathcal{BI}_P$. By Corollary~\ref{Corollary: Opening and Closing ordering relations}, $O_{B_2} \leq O_{B_1}$. For (b), if $B_1 \subseteq_{S,P,+} B_2$, then $-B_1 \subseteq_{S,P,-} -B_2$. By (a), we have $O_{-B_2} \leq O_{-B_1}$. By Proposition \ref{Proposition: O2 <= O1 iff C-1 <= C-2} and Corollary~\ref{Corollary: Opening and Closing ordering relations}, we deduce that $C_{B_1} \leq C_{B_2}$.
\end{proof}

Theorem~\ref{Theorem: Shift Inclusion as a sufficient condition, strong form} shows that shift inclusion is a sufficient condition for the absorption property. However, as illustrated in Example~\ref{ExampleA: desired property doesn't imply subseteq_S,P}, it is not necessary; in general, the two properties are not equivalent. Nevertheless, when $P = \mathbb{Z}^2$, Theorem~\ref{TheoremA: Shift Inclusion over the entire space is equivalent to the absorption property} shows that they become equivalent.

%%%%%%%%%%%%%%%%%%%%%%%%%%%%%%%%%%%%%%%%%%%%%%
\subsubsection{Comparison to Existing Results}
\label{Section: Comparison to Existing Results}
%%%%%%%%%%%%%%%%%%%%%%%%%%%%%%%%%%%%%%%%%%%%%%
At the end of Section~\ref{Section: Shift Inclusion for Topological Filtration Construction}, we discuss how the proposed shift inclusion framework relates to existing results on the absorption property of opening and closing operations. In this setting, classical constructions and the verification of the openness property for SE sequences can be viewed as a special case of shift inclusion under $\subseteq_{S,\mathbb{Z}^2}$ (see, e.g., Theorem~\ref{Decomposition theorem of shift inclusion}).

Specifically, by inductively generating \textit{periodic line segments} (cf.~\cite{jonesSoille1996}) through Minkowski addition in $\mathbb{Z}^2$, one obtains a sequence $D_1, D_2, ...$ of discrete disks. This construction ensures that the induced opening and closing operations satisfy the absorption property. Specifically, by defining the periodic line
\begin{equation*}
P_{n,\mathbf{v}} = \bigcup_{i = 0}^{n-1}~\{ i\mathbf{v} \} 
\end{equation*}
for a given $n \in \mathbb{N}$ and $\mathbf{v} \in \mathbb{Z}^2$, one can construct the following SE sequence of discrete disks, which satisfies the absorption property:
\begin{equation*}
\begin{split}
D_1 &= P_{2,\mathbf{e}_1} + P_{2,\mathbf{e}_2}, \\ 
D_2 &= P_{2,\mathbf{e}_1} + P_{2,\mathbf{e}_2} + P_{2,\mathbf{e}_1 + \mathbf{e}_2} + P_{2,\mathbf{e}_1 - \mathbf{e}_2}, \\
D_3 &= P_{3,\mathbf{e}_1} + P_{3,\mathbf{e}_2} + P_{2,\mathbf{e}_1 + \mathbf{e}_2} + P_{2,\mathbf{e}_1 - \mathbf{e}_2},
\end{split}    
\end{equation*}
and so on \cite[Fig.~11.8]{soille2013}. In particular, since $P_{n,\mathbf{v}} = P_{2,\mathbf{v}} + P_{n-1,\mathbf{v}}$ for all $n \geq 2$, one obtains
\begin{equation*}
\begin{split}
D_2 &= D_1 + P_{2,\mathbf{e}_1 + \mathbf{e}_2} + P_{2,\mathbf{e}_1 - \mathbf{e}_2}, \\
D_3 &= P_{2,\mathbf{e}_1} + P_{1,\mathbf{e}_1} + P_{2,\mathbf{e}_2} + P_{1,\mathbf{e}_2} + P_{2,\mathbf{e}_1 + \mathbf{e}_2} + P_{2,\mathbf{e}_1 - \mathbf{e}_2} = D_2 + P_{1,\mathbf{e}_1} + P_{1,\mathbf{e}_2}.
\end{split}    
\end{equation*}
In particular, $D_2$ and $D_3$ are translated copies of $D_1$. In the context of shift inclusion, we have $D_1 \subseteq_{S,\mathbb{Z}^2} D_2 \subseteq_{S,\mathbb{Z}^2} D_3$ by Theorem \ref{Decomposition theorem of shift inclusion}, showing that $O_{D_3} \leq O_{D_2} \leq O_{D_1}$ and $C_{D_1} \leq C_{D_2} \leq C_{D_3}$ by Theorem \ref{Theorem: Shift Inclusion as a sufficient condition, strong form}.

More generally, the openness property of an SE sequence $\{ B_i \}_{i = 1}^n$ can also be employed to verify whether the absorption property holds for the induced opening and closing operations on images defined over the entire lattice (e.g., $\mathbb{Z}^m$)~\cite{heijmans1994MorImgOperators}. Mathematically, an SE $B \subseteq \mathbb{Z}^2$ is called $A$\textit{-open} with respect to an SE $A \subseteq \mathbb{Z}^2$ if 
\begin{equation*}
B = A + X = \bigcup_{\mathbf{x} \in X}~\mathbf{x} + A   
\end{equation*}
for some SE $X \subseteq \mathbb{Z}^2$~\cite[Proposition~4.21]{heijmans1994MorImgOperators}. Within the proposed shift inclusion framework, $B$ is an $A$-open SE if and only if 
\begin{equation*}
A \subseteq_{S,\mathbb{Z}^2} B.
\end{equation*}
In particular, the induced sequences $\{ O_{B_i} \}$ and $\{ C_{B_i} \}$ satisfy the absorption property if $B_{i+1}$ is $B_i$-open for $i = 1,2,\ldots,n-1$, that is, $B_1 \subseteq_{S,\mathbb{Z}^2} B_2 \subseteq_{S,\mathbb{Z}^2} \cdots \subseteq_{S,\mathbb{Z}^2} B_n$~\cite[Proposition~4.22]{heijmans1994MorImgOperators}. This result is consistent with \cite[Proposition~4.21]{heijmans1994MorImgOperators} in the special case $P = \mathbb{Z}^2$.

However, although the openness property of an SE sequence ensures the absorption property in $\mathbb{Z}^2$, it does not extend to images defined on general domains $P \subseteq \mathbb{Z}^2$ (see, e.g., Example~\ref{ExampleA: B_1 subseteq_S,Zm B_2 doesn't imply B_1 subseteq_S,P B_2-2}). To address this, we consider the relation $\subseteq_{S,P}$ for arbitrary $P$, under which Theorems~\ref{Theorem: Shift Inclusion as a sufficient condition, strong form} and \ref{Theorem: Shift Inclusion as a sufficient condition} establish shift inclusion as a systematic criterion for verifying absorption. Furthermore, the weak shift inclusion $\subseteq_{WS,P}$ (Section~\ref{Section: Weak Shift Inclusion}) provides an equivalent condition, offering a robust theoretical foundation for investigating the absorption property of opening and closing operations on general image domains. Together, the relations $\subseteq_{S,P}$ and $\subseteq_{WS,P}$ support the construction of PH-based filtrations and provide a solid framework for advancing image-based PH research within the TDA community. 

\subsection{Firn Data Analysis}\label{Section: Firn Data Analysis}

Firn is a porous material intermediate between snow and glacial ice, composed of interconnected ice grains and pore spaces that evolve with depth and time through compaction, metamorphism, and recrystallization. Quantifying this evolving microstructure is essential for understanding firn densification, air and vapor transport, and other processes central to glaciological applications, including the interpretation of climate records from polar ice cores. Micro-computed tomography (micro-CT) provides a stack of two-dimensional cross-sectional images, or \textit{slices}, that together represent the three-dimensional microstructure of firn. These data enable direct analysis of geometric and topological features such as grain connectivity, pore structure, and interfacial complexity. Here, we demonstrate the utility of using morphological PH on firn micro-CT data to quantify and define the microstructures of specific layers in the firn stratigraphy. %, and effectively remove noisy features caused bys thresholding to reveal the original microstructure for analysis.

\begin{figure}
\centering
\includegraphics[width=\linewidth]{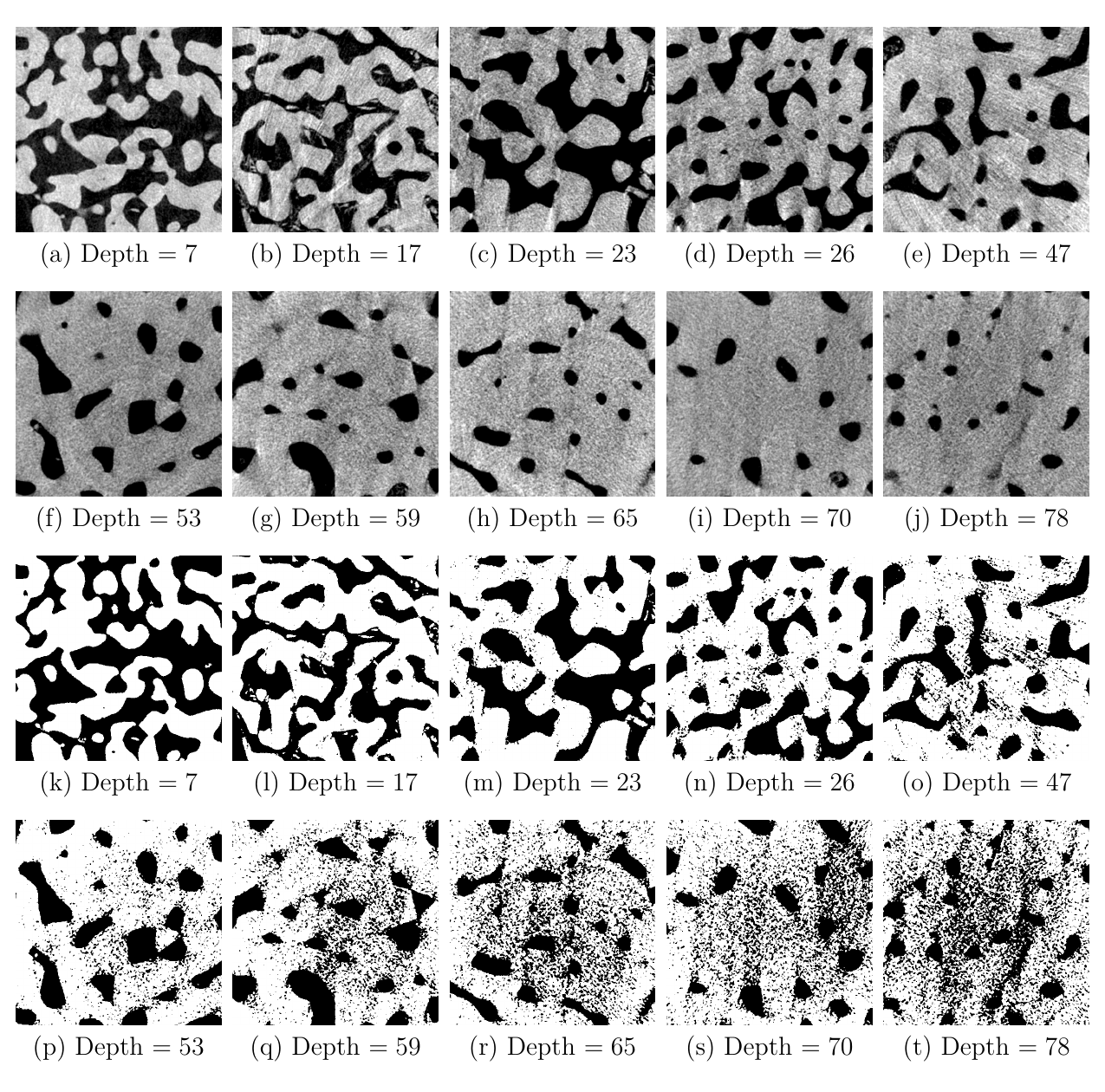}
\caption{Sampled micro-CT firn images from different layers at various depths of the NEEM firn layer. (a)--(j): Sampled images at depths \( 7, 17, 23, 26, 47, 53, 59, 65, 70, \) and \( 78 \) m. Each firn image is a grayscale image of size \( 500 \times 500 \) pixels, with pixel values ranging from \( 0 \) to \( 255 \).(k)--(t): Corresponding binary images for the same depth levels as in (a)--(j), where each image is obtained by applying a thresholding operation, where the threshold value is set to the average pixel value of the corresponding grayscale image.}
\label{Fig. Sampled firn image with depth information}
\end{figure}

\subsubsection{Dataset}\label{Section: Dataset}

The firn dataset used in this study originates from a firn core drilled at the NEEM Drilling Camp, Greenland, in 2009. From the main core, $1 \times 1 \times 1.5$ cm samples were extracted at selected depths and scanned using micro-computed tomography (micro-CT). The reconstruction of the micro-CT data yielded stacks of approximately 900 cross-sectional images per sample, representing their full volumes. Each cross-sectional image has a resolution of approximately $500 \times 500$ pixels, with a pixel size of 15~\textmu m. In this study, we analyze samples taken from depths of 7, 17, 23, 26, 47, 53, 59, 65, 70, and 78 meters, covering the full depth range of the NEEM firn layer.

Fig.~\ref{Fig. Sampled firn image with depth information} presents a series of sampled micro-CT firn images taken from layers at various depths, ranging from 7 to 78 m. Specifically, images (a)--(j) show the original grayscale cross-sections corresponding to increasing depth levels, while images (k)--(t) display the corresponding binarized images, thresholded using the average pixel values of the original grayscale images. These visualizations highlight the topological evolution and increasing fragmentation of the pore spaces (\textit{black pixels}) within the ice matrix (\textit{gray pixels}) with depth.

\begin{figure}
\centering
\includegraphics[width=\linewidth]{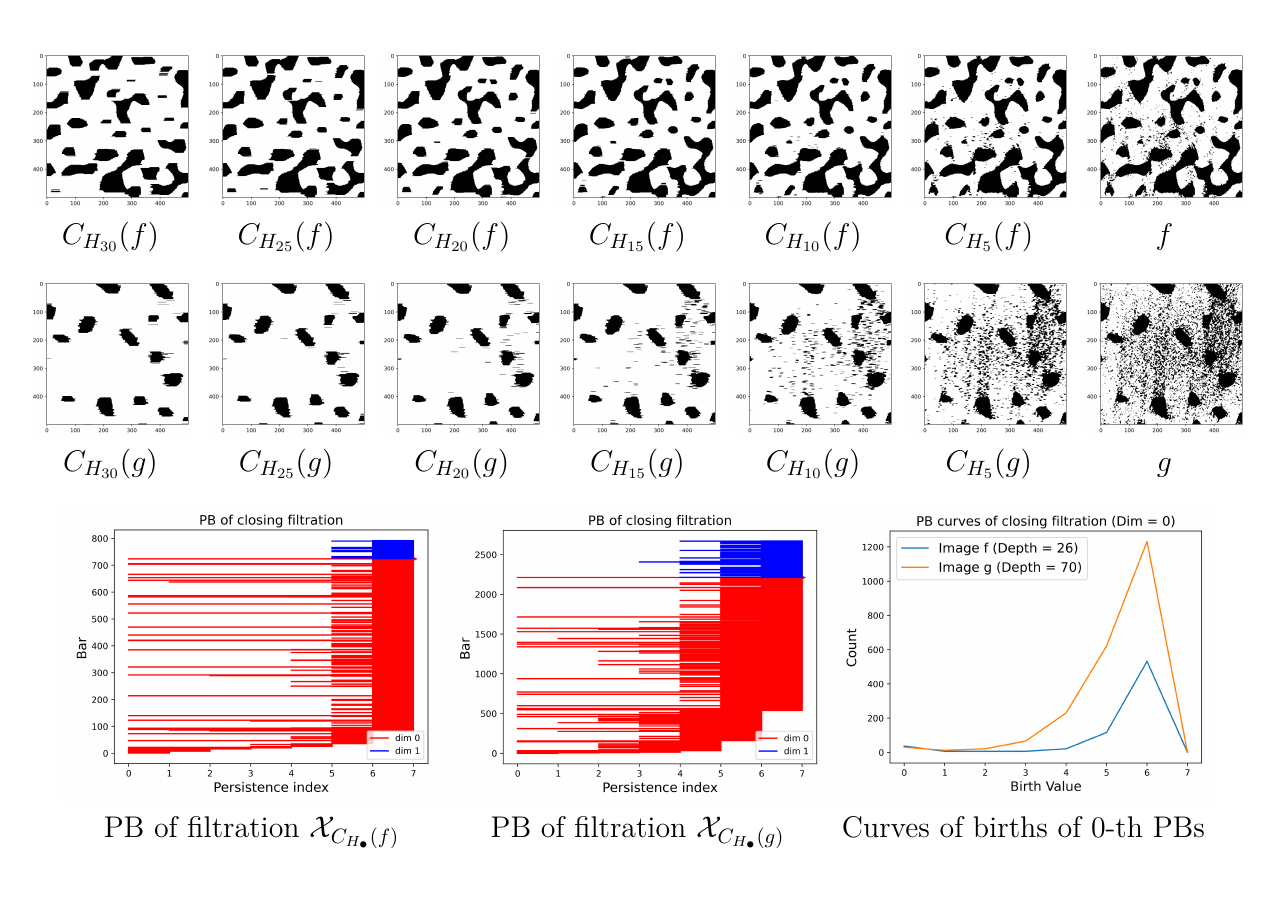}
\caption{Sampled micro-CT firn images $f$ and $g$ from layers at depths 26 and 70 m, respectively, and examples of closing filtrations based on horizontal SEs $H_5 \subseteq H_{10} \subseteq \cdots \subseteq H_{30}$ (see, e.g., Example \ref{Example: Horizontal and vertical lines}). Black pixels indicate the included regions. The first and second rows represent the topological filtrations $\mathcal{X}_{C_{H_{30}}(f)} \subseteq \mathcal{X}_{C_{H_{25}}(f)} \subseteq \cdots \subseteq \mathcal{X}_{C_{H_{5}}(f)} \subseteq \mathcal{X}_f$ and $\mathcal{X}_{C_{H_{30}}(g)} \subseteq \mathcal{X}_{C_{H_{25}}(g)} \subseteq \cdots \subseteq \mathcal{X}_{C_{H_{5}}(g)} \subseteq \mathcal{X}_g$, respectively. The third row shows the induced PBs along with the birth-value curves of the \( 0 \)-th PBs.}
\label{Fig. closing filtrations and curves}
\end{figure}

\subsubsection{Firn Layer Analysis Based on Closing Filtrations}
\label{Section: Firn Layer Analysis Based on Closing Filtrations}

Based on the topological structures of thresholded firn images, we demonstrate how the closing filtration and its PBs capture topological changes across different layers. In particular, the resulting PH and PBs reveal distinctive spatial features of the firn stratigraphy. As illustrated in Fig.~\ref{Fig. Sampled firn image with depth information}, sampled images at greater depths typically contain more scattered black pixel regions, corresponding to connected components in the associated cubical complexes and captured by the \(0\)-th Betti number \( \beta_0 \).

In order to extract the distribution of black fragments, we utilize a closing filtration along with its corresponding PB and compute the induced PH and PB. As demonstrated in Fig.~\ref{Fig. closing filtrations and curves}, for a binarized image obtained from a sampled cross-section at a given depth, such as the image \( f \) shown in Fig.~\ref{Fig. closing filtrations and curves}, a closing filtration of cubical complexes $\mathcal{X}_f \supseteq \mathcal{X}_{C_{B_1}(f)} \supseteq \cdots \supseteq \mathcal{X}_{C_{B_n}(f)}$ can be constructed from any shift inclusion sequence \( B_1 \subseteq B_2 \subseteq \cdots \subseteq B_n \). 

Note that the inclusion direction of the complexes is reversed relative to the ordering of the SEs, since the closing operation focuses on the geometry of black pixel regions, and larger SEs result in greater smoothing of these features. By re-indexing the closing filtration $\mathcal{X}_{C_{B_n}(f)} \subseteq \mathcal{X}_{C_{B_{n-1}}(f)} \subseteq \cdots \subseteq \mathcal{X}_{C_{B_1}(f)} \subseteq \mathcal{X}_f$ as $\mathcal{Y}_0 \subseteq \mathcal{Y}_1 \subseteq \cdots \subseteq \mathcal{Y}_{n-1} \subseteq \mathcal{Y}_{n} = \mathcal{X}_f$ with $\mathcal{Y}_{i} = \mathcal{X}_{C_{B_{n-i}}(f)}$, the re-defined birth and death information can be obtained based on the PH $\operatorname{PH}(\mathcal{Y}_\bullet)$.  For simplicity, we still denote the barcode $\operatorname{PB}(\mathcal{Y}_\bullet)$ by \( \operatorname{PB}(\mathcal{X}_{C_{B_\bullet}(f)}) \). 

Based on this PB, the birth values of the persistence intervals reflect the appearance of connected components formed by black pixels. In particular, the curve \( \gamma_f: \{ 0, 1, \ldots, 5 \} \rightarrow \mathbb{Z}_{\geq 0} \), referred to as the \textit{birth value curve}, is formally defined as
\begin{equation}\label{Eq. birth value curves}
\gamma_f(b) = \# \left\{ (b, d) \in \operatorname{PB}_0(\mathcal{X}_{C_{B_\bullet}(f)}) \right\},
\end{equation}
which counts the number of persistence intervals with birth value \( b \) in the 0-dimensional barcode of the closing filtration \( \mathcal{X}_{C_{B_\bullet}(f)} \). This curve effectively captures the distribution of black pixel fragments across multiple scales, as determined by the underlying SEs. 

Fig.~\ref{Fig. closing filtrations and curves} presents two sampled binary firn images from depths 26 and 70 m. In this demonstration, the chosen SE sequence \( B_1 \subseteq B_2 \subseteq \cdots \subseteq B_6 \) consists of horizontal lines \( H_\bullet \), with each element defined as \( B_n = H_{5n} \) (see Example~\ref{Example: Horizontal and vertical lines}). The induced 0-dimensional PBs and their corresponding birth value curves are depicted in the third row of Fig.~\ref{Fig. closing filtrations and curves}.

\begin{figure}
\centering
\includegraphics[width=\linewidth]{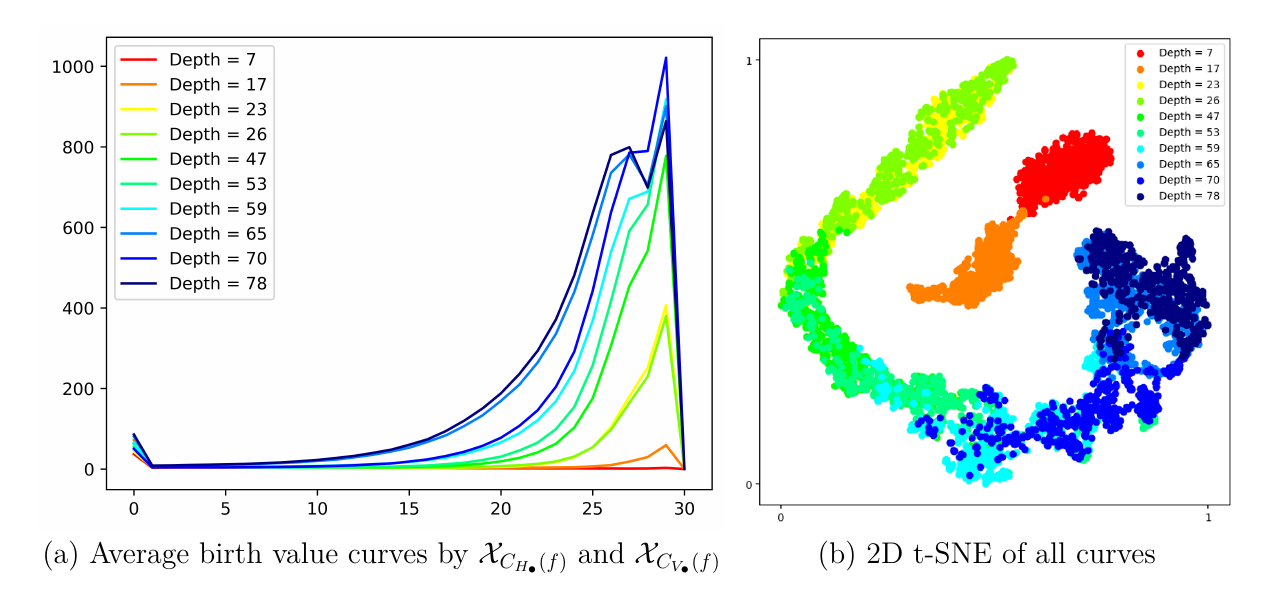}
\caption{(a) Demonstration of the average birth value curves of firn images from 10 different depth categories. Each curve is obtained by averaging 500 birth value curves computed from binarized firn images, where each individual curve is the sum of the birth value curves derived from \( \operatorname{PB}_0(\mathcal{X}_{C_{H_\bullet}(f)}) \) and \( \operatorname{PB}_0(\mathcal{X}_{C_{V_\bullet}(f)}) \).  (b) 2D t-SNE visualization of the full set of 5000 birth value curves. The curves are projected and rescaled into the domain \( [0,1] \times [0,1] \subseteq \mathbb{R}^2 \) for clustering and visualization purposes. }
\label{Fig. tSNE demonstration}
\end{figure}
 
\paragraph{Analysis based on the birth value curves}

To investigate the distributional characteristics of firn microstructures across different depths and to mitigate the effects of rotational variance in the input images, we designed an experiment based on both horizontal and vertical morphological closing filtrations. Specifically, for each binarized firn image \( f \), we compute two closing filtrations: \( \mathcal{X}_{C_{H_\bullet}(f)} \) and \( \mathcal{X}_{C_{V_\bullet}(f)} \), where \( H_2 \subseteq \cdots \subseteq H_{30} \) and \( V_2 \subseteq \cdots \subseteq V_{30} \) denote nested sequences of horizontal and vertical line SEs, respectively, as introduced in Example~\ref{Example: Horizontal and vertical lines}. Each filtration induces a $0$-dimensional PB, from which we extract a birth value curve that counts the number of connected components appearing at each filtration step. To obtain a rotation-robust descriptor, we define the final curve associated with image \( f \) as the sum of the birth value curves derived from \( \operatorname{PB}_0(\mathcal{X}_{C_{H_\bullet}(f)}) \) and \( \operatorname{PB}_0(\mathcal{X}_{C_{V_\bullet}(f)}) \).

Using a total of 500 binarized firn images per depth category across 10 depth levels, we compute 5000 such birth value curves. In Fig.~\ref{Fig. tSNE demonstration}(a), we illustrate the average birth value curves for each depth. These curves reveal systematic differences in topological structure. Shallower layers tend to exhibit a single dominant peak at later filtration steps, indicating the presence of smaller and more fragmented ice structures. In contrast, deeper layers often display multiple peaks: an early peak suggests the existence of larger black components that appear early in the filtration, while a secondary peak---typically located at a filtration step similar to that of shallower layers---suggests the continued presence of small black fragments. This bimodal pattern observed in deeper layers reflects the coexistence of both large-scale and fine-scale structures, indicating more complex, dense, or interconnected features. These conclusions are based on the binarization of the original firn images using their average pixel values as thresholding criteria. Furthermore, as depicted by the average birth value curves, deeper layers tend to exhibit higher peak amplitudes, indicating a greater number of connected components appearing at specific filtration steps.

To further examine the global structure and potential clustering of these descriptors, we apply 2D t-SNE to the 5000 birth value curves. The results, shown in Fig.~\ref{Fig. tSNE demonstration}(b), are projected into the normalized domain \( [0,1] \times [0,1] \subseteq \mathbb{R}^2 \). The resulting visualization reveals clear gradients and cluster patterns corresponding to depth, suggesting that the birth value curves effectively capture geometric and topological information that varies with firn stratigraphy. 

As shown in Fig.~\ref{Fig. tSNE demonstration}(a), the peak amplitudes of the average birth value curves generally follow the ordering of the depth levels, with deeper layers tending to exhibit higher peaks. However, this trend is not strictly monotonic, and some deviations in the ordering can be observed across different layers. Similarly, in Fig.~\ref{Fig. tSNE demonstration}(b), the 2D projections of the birth value curves via the t-SNE clustering algorithm reveal a clear depth-dependent organization. The points are arranged gradually according to depth, with clusters corresponding to neighboring depths positioned adjacently. Notably, the curves from depths 7 and 17 are well separated, indicating distinct structural differences, while curves from closely spaced depth levels occasionally overlap due to similar morphological characteristics (e.g., collections $\{ 23, 26 \}$ and $\{ 59, 65 \}$ of depth levels), which can be an indicator of specific depositional or metamorphic conditions acting on those firn layers. This experiment highlights the potential of morphological PH for structural analysis and offers practical guidance for selecting suitable horizontal and vertical SEs when analyzing firn images.

\section{Discussion and Conclusion}\label{Section: Discussion and Conclusion}

In this work, we introduced the concept of shift inclusion as a unifying framework for studying absorption properties in mathematical morphology. By formulating the definitions entirely within set-theoretic and topological language, we bridged the notational and conceptual gap between the MM and TDA communities. This abstraction not only clarifies the underlying structure of morphological operations but also enables their extension to general, non-rectangular image domains.

Our main theoretical contribution is the equivalence between shift inclusion and the absorption property for opening and closing operators, providing a rigorous and generalizable condition for their validity. Furthermore, the introduction of weak shift inclusion extends this equivalence to a broader class of structuring elements and domains, including irregular image supports, thereby increasing the flexibility of MM in real-world applications. From an applied perspective, the proposed framework offers a systematic way to verify or design structuring elements that ensure desirable morphological properties, even in settings where classical rectangular-domain assumptions do not hold. This is particularly relevant in digital image analysis scenarios with irregular boundaries, anisotropic pixel arrangements, or domain constraints dictated by the application.

As a case study, we applied the proposed MM-based persistent homology framework to the analysis of firn-- compacted old snow that evolves into glacial ice over time. Firn exhibits a highly porous structure, with complex networks of interconnected air pockets embedded within an ice matrix. The topology and geometry of these pore spaces directly influence firn densification, gas transport, and long-term climate signal preservation in ice cores. By integrating MM-based pre-filtering with persistent Betti number analysis, we quantified multiscale topological features of pores and ice grains, enabling the characterization of structural differences across firn layers.

Looking forward, our results suggest several promising research directions. First, integrating shift inclusion with persistent homology could yield multi-scale topological descriptors informed by morphological stability conditions. Second, extending the framework to grayscale and multichannel images could broaden its practical utility. Finally, exploring connections with sheaf-theoretic and algebraic-topological formulations may further solidify the theoretical links between MM and TDA, paving the way for cross-disciplinary methods that combine local morphological filtering with global topological invariants. By unifying MM's structuring-element theory with topological formalism, this work lays the groundwork for robust, topology-aware morphological analysis applicable to both classical image processing and emerging data-analysis domains.

\backmatter

%\bmhead{Supplementary information}
%If your article has accompanying supplementary file/s please state so here. 
%Authors reporting data from electrophoretic gels and blots should supply the full unprocessed scans for key as part of their Supplementary information. This may be requested by the editorial team/s if it is missing.
%Please refer to Journal-level guidance for any specific requirements.

\bmhead{Acknowledgements}
Most of the theoretical and experimental results presented in this paper were completed during CSH's Ph.D. studies in the Department of Mathematics at National Taiwan Normal University. The authors used Grammarly and OpenAI's ChatGPT-5 to improve the clarity and consistency of the writing. All technical content, analyses, interpretations, and conclusions are solely the responsibility of the authors. %% Parts of the experimental extensions presented in this paper were conducted during CHS's postdoctoral research at the Department of Mathematics, National Central University (NCU). CSH sincerely thanks Dr. John M. Hong (NCU) for his generous support of this work. 

\section*{Declarations}

\bmhead{Funding} 
CSH was partially supported in conducting this research during his Ph.D. program in the Department of Mathematics at National Taiwan Normal University by the research projects MOST 108-2119-M-002-031 and MOST 108-2115-M-003-005-MY2, funded by the Ministry of Science and Technology (MOST), Taiwan.

\bmhead{Consent for publication} Not applicable.

\bmhead{Conflict of interest} The Authors declare no competing financial or non-financial interests.

\bmhead{Data availability}
The datasets generated and analyzed during the current study are not publicly available due to restrictions but can be obtained from the corresponding author upon reasonable request and subject to approval.

\bmhead{Code availability} 
The full Python code for the implementations of the MM-based persistent homology computation on digital images is available on our GitHub repository. Historic versions and future updates will be maintained and announced on the same platform. GitHub: \url{https://github.com/ChuanShenHu/Firn_Projects_with_MM_TDA}.

\bmhead{Author contribution}  
This work was inspired by the collaboration between YMC and KK on the integration and application of topological data analysis for firn image analysis. YMC and CSH proposed the mathematical framework and introduced the notions of shift inclusion and weak shift inclusion. CSH formulated and proved the main theorems and implemented the computational framework. 
YMC and CSH jointly verified the theoretical results. KK provided the dataset, evaluated the experimental outcomes, and interpreted the physical implications in the context of firn structure analysis. KSY conducted extended MM-based filtration experiments on firn data and analyzed the geometric meaning of the resulting persistence information. All authors contributed to the research and the writing of the manuscript.

%%%%%%%%%%%%%%%%%%%%%%%%%%%%%%%%%%%%%%%%
%%%%%%%%%%%%%%%%%%%%%%%%%%%%%%%%%%%%%%%%
%%%%%%%%%%%%%%%%%%%%%%%%%%%%%%%%%%%%%%%%
%%%%%%%%%%%%%%%%%%%%%%%%%%%%%%%%%%%%%%%%
%%%%%%%%%%%%%%%%%%%%%%%%%%%%%%%%%%%%%%%%
%%%%%%%%%%%%%%%%%%%%%%%%%%%%%%%%%%%%%%%%
\begin{appendices}

%%%%%%%%%%%%%%%%%%%%%%%%%%%%%%%%%%%%%%%%
%%%%%%%%%%%%%%%%%%%%%%%%%%%%%%%%%%%%%%%% 
\section{Digital Images and Cubical Complexes}
\label{Appendix: Digital Images and Cubical Complexes}
%%%%%%%%%%%%%%%%%%%%%%%%%%%%%%%%%%%%%%%%
%%%%%%%%%%%%%%%%%%%%%%%%%%%%%%%%%%%%%%%%

In Appendix~\ref{Appendix: Digital Images and Cubical Complexes}, we concisely present the mathematical foundations underlying the correspondence between digital images and (regular) cubical complexes, providing a rigorous basis for homology and persistent homology computations. For a more comprehensive and detailed introduction to digital images and cubical complexes, such as more general cubical complexes allowing irregular cubes or higher-dimensional cubical complexes, see~\cite{kaczynski2004computational,Kovalevsky2006}.

\bigskip
\begin{definitionA}[Elementary cubes in $\mathbb{R}^2$]
An \textbf{elementary interval} in \( \mathbb{R} \) is a set of the form \( [l, l+1] \) or \( \{l\} \), for some \( l \in \mathbb{R} \). An \textbf{elementary cube} in \( \mathbb{R}^2 \) is a set of the form \( I \times J \), where $I$ and $J$ are elementary intervals in $\mathbb{R}$. 
\end{definitionA}
\bigskip

Elementary cubes of the form \( I \times J \), where \( I \) and \( J \) are elementary intervals, are compact subsets of \( \mathbb{R}^2 \) and serve as the basic building blocks for constructing cubical complexes in \( \mathbb{R}^2 \). In particular, an elementary cube \( C = I \times J \), where \( I = [a, b] \) and \( J = [c, d] \) with \( a \leq b \) and \( c \leq d \), is called \textit{non-degenerate} if \( a < b \) and \( c < d \); otherwise, \( C \) is called \textit{degenerate}. Furthermore, the boundary of \( C \) is precisely the set
\begin{equation*}
\begin{split}
\operatorname{bd}(C) &= \operatorname{bd}(I \times J) = (\operatorname{bd}(I) \times J) \cup (I \times \operatorname{bd}(J)) \\
&= (\{ a \} \times J) \cup (\{ b \} \times J) \cup (I \times \{ c \}) \cup (I \times \{ d \}).
\end{split}
\end{equation*}
When $C$ is a non-degenerate cube, the sets $\{ a \} \times J$, $\{ b \} \times J$, $I \times \{ c \}$, and $I \times \{ d \}$ are non-degenerate line segments in $\mathbb{R}^2$, which are called the $1$\textit{-dimensional faces} of $C$. On the other hand, the four vertices $(a, c), (a, d), (b, c)$, and $(b, d)$ are the $0$\textit{-dimensional faces} of $C$.

\bigskip
\begin{definitionA}[Regular cubical complexes in $\mathbb{R}^2$]
A \textbf{regular cubical complex} in \( \mathbb{R}^2 \) is a union of a family \( \{ C_\alpha \mid \alpha \in \mathcal{A} \} \) of non-degenerate elementary cubes in \( \mathbb{R}^2 \), where for any distinct \( \alpha, \beta \in \mathcal{A} \), the intersection \( C_\alpha \cap C_\beta \) is a common face of both elementary cubes \( C_\alpha \) and \( C_\beta \).
\end{definitionA}
\bigskip

As shown in  \eqref{Eq. Cubical complex representation} and Fig. \ref{Fig. Definition of images}, this paper focuses on using regular cubical complexes to model the geometric structure of black pixels in binary images.  Specifically, for a given image domain $P \subseteq \mathbb{Z}^2$ and $X \subseteq P$ representing the set of positions of the black pixels with $P$, the corresponding cubical complex can be formalized by the following mapping:
\begin{equation}
\label{Eq. Mapping between black pixels to elementary cubes}
X \longrightarrow \mathcal{P}(\mathbb{R}^2), \quad (a,b) \longmapsto [a, a+1] \times [b, b+1], 
\end{equation}
where $\mathcal{P}(\mathbb{R}^2)$ denotes the power set of $\mathbb{R}^2$, collecting all the subsets of $\mathbb{R}^2$. More precisely, for every pixel $\mathbf{x} = (a,b) \in X \subseteq P$, the mapping in  \eqref{Eq. Mapping between black pixels to elementary cubes} to the elementary cube in $\mathbb{R}^2$ that has the barycenter $(a + \frac{1}{2}, b + \frac{1}{2})$. The corresponding (regular) cubical complex $\mathcal{X}$ is then defined as the union of all associated elementary cubes, forming a compact subset of $\mathbb{R}^2$.

\begin{figure}%[ht]
\centering
\includegraphics[width=0.9\linewidth]{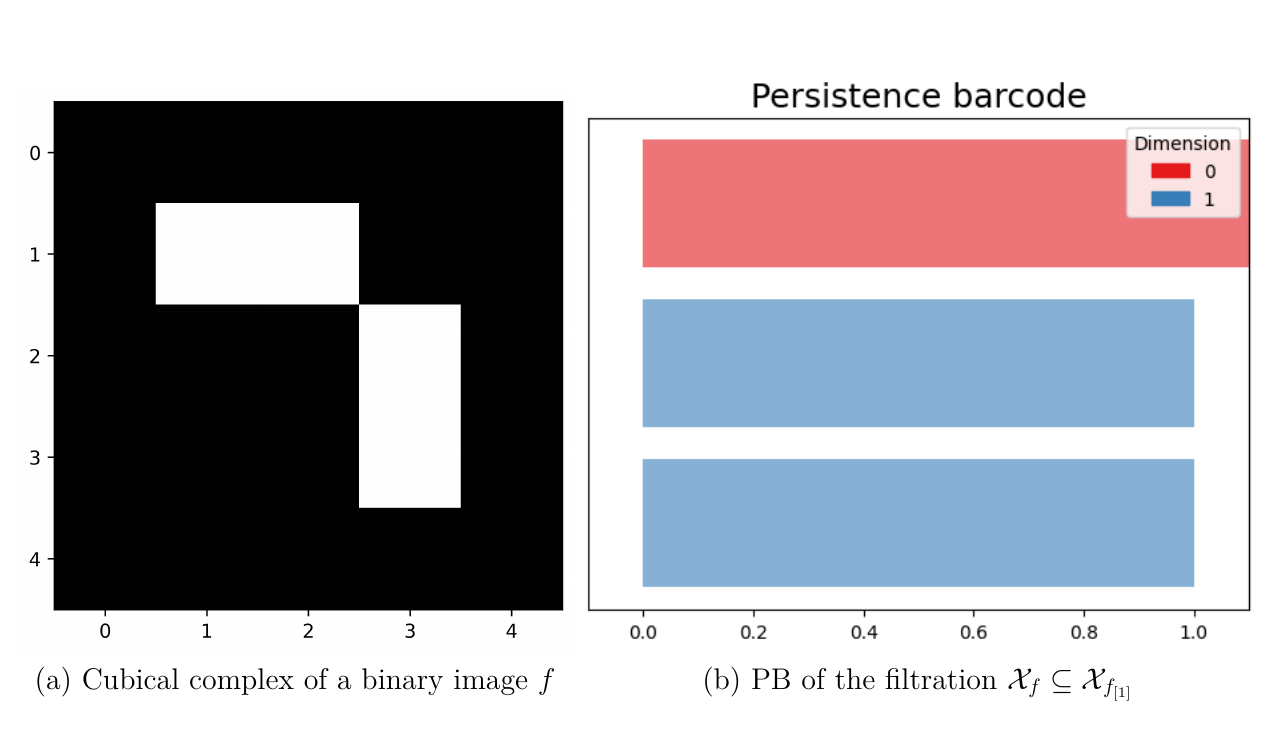}
\caption{(a) The cubical complex representation of a $5 \times 5$ binary image $f$, i.e., $\mathcal{X}_f$.  (b) The persistence barcode corresponding to the filtration $\mathcal{X}_f \subseteq \mathcal{X}_{f_{[1]}}$, computed using the GUDHI library's cubical complex package~\cite{gudhi:CubicalComplex}. In detail, the calculated $\operatorname{PB}_0$ and $\operatorname{PB}_1$ are the multisets $\{ (0,+\infty) \}$ and $\{ (0,1), (0,1) \}$, respectively.}
\label{Fig. Appendix B cubical complex demo}
\end{figure}

Fig. \ref{Fig. Appendix B cubical complex demo} illustrates a cubical complex representation of a binary image along with its topological characteristics, computed using the GUDHI library's cubical complex package~\cite{gudhi:CubicalComplex}. Specifically, by considering the filtration $\mathcal{X}_f \subseteq \mathcal{X}_{f_{[1]}}$ based on the thresholding operation introduced in  \eqref{Eq. Thresholding operation}, the persistence barcode is presented in Fig. \ref{Fig. Appendix B cubical complex demo}(b), where the persistence intervals recode the homological information of the cubical complex $\mathcal{X}_f$.
 
In particular, the $0$-th and $1$-st persistence barcodes of the cubical complex in Fig. \ref{Fig. Appendix B cubical complex demo} are the multisets $\{ (0,+\infty) \}$ and $\{ (0,1), (0,1) \}$, respectively. They depict the connected component and $1$-dimensional hole information enclosed by the elementary cubes. In particular, according to the mapping of elementary cubes defined in  \ref{Eq. Mapping between black pixels to elementary cubes}, the barcodes indicate that the cubical complex $\mathcal{X}_f$ contains exactly one connected component and two holes.

\section{Essential Properties of Image Operations}\label{Appendix: Essential Properties of Image Operations}

In Appendix~\ref{Appendix: Essential Properties of Image Operations}, we present several essential properties of image operations that are instrumental in establishing the main results discussed in the paper. These properties not only simplify various arguments but also play a pivotal role in the proofs of key theorems, such as Theorem~\ref{Theorem: Shift Inclusion as a sufficient condition}.

While similar results may be found in standard references on mathematical morphology (e.g.,~\cite{soille2013,Najman-Mathematical-Morphology}), they are often stated using different notations or within slightly different contexts. To align with the digital image framework and notation adopted in this work, we provide concise statements and proofs of these properties for completeness and clarity.

\subsection{Alternative Definition of Morphological Dilation}
\label{Appendix Section: Alternative Definition of Morphological Dilation}
In Section \ref{Section: Mathematical Morphology}, we see that the discrete neighborhoods of $\mathbf{x}$ in the definitions of erosion and dilation in \eqref{Equation: Erosion and dilation} are considered with opposite directions, i.e., $\mathbf{x} + B$ and $\mathbf{x} - B$. In this appendix, we explain the reason by presenting a visual example. Specifically, consider the $2$-dimensional digital image $g \in \mathcal{I}_P$ defined as
\begin{equation*}
\ytableausetup{centertableaux}
g =
\begin{ytableau}
0 & 0 & 0 & 0 & 0 & 0\\
0 & 0 & 1 & 1 & 0 & 0\\
0 & 0 & 1 & 1 & 0 & 0\\
0 & 0 & 0 & 0 & 0 & 0\\
0 & 0 & 0 & 0 & 0 & 0\\
0 & 0 & 0 & 0 & 0 & 0\\
\end{ytableau}
\text{ \quad and \quad }
P = \begin{ytableau}
 \ &  \ & \  &  \ & \  & \ \\
  &   &   &   &   &  \\
  &   &   &   &   &  \\
  &   &   &   &   &  \\
  &   &   &   &   &  \\
  &   &   &   &   &  \\  
\end{ytableau} \ \subseteq \mathbb{Z}^2.
\end{equation*}
Furthermore, by considering the non-symmetric SE
\begin{equation*}
B = \begin{ytableau}
\ & \ \\
\mathbf{0} & \ \\
\end{ytableau} = \{ (0,0), (1,0), (0,1), (1,1) \} \subseteq \mathbb{Z}^2,
\end{equation*}
then the corresponding eroded and opened images are
\begin{equation*}
\ytableausetup{centertableaux}
\epsilon_B(g)=
\begin{ytableau}
0 & 0 & 0 & 0 & 0 & 0\\
0 & 0 & 0 & 0 & 0 & 0\\
0 & 0 & 1 & 0 & 0 & 0\\
0 & 0 & 0 & 0 & 0 & 0\\
0 & 0 & 0 & 0 & 0 & 0\\
0 & 0 & 0 & 0 & 0 & 0\\
\end{ytableau}
\text{ \quad  and  \quad }
O_B(g)=
\begin{ytableau}
0 & 0 & 0 & 0 & 0 & 0\\
0 & 0 & 1 & 1 & 0 & 0\\
0 & 0 & 1 & 1 & 0 & 0\\
0 & 0 & 0 & 0 & 0 & 0\\
0 & 0 & 0 & 0 & 0 & 0\\
0 & 0 & 0 & 0 & 0 & 0\\
\end{ytableau} = g.
\end{equation*}
Suppose a new ``dilation'' operation $\widetilde{\delta_B}: \mathcal{I}_P \rightarrow \mathcal{I}_P$ is defined as in~\eqref{Equation: Opposite dilation}, i.e., 
\begin{equation}
\label{Equation: Opposite dilation}
\widetilde{\delta_B}(\bfx) = \max_{\mathbf{y} \in \bfx + B} g(\mathbf{y}) = \max \{ g(\bfx+\bfb) \ | \ \bfb \in B, \bfx+\bfb \in P \},
\end{equation}
then the corresponding images obtained from applying $(\widetilde{\delta_B} \circ \epsilon_B)$ on image $g$ is
\begin{equation*}
\ytableausetup{centertableaux}
(\widetilde{\delta_B} \circ \epsilon_B)(g)=
\begin{ytableau}
0 & 0 & 0 & 0 & 0 & 0\\
0 & 0 & 0 & 0 & 0 & 0\\
0 & 1 & 1 & 0 & 0 & 0\\
0 & 1 & 1 & 0 & 0 & 0\\
0 & 0 & 0 & 0 & 0 & 0\\
0 & 0 & 0 & 0 & 0 & 0\\
\end{ytableau}
\neq (\delta_B \circ \epsilon_B)(g).
\end{equation*}
Dually, by defining the image $h$ on $P$ by
\begin{equation*}
\ytableausetup{centertableaux}
h=
\begin{ytableau}
1 & 1 & 1 & 1 & 1 & 1\\
1 & 1 & 0 & 0 & 1 & 1\\
1 & 1 & 0 & 0 & 1 & 1\\
1 & 1 & 1 & 1 & 1 & 1\\
1 & 1 & 1 & 1 & 1 & 1\\
1 & 1 & 1 & 1 & 1 & 1\\
\end{ytableau}\quad,
\end{equation*}
then the corresponding images from applying $C_B$ and $(\widetilde{\delta_B} \circ \epsilon_B)$ on image $g$ are 
\begin{equation*}
\ytableausetup{centertableaux}
C_B(h)=
\begin{ytableau}
1 & 1 & 1 & 1 & 1 & 1\\
1 & 1 & 0 & 0 & 1 & 1\\
1 & 1 & 0 & 0 & 1 & 1\\
1 & 1 & 1 & 1 & 1 & 1\\
1 & 1 & 1 & 1 & 1 & 1\\
1 & 1 & 1 & 1 & 1 & 1\\
\end{ytableau}
\text{ \quad and \quad }
(\epsilon_B \circ \widetilde{\delta_B})(h)=
\begin{ytableau}
1 & 1 & 1 & 1 & 1 & 1\\
1 & 1 & 1 & 1 & 1 & 1\\
1 & 0 & 0 & 1 & 1 & 1\\
1 & 0 & 0 & 1 & 1 & 1\\
1 & 1 & 1 & 1 & 1 & 1\\
1 & 1 & 1 & 1 & 1 & 1\\
\end{ytableau} \quad .
\end{equation*}
%%%%%%%%%%%%%%%%%%%%%%%%%%%%%%%%%%%%%%%%
%%%%%%%%%%%%%%%%%%%%%%%%%%%%%%%%%%%%%%%%
%%%%%%%%%%%%%%%%%%%%%%%%%%%%%%%%%%%%%%%%
In particular, as depicted in the images $C_B(g)$ and $O_B(g)$, the operations $\widetilde{\delta_B} \circ \epsilon_B$ and $\epsilon_B \circ \widetilde{\delta_B}$ shift the blocks 
\begin{equation*}
\begin{ytableau}
1 & 1 \\
1 & 1 \\
\end{ytableau}
\text{ \quad and \quad }
\begin{ytableau}
0 & 0 \\
0 & 0 \\
\end{ytableau}    
\end{equation*}
within the original images $g$ and $h$. To effectively erase white/black pixels while preserving the main image structure, we prefer to define dilation as in~\eqref{Equation: Erosion and dilation} rather than the definition in~\eqref{Equation: Opposite dilation}.

\bigskip
\begin{remarkA}
When \( B \) is symmetric, the two dilation operations defined in~\eqref{Equation: Erosion and dilation} and~\eqref{Equation: Opposite dilation} coincide.  
\end{remarkA}
\bigskip

These definitions follow a typical manner of opening and closing operations in mathematical morphology and are also implemented in Matlab (\texttt{imopen} and \texttt{imclose}) \cite{Matlab}. On the other hand, the OpenCV package in Python (\texttt{cv2.MORPH\_OPEN} and \texttt{cv2.MORPH\_CLOSE}) follows the same composition ordering of erosion and dilation \cite{opencv_library}.

%%%%%%%%%%%%%%%%%%%%%%%%%%%%%%%%%%%%%%%%
%%%%%%%%%%%%%%%%%%%%%%%%%%%%%%%%%%%%%%%%
\subsection{Counterexamples to the Absorption Property}
\label{Appendix Section: Counterexamples to the Absorption Property}
%%%%%%%%%%%%%%%%%%%%%%%%%%%%%%%%%%%%%%%%
%%%%%%%%%%%%%%%%%%%%%%%%%%%%%%%%%%%%%%%%

\begin{exampleA}
\label{ExampleA: B1 subseteq B2, but not opening decreasing}
Let $g: P \rightarrow \{ 0, 1 \}$ be a binary image defined by
\begin{equation}
\label{Eq. Counterexample image}
\ytableausetup{centertableaux}
g=
\begin{ytableau}
0 & 0 & 0 & 0 & 0 & 0\\
0 & 0 & 1 & 1 & 1 & 0\\
0 & 0 & 1 & 1 & 1 & 0\\
0 & 1 & 1 & 1 & 1 & 0\\
0 & 0 & 1 & 0 & 0 & 0\\
0 & 0 & 0 & 0 & 0 & 0\\
\end{ytableau}
\text{ \ \ \ with \ \ \ }
\ytableausetup{centertableaux}
P = \begin{ytableau}
 \ &  \ & \  &  \ & \  & \ \\
  &   &   &   & \bfx &  \\
  &   &   &   &   &  \\
  &   &   &   &   &  \\
  &   &   &   &   &  \\
  &   &   &   &   &  \\  
\end{ytableau} \ \ \subseteq \mathbb{Z}^2,
\end{equation}
where $\bfx \in P$ is a specified point in the image domain representation. Consider symmetric SEs \( B_1, B_2 \subseteq \mathbb{Z}^2 \) with \( B_1 \subseteq B_2 \), defined by
\begin{equation}
\label{Eq. Counterexample SEs}
B_1 = 
\ytableausetup{centertableaux}
\begin{ytableau}
\none & \ & \none \\
 & \mathbf{0} & \\
\none & \ & \none \\
\end{ytableau}
\quad \text{and} \quad
B_2 = 
\ytableausetup{centertableaux}
\begin{ytableau}
\ & \ & \ \\
 & \mathbf{0} & \\
& \ & \\
\end{ytableau} \quad.
\end{equation}
Then, the images \( O_{B_1}(g) \) and \( O_{B_2}(g) \) are given by:
\begin{equation*}
O_{B_1}(g) =
\begin{ytableau}
0 & 0 & 0 & 0 & 0 & 0\\
0 & 0 & 0 & 1 & 0 & 0\\
0 & 0 & 1 & 1 & 1 & 0\\
0 & 1 & 1 & 1 & 0 & 0\\
0 & 0 & 1 & 0 & 0 & 0\\
0 & 0 & 0 & 0 & 0 & 0\\
\end{ytableau} 
\quad \text{and} \quad
O_{B_2}(g) = 
\begin{ytableau}
0 & 0 & 0 & 0 & 0 & 0\\
0 & 0 & 1 & 1 & 1 & 0\\
0 & 0 & 1 & 1 & 1 & 0\\
0 & 0 & 1 & 1 & 1 & 0\\
0 & 0 & 0 & 0 & 0 & 0\\
0 & 0 & 0 & 0 & 0 & 0\\
\end{ytableau}
\quad .
\end{equation*}
By comparing the pixel values at \( \mathbf{x} \), we observe that \( O_{B_2}(g) \nleq O_{B_1}(g) \) since \( O_{B_1}(g)(\mathbf{x}) = 0 \) while \( O_{B_2}(g)(\mathbf{x}) = 1 \). Dually, we observe that \( C_{B_1}(f) \nleq C_{B_2}(f) \), where \( f \) is the complement of \( g \), i.e., \( f(\mathbf{x}) = 1 - g(\mathbf{x}) \) for all \( \mathbf{x} \in P \) (see Appendix~\ref{Appendix Section: Complement of a Binary Image}).
\end{exampleA}

By the definitions of erosion and dilation~\eqref{Equation: Erosion and dilation}, it is evident that $\mathcal{X}_{\epsilon_{B_1}} \subseteq \mathcal{X}_{\epsilon_{B_2}}$ and $\mathcal{X}_{\delta_{B_2}} \subseteq \mathcal{X}_{\delta_{B_1}}$ whenever $B_1 \subseteq B_2$. This makes it possible to study PH on filtrations constructed by erosion and dilation. However, as shown in Fig. \ref{Fig. Demonstration of erosion, dilation, opening, and closing}, opening and closing operations better preserve the primary structure of porous data. Therefore, it is crucial to investigate the conditions under which $B_1 \subseteq B_2$ preserves the desired absorption property.

%%%%%%%%%%%%%%%%%%%%%%%%%%%%%%%%%%%%%%%%
%%%%%%%%%%%%%%%%%%%%%%%%%%%%%%%%%%%%%%%%
\subsection{Complement of a Binary Image}
\label{Appendix Section: Complement of a Binary Image}
%%%%%%%%%%%%%%%%%%%%%%%%%%%%%%%%%%%%%%%%
%%%%%%%%%%%%%%%%%%%%%%%%%%%%%%%%%%%%%%%%
Let $P \subseteq \mathbb{Z}^2$ be an image domain and $g \in \mathcal{BI}_P$ be a binary image defined on $P$. Then \textit{complement} of $g$ is defined as the binary image $\widehat{g} \in \mathcal{BI}_P$ which assigns each $\bfx \in P$ to the value $1 - g(\bfx)$; in other words, for any $\mathbf{x} \in P$, 
\begin{equation}
\widehat{g}(\bfx) = \begin{cases}
1 & \text{if \ } g(\bfx) = 0,  \\
0 & \text{if \ } g(\bfx) = 1.  \\
\end{cases}    
\end{equation}
In particular, this process defines a mapping $g \mapsto \widehat{g}$ from $\mathcal{BI}_P$ to itself. In the following equation, we exhibit a pair $(g, \widehat{g})$ of a $3 \times 3$ binary image and its complement image:
\begin{equation}\label{Eq. A binary image and its complement}
g = \begin{ytableau}
 1 & 1 & 1 \\
 1 & 0 & 0 \\
 0 & 1 & 0 \\
\end{ytableau} \text{ \ and \ } \mathcal{X}_g = \begin{ytableau}
 \none[] & \none[] & \none[] \\
 %\none[1] & *(gray) & *(gray) \\
 \none[] & *(gray) & *(gray) \\
 *(gray) & \none[] & *(gray) \\
\end{ytableau}~.  
\end{equation}
Then, the induced complement image $\widehat{g}$ and the induced cubical complex are
\begin{equation}\label{Eq. A binary image and its complement-2}
g = \begin{ytableau}
 0 & 0 & 0 \\
 0 & 1 & 1 \\
 1 & 0 & 1 \\
\end{ytableau} \text{ \ and \ } \mathcal{X}_g = \begin{ytableau}
*(gray) & *(gray) & *(gray) \\
*(gray) & \none[] & \none[] \\
\none[] & *(gray) & \none[] \\
\end{ytableau}~.  
\end{equation}

According to the dual relationship within the definitions of erosion and dilation, this mapping provides a connection to these two operations, making the exploration and explanation of these operations more convenient and concise. Specifically, for any
binary image $f \in \mathcal{BI}_P$ and subset $Q$ of $P$,
\begin{equation*}
\min f(Q) = 1 - \max \widehat{f}(Q) \quad \text{and} \quad  \max f(Q) = 1 - \min \widehat{f}(Q). 
\end{equation*}
In addition, if $B \subseteq \mathbb{Z}^2$ is an SE and $\mathbf{x} \in P$, the following two equations hold:
\begin{equation*}
\begin{split}
\epsilon_B(f)(\bfx) &= \min f((\bfx + B) \cap P) = 1 - \max \widehat{f}((\bfx + B) \cap P) = 1 - \delta_{-B}(\widehat{f})(\bfx), \\
\delta_B(f)(\bfx) &= \max f((\bfx - B) \cap P) = 1 - \min \widehat{f}((\bfx - B) \cap P) = 1 - \epsilon_{-B}(\widehat{f})(\bfx).
\end{split}    
\end{equation*}

This demonstrates that, for any binary image $f \in \mathcal{BI}_P$, the erosion $\epsilon_B(f)$ is the complement of the dilation $\delta_{-B}(\widehat{f})$, and the dilation $\delta_B(f)$ is the complement of the erosion $\epsilon_{-B}(\widehat{f})$. i.e.,
\begin{equation}
\label{Eq. Equations of complement and erosion/dilation}
\epsilon_B(f) = \widehat{\delta_{-B}(\widehat{f})} \quad \text{and} \quad \delta_B(f) = \widehat{\epsilon_{-B}(\widehat{f})}.    
\end{equation}

Furthermore, by applying the operations $\delta_B$ and $\epsilon_B$ to the equations in~\eqref{Eq. Equations of complement and erosion/dilation}, respectively, the following proposition follows.

\bigskip
\begin{propositionA}
Let $f$ be a binary image defined on a domain $P \subseteq \mathbb{Z}^2$. Then,
\begin{equation}
O_B(f) = \widehat{C_{-B}(\widehat{f})}  \quad \text{and} \quad C_B(f) = \widehat{O_{-B}(\widehat{f})}.    
\end{equation}
That is, $O_B(f)$ is the complement of $C_{-B}(\widehat{f})$ and $C_B(f)$ is the complement of $O_{-B}(\widehat{f})$.
\end{propositionA}
\begin{proof}
We prove that $O_B(f) = \widehat{C_{-B}(\widehat{f})}$, while the equation $C_B(f) = \widehat{O_{-B}(\widehat{f})}$ follows by a dual argument. By  \eqref{Eq. Equations of complement and erosion/dilation}, set $g = \epsilon_B(f) = \widehat{\delta_{-B}(\widehat{f})}$. Then, 
\begin{equation*}
\begin{split}
O_B(g)(\mathbf{x}) = \delta_B(g)(\mathbf{x}) &= 1 - \epsilon_{-B} \left( \widehat{g} \right)(\mathbf{x}) = 1 - \epsilon_{-B} \left( \widehat{\widehat{\delta_{-B}(\widehat{f})}} \right)(\mathbf{x}) 
\end{split}
\end{equation*}
for any $\mathbf{x} \in P$. Because $\widehat{\widehat{h}} = h$ whenever $h \in \mathcal{BI}_P$, the proposition follows.
\end{proof}
In particular, suppose $B_1$ and $B_2$ are SEs in $\mathbb{Z}^2$, then $O_{B_2}(f) \leq O_{B_1}(f)$ whenever $f \in \mathcal{BI}_P$ if and only if $C_{-B_1}(\widehat{f}) \leq C_{-B_2}(\widehat{f})$ whenever $f \in \mathcal{BI}_P$. Because $\widehat{f}$ goes through all the elements in $\mathcal{BI}_P$, it is equivalent to $C_{-B_1}(f) \leq C_{-B_2}(f)$ whenever $f \in \mathcal{BI}_P$. We summarize this as the following proposition.
\bigskip
\begin{propositionA}\label{Proposition: O2 <= O1 iff C-1 <= C-2}
Let $P \subseteq \mathbb{Z}^2$ be an image domain, let $f \in \mathcal{I}_P$, and let $B_1, B_2 \subseteq \mathbb{Z}^2$ be SEs. Then, $O_{B_2}(f) \leq O_{B_1}(f)$ whenever $f \in \mathcal{BI}_P$ if and only if $C_{-B_1}(f) \leq C_{-B_2}(f)$ whenever $f \in \mathcal{BI}_P$.
\end{propositionA}
\bigskip

Using Proposition~\ref{Proposition: O2 <= O1 iff C-1 <= C-2}, many of the theorem proofs in this paper can be reduced to verifying whether $O_{B_2}(f) \leq O_{B_1}(f)$ for all $f \in \mathcal{BI}_P$, given SEs $B_1$ and $B_2$. This simplification streamlines the argumentation in proofs such as Theorems~\ref{Theorem: Shift Inclusion as a sufficient condition} and~\ref{Theorem: Equivalence of WS,P and Desired property-form 1}.

%%%%%%%%%%%%%%%%%%%%%%%%%%%%%%%%%%%%%%%%
%%%%%%%%%%%%%%%%%%%%%%%%%%%%%%%%%%%%%%%%

%%%%%%%%%%%%%%%%%%%%%%%%%%%%%%%%%%%%%%%%
%%%%%%%%%%%%%%%%%%%%%%%%%%%%%%%%%%%%%%%%

%%%%%%%%%%%%%%%%%%%%%%%%%%%%%%%%%%%%%%%%
%%%%%%%%%%%%%%%%%%%%%%%%%%%%%%%%%%%%%%%%

\subsection{Thresholding Operations}\label{Section: Thresholding Operations}

The thresholding operation defined in Equation~\eqref{Eq. Thresholding operation} is widely used to analyze the topological structure of grayscale images via their binarized representations at varying threshold levels. This section outlines key properties of the thresholding operation that are essential for subsequent analysis. 

\bigskip
\begin{propositionA}
\label{Proposition: Threshold inclusion relation}
For $f, g \in \mathcal{I}_P$, $f \leq g$ if and only if $f_{[t]} \leq g_{[t]}$ for every $t \in \mathbb{R}_{\geq 0}$.
\end{propositionA}
\begin{proof}
It is clear that $f \leq g$ implies $f_{[t]} \leq g_{[t]}$ for every $t \in \mathbb{R}_{\geq 0}$. Conversely, suppose $0 \leq g(\bfx) < f(\bfx)$ for some $\bfx \in P$. Set $t = g(\bfx)$, then $t \in \mathbb{R}_{\geq 0}$ and $f_{[t]}(\bfx) = 1$, $g_{[t]}(\bfx) = 0$, and this shows that $f_{[t]} \nleq g_{[t]}$.
\end{proof}
\bigskip

In this paper, we focus on image operations that are commutative with the thresholding operation. Mathematically, we consider image operations $\xi: \mathcal{I}_P \rightarrow \mathcal{I}_P$ satisfying the following property: for every $t \in \mathbb{R}_{\geq 0}$, the diagram
\begin{equation}
\label{Eq. Image operations that are commutative with the thresholding operation}
\xymatrix@+1.0em{
		      & \mathcal{I}_P
			\ar[d]_{\xi}
			\ar[r]^{\tau_t}
                & \mathcal{I}_P
                \ar[d]^{\xi}
        	\\
        	& \mathcal{I}_P
			\ar[r]^{\tau_t}
                & \mathcal{I}_P
}    
\end{equation}
commutes; that is, $\xi \circ \tau_t = \tau_t \circ \xi$, where $\tau_t: \mathcal{I}_P \rightarrow \mathcal{I}_P$ is the thresholding operation defined in Equation~\eqref{Eq. Thresholding operation}. The following proposition presents an important observation about such image operations, which will be used frequently throughout this paper.

\bigskip
\begin{propositionA}
\label{Proposition: Threshold inclusion relation--key prop}
Let $P \subseteq \mathbb{Z}^2$ be an image domain, and let $\xi, \zeta: \mathcal{I}_P \rightarrow \mathcal{I}_P$ be image operations such that $\xi \circ \tau_t = \tau_t \circ \xi$ and $\zeta \circ \tau_t = \tau_t \circ \zeta$ for every $t \in \mathbb{R}_{\geq 0}$. Then, the following conditions are equivalent:
\begin{enumerate}[label={\rm (\alph*)}]
\item $\xi(f) \leq \zeta(f)$ whenever $f \in \mathcal{I}_P$.
\item $\xi(f) \leq \zeta(f)$ whenever $f \in \mathcal{BI}_P$.
\end{enumerate}
\end{propositionA}
\begin{proof}
Assertion (a) implies (b) since $\mathcal{BI}_P \subseteq \mathcal{I}_P$. Conversely, suppose assertion~(b) holds. We will show that~(a) also holds.  
Assume, for contradiction, that~(a) fails. Then there exists an $f \in \mathcal{I}_P$ such that $\zeta(f) < \xi(f)$. Then, there is a $t \in \mathbb{R}_{\geq 0}$ such that $\zeta(f)_{[t]} < \xi(f)_{[t]}$. By Proposition \ref{Proposition: Threshold inclusion relation} and the assumption of operations $\zeta$ and $\xi$,
\begin{equation*}
\zeta(f_{[t]}) = \zeta(f)_{[t]} < \xi(f)_{[t]} = \xi(f_{[t]}).     
\end{equation*}
Because $f_{[t]}$ is a binary image, assertion (b) implies that
\begin{equation*}
\xi(f_{[t]}) \leq \zeta(f_{[t]}) < \xi(f_{[t]}).
\end{equation*}
This is a contradiction, and thus we have shown that $\xi(f) \leq \zeta(f)$ for all $f \in \mathcal{I}_P$.
\end{proof}
 
The following theorem shows that the morphological erosion and dilation operators, $\epsilon_B$ and $\delta_B: \mathcal{I}_P \rightarrow \mathcal{I}_P$, defined for any SE $B$, satisfy the commutativity property given in Equation~\eqref{Eq. Image operations that are commutative with the thresholding operation}.

\bigskip
\begin{theoremA}\label{Theorem: Restriction and opening/closing are commutative}   
Let $P \subseteq \mathbb{Z}^2$ be an image domain and $g \in \mathcal{I}_P$. Then, For any threshold $t \in \mathbb{R}_{\geq 0}$ and SE $B$ in $\mathbb{Z}^2$, the following diagrams are commutative:
\begin{equation}
\label{Commutative diagrams of thresholding and erosion, dilation}
\begin{split}
\xymatrix@+1.0em{
				& \mathcal{I}_P
				\ar[d]_{\epsilon_B}
				\ar[r]^{\tau_t}
                & \mathcal{I}_P
                \ar[d]^{\epsilon_B}
        		\\
        		& \mathcal{I}_P
				\ar[r]^{\tau_t}
                & \mathcal{I}_P
}
\ \
\xymatrix@+1.0em{
				& \mathcal{I}_P
				\ar[d]_{\delta_B}
				\ar[r]^{\tau_t}
                & \mathcal{I}_P
                \ar[d]^{\delta_B}
        		\\
        		& \mathcal{I}_P
				\ar[r]^{\tau_t}
                & \mathcal{I}_P
}
\end{split}.
\end{equation}
In other words, $\epsilon_B \circ \tau_t = \tau_t \circ \epsilon_B$ and $\delta_B \circ \tau_t = \tau_t \circ \delta_B$ for every $t \in \mathbb{R}_{\geq 0}$ and SE $B$. Furthermore, the opening and closing operations also commute with thresholding: $O_B \circ \tau_t = \tau_t \circ O_B$ and $C_B \circ \tau_t = \tau_t \circ C_B$.
\end{theoremA}
\begin{proof}
Suppose the diagrams in Equation~\eqref{Commutative diagrams of thresholding and erosion, dilation} are commutative for every $t \in \mathbb{R}_{\geq 0}$ and SE $B$, then the two rectangles
\begin{equation}
\label{Commutative diagrams of thresholding and erosion, dilation-2}
\begin{split}
\xymatrix@+1.0em{
				& \mathcal{I}_P
				\ar[d]_{\epsilon_B}
				\ar[r]^{\tau_t}
                & \mathcal{I}_P
                \ar[d]^{\epsilon_B}
        		\\
        		& \mathcal{I}_P
				\ar@{.>}[r]^{\tau_t}
				\ar[d]_{\delta_B}
                & \mathcal{I}_P
                \ar[d]^{\delta_B}
                \\
                & \mathcal{I}_P
				\ar[r]^{\tau_t}
                & \mathcal{I}_P
}
\ \
\xymatrix@+1.0em{
				& \mathcal{I}_P
				\ar[d]_{\delta_B}
				\ar[r]^{\tau_t}
                & \mathcal{I}_P
                \ar[d]^{\delta_B}
        		\\
        		& \mathcal{I}_P
        		\ar[d]_{\epsilon_B}
				\ar@{.>}[r]^{\tau_t}
                & \mathcal{I}_P
                \ar[d]^{\epsilon_B}
                \\
                & \mathcal{I}_P
				\ar[r]^{\tau_t}
                & \mathcal{I}_P
}
\end{split}
\end{equation}
are also commutative due to the associativity of function composition, i.e., $O_B \circ \tau_t = \tau_t \circ O_B$ and $C_B \circ \tau_t = \tau_t \circ C_B$. Therefore, it remains to show that the rectangles in Equation~\eqref{Commutative diagrams of thresholding and erosion, dilation} commute.
 
By the definition of erosion, for $\bfx = (x,y) \in P \subseteq \mathbb{Z}^2$, we have
\begin{equation*}
\begin{split}
    \epsilon_B(g)(\bfx) = \min g \bigg((\bfx + B) \cap P\bigg) \ \ {\rm and} \ \
    \epsilon_B(g_{[t]})(\bfx) = \min g_{[t]}\bigg((\bfx + B) \cap P\bigg).
\end{split}
\end{equation*}
If there is a $\bfb \in B(\bfx;P,+)$ such that $g(\bfx + \bfb) \leq t$, then $g_{[t]}(\bfx + \bfb) = 0$ and $\epsilon_B(g)(\bfx) \leq t$, and this shows that $\epsilon_B(g)_{[t]}(\bfx) = 0 = \epsilon_B(g_{[t]})(\bfx)$. On the other hand, if $g(\bfx + \bfb) > t$ for all $\bfb \in B(\bfx;P,+)$, then $\epsilon_B(g)(\bfx) > t$ and $g_{[t]}(\bfx + \bfb) = 1$ for all $\bfb \in B(\bfx;P,+)$. This shows that $\epsilon_B(g)_{[t]}(\bfx) = 1 = \epsilon_B(g_{[t]})(\bfx)$.
 
Similarly, by the definition of dilation, for $\bfx \in P$, we have
\begin{equation*}
\begin{split}
    \delta_B(g)(\bfx) = \max g\bigg((\bfx - B) \cap P\bigg) \ \ {\rm and} \ \ \delta_B(g_{[t]})(\bfx) = \max g_{[t]}\bigg((\bfx - B) \cap P\bigg).
\end{split}
\end{equation*}
If $g(\bfx - \bfb) \leq t$ for all $\bfb \in B(\bfx;P,-)$, then $\delta_B(g_{[t]})(\bfx) = 0$ and $\delta_B(g)(\bfx) \leq t$. This shows that $\delta_B(g)_{[t]}(\bfx) = 0 = \delta_B(g_{[t]})(\bfx)$.  On the other hand, if $g(\bfx - \bfb) > t$ for some $\bfb \in B(\bfx;P,-)$, then $\delta_B(g)(\bfx) > t$ and $g_{[t]}(\bfx -  \bfb) = 1$. This shows that $\delta_B(g)_{[t]}(\bfx) = 1 = \delta_B(g_{[t]})(\bfx)$.
\end{proof}
\bigskip
\begin{corollaryA}
\label{Corollary: Opening and Closing ordering relations}
Let $P \subseteq \mathbb{Z}^2$ be an image domain, and let $B_1, B_2$ be SEs. Then the following assertions hold:
\begin{enumerate}[label={\rm (\alph*)}]
\item $O_{B_1} \leq O_{B_2}$ if and only if $O_{B_1}(f) \leq O_{B_2}(f)$ whenever $f \in \mathcal{BI}_P$.
\item $C_{B_1} \leq C_{B_2}$ if and only if $C_{B_1}(f) \leq C_{B_2}(f)$ whenever $f \in \mathcal{BI}_P$.
\end{enumerate}
\end{corollaryA}
\begin{proof}
Since $O_B \circ \tau_t = \tau_t \circ O_B$ and $C_B \circ \tau_t = \tau_t \circ C_B$ by Theorem~\ref{Theorem: Restriction and opening/closing are commutative}, the hypothesis of Proposition~\ref{Proposition: Threshold inclusion relation--key prop} is satisfied. Consequently, assertions~(a) and~(b) follow.   
\end{proof}

Corollary~\ref{Corollary: Opening and Closing ordering relations} provides an efficient tool for verifying whether two opening (resp. closing) operations, $O_{B_1}$ and $O_{B_2}$ (resp. $C_{B_1}$ and $C_{B_2}$), defined by SEs $B_1$ and $B_2$, can be compared under the partial order $\leq$ on $\mathcal{I}_P$.  For example, rather than checking whether $O_{B_1}(f) \leq O_{B_2}(f)$ holds for every grayscale image $f$ on $P$, it suffices to verify the relation for all binary images.

%%%%%%%%%%%%%%%%%%%%%%%%%%%%%%%%%%%%%%%%%%%%%%
%%%%%%%%%%%%%%%%%%%%%%%%%%%%%%%%%%%%%%%%%%%%%%

%%%%%%%%%%%%%%%%%%%%%%%%%%%%%%%%%%%%%%%%%%%%%%
%%%%%%%%%%%%%%%%%%%%%%%%%%%%%%%%%%%%%%%%%%%%%%

%%%%%%%%%%%%%%%%%%%%%%%%%%%%%%%%%%%%%%%%%%%%%%
%%%%%%%%%%%%%%%%%%%%%%%%%%%%%%%%%%%%%%%%%%%%%%
 
%%%%%%%%%%%%%%%%%%%%%%%%%%%%%%%%%%%%%%%%%%%%%%
%%%%%%%%%%%%%%%%%%%%%%%%%%%%%%%%%%%%%%%%%%%%%%

% In particular, $\xi(f) \leq \zeta(f)$ whenever $f \in \mathcal{I}_P$ if and only if $\zeta(f)^{-1}(0) \subseteq \xi(f)^{-1}(0)$ whenever $f \in \mathcal{BI}_P$.
%%%%%%%%%%%%%%%%%%%%%%%%%%%%%%%%%%%%%%%%%%%%%%
%%%%%%%%%%%%%%%%%%%%%%%%%%%%%%%%%%%%%%%%%%%%%%

%%%%%%%%%%%%%%%%%%%%%%%%%%%%%%%%%%%%%%%%%%%%%%
%%%%%%%%%%%%%%%%%%%%%%%%%%%%%%%%%%%%%%%%%%%%%%

\section{Further Discussion on Shift Inclusion}\label{Appendix: Further Discussion on Shift Inclusion}

In Section~\ref{Section: Shift Inclusion for Topological Filtration Construction}, shift inclusion was introduced and shown to ensure the absorption property. Although the main text focuses on rectangular domains, the results extend to general $P \subseteq \mathbb{Z}^2$ (e.g., Theorems~\ref{Theorem: Shift Inclusion as a sufficient condition} and~\ref{Theorem: Shift Inclusion as a sufficient condition, strong form}). This appendix develops the theory in this broader setting and summarizes key properties of the thresholding operation~\eqref{Eq. Thresholding operation}. Parts of this appendix are adapted from~\cite{Hu_PhD_Dissertation}, with reorganization, refinements, and simplifications.

%%%%%%%%%%%%%%%%%%%%%%%%%%%%%%%%%%%%%%%%%%%%%%
%%%%%%%%%%%%%%%%%%%%%%%%%%%%%%%%%%%%%%%%%%%%%%
\subsection{Shift Inclusion over the Entire Space}
\label{Section: Shift Inclusion over the Entire Space}
%%%%%%%%%%%%%%%%%%%%%%%%%%%%%%%%%%%%%%%%%%%%%%
%%%%%%%%%%%%%%%%%%%%%%%%%%%%%%%%%%%%%%%%%%%%%%

In Appendix~\ref{Section: Shift Inclusion over the Entire Space}, we discuss the shift inclusion property of SEs in the space \(\mathbb{Z}^2\), assuming the image domain \(P\) is the entire space \(\mathbb{Z}^2\). Under this assumption, the shift inclusion property admits a cleaner and more elegant formulation than in the case where \(P \neq \mathbb{Z}^2\). Specifically, as presented in Theorem \ref{Decomposition theorem of shift inclusion}, we establish a precise, compact, and computationally feasible condition on SEs \(B_1\) and \(B_2\) that is equivalent to \(B_1 \subseteq_{\mathbb{Z}^2} B_2\) for any \(B_1, B_2 \subseteq \mathbb{Z}^2\).

\bigskip
\begin{lemmaA}\label{LemmaA: positive and negative properties are equivalent when P is the entire space}
Let $P = \mathbb{Z}^2$ be the entire space, and let $B_1, B_2$ be SEs in $\mathbb{Z}^2$. Then, $B_1 \subseteq_{S,P,+} B_2$ if and only if $B_1 \subseteq_{S,P,-} B_2$.    
\end{lemmaA}
\begin{proof}
Let $B$ be an arbitrary SE in $\mathbb{Z}^2$. Then, for any $\mathbf{x} \in P$, the sets $B(\mathbf{x};P,+)$ and $B(\mathbf{x};P,-)$ are exactly the set $B$, as $\mathbf{x} + \mathbf{b} \in \mathbb{Z}^2$ whenever $\mathbf{x} \in P$ and $\mathbf{b} \in B \subseteq P$. This identifications show that the conditions $B_1 \subseteq_{S,P,+} B_2$ and $B_1 \subseteq_{S,P,-} B_2$ are actually equivalent.
\end{proof}
\bigskip

In other words, to verify whether two SEs \( B_1 \) and \( B_2 \) in \( \mathbb{Z}^2 \) satisfy the shift inclusion \( B_1 \subseteq_{S,\mathbb{Z}^2} B_2 \), it is sufficient to check whether \( B_1 \subseteq_{S,\mathbb{Z}^2,+} B_2 \) or \( B_1 \subseteq_{S,\mathbb{Z}^2,-} B_2 \). In particular, the relation \( B_1 \subseteq_{S,\mathbb{Z}^2} B_2 \) can be equivalently redefined as follows: for every \( \mathbf{b}_2 \in B_2 \), there exists \( \mathbf{b}_1 \in B_1 \) such that \( B_1 + (\mathbf{b}_2 - \mathbf{b}_1) \subseteq B_2 \).

\bigskip
\begin{theoremA}
[Decomposition theorem of shift inclusion]
\label{Decomposition theorem of shift inclusion}
Let $P = \mathbb{Z}^2$, and let $B_1 \subseteq B_2 \subseteq P$ be SEs. Then $B_1 \subseteq_{S,P} B_2$ if and only if there are finitely many $\bfv_1, ..., \bfv_n \in \mathbb{Z}^2$ such that
\begin{equation}
\label{Equation: Decomposition thm for shift inclusion}
    B_2 = B_1 \cup \left( \bigcup_{i = 1}^n B_1 + \bfv_i \right).
\end{equation}
\end{theoremA}
\begin{proof}
Suppose there are finitely many $\bfv_1, ..., \bfv_n \in \mathbb{Z}^2$ such that Equation~\eqref{Equation: Decomposition thm for shift inclusion} holds, then each $\bfb_2 \in B_2$ admits a $\bfb_1 \in B_1$ such that $\bfb_2 = \bfb_1$ or $\bfb_2 = \bfb_1 + \bfv_i$ for some $i \in \{ 1,2, ..., n \}$. In particular, $B_1 + (\bfb_2 - \bfb_1) = B_1$ or $B_1 + \bfv_i$, which is a subset of $B_2$. Therefore, we conclude that $B_1 \subseteq_{S,P} B_2$.   

Conversely, suppose $B_1 \subseteq_{S,P} B_2$. Then, every $\bfb_2 \in B_2$ admits a $\bfb_1 \in B_1$ such that $B_1 + (\bfb_2 - \bfb_1) \subseteq B_2$. For each $\bfb_2$, we fix one $\bfb_1 \in B_1$ satisfying this property, and then we get a function $\phi: B_2 \rightarrow B_1$ with $\mathbf{b}_2 \mapsto \phi(\mathbf{b}_2) := \mathbf{b}_1$. In particular, we have $B_1 + (\bfb_2 - \phi(\bfb_2)) \subseteq B_2$ for every $\bfb_2 \in B_2$. Therefore,
\begin{equation*}
B_1 \cup \left( \bigcup_{\bfb_2 \in B_2} B_1 + (\bfb_2 - \phi(\bfb_2)) \right) \subseteq B_2.
\end{equation*}
On the other hand, if $\bfb_2 \in B_2$, then $\bfb_2 = \phi(\bfb_2) + (\bfb_2 - \phi(\bfb_2)) \in B_1 + (\bfb_2 - \phi(\bfb_2))$, and this shows that
\begin{equation*}
B_2 \subseteq \bigcup_{\bfb_2 \in B_2} B_1 + (\bfb_2 - \phi(\bfb_2)) \subseteq  B_1 \cup \left( \bigcup_{\bfb_2 \in B_2} B_1 + (\bfb_2 - \phi(\bfb_2)) \right).
\end{equation*}
Finally, since $B_2$ is a finite set, we may choose finitely many  $\bfv_1, \bfv_2, ..., \bfv_n \in B_2$ such that Equation~\eqref{Equation: Decomposition thm for shift inclusion} holds. 
\end{proof}
\bigskip

The following theorem shows that, when $P = \mathbb{Z}^2$, the shift inclusion relation $B_1 \subseteq_{S,P} B_2$ holds if and only if $B_1$ and $B_2$ satisfy the absorption property stated in Equation~\eqref{Equation: Absorption property-X-form}.

\bigskip
\begin{theoremA}
\label{TheoremA: Shift Inclusion over the entire space is equivalent to the absorption property}
Let $B_1 \subseteq B_2$ be SEs in $\mathbb{Z}^2$. Then the following assertions hold:
\begin{itemize}
\item[$({\rm a})$] $O_{B_1}(g)^{-1}(0) \subseteq O_{B_2}(g)^{-1}(0)$ whenever $g \in \mathcal{I}_{\mathbb{Z}^2}$ if and only if $B_1 \subseteq_{S,\mathbb{Z}^2} B_2$.
\item[$({\rm b})$] $C_{B_2}(g)^{-1}(0) \subseteq C_{B_1}(g)^{-1}(0)$ whenever $g \in \mathcal{I}_{\mathbb{Z}^2}$ if and only if $B_1 \subseteq_{S,\mathbb{Z}^2} B_2$.
\end{itemize}
\end{theoremA}
\begin{proof}
By a symmetric argument, it is sufficient to prove assertion (a). The ``if'' part of (a) follows from Theorem \ref{Theorem: Shift Inclusion as a sufficient condition}.  Conversely, for the ``only if'' part of assertion (a), we suppose $B_1 \nsubseteq_{S,\mathbb{Z}^2} B_2$ and prove that there is a image $g \in \mathcal{I}_{\mathbb{Z}^2}$ such that
\begin{equation}
\label{Eq. Goal for the Thm: Shift Inclusion over the entire space is equivalent to the absorption property}
O_{B_1}(g)^{-1}(0) \nsubseteq O_{B_2}(g)^{-1}(0).    
\end{equation}
By the assumption of $B_1 \nsubseteq_{S,\mathbb{Z}^2} B_2$, there is a $\bfb_2 \in B_2$ such that $B_1 + (\bfb_2 - \bfb_1) \nsubseteq B_2$ for every $\bfb_1 \in B_1$.  For each $\bfb_1$, we fix one $\bfb_1' \in B_1$ such that $\bfb_1' + (\bfb_2 - \bfb_1) \notin B_2$. This induces a function $\phi: B_1 \rightarrow B_1$ such that $\mathbf{b}_1 \mapsto \phi(\mathbf{b}_1) := \mathbf{b}_1'$. In particular, $\phi(\bfb_1) + (\bfb_2 - \bfb_1) \notin B_2$ for every $\bfb_1 \in B_1$.

We consider the characteristic function $\chi_{B_2}: \mathbb{Z}^2 \rightarrow \{0, 1\}$, defined by $\chi_{B_2}(\bfx) = 1$ if $\bfx \in B_2$, and $\chi_{B_2}(\bfx) = 0$ otherwise. Then $\chi_{B_2}(\bfb_2 -\bfb_1 + \phi(\bfb_1)) = 0$ for every $\bfb_1 \in B_1$. This shows that $O_{B_1}(\chi_{B_2})(\bfb_2) = 0$ by Lemma \ref{Lemma: crucial observation for shift inclusion}(a). On the other hand, we have $\chi_{B_2}(\bfb_2 - \bfb_2 + \bfb_2') = \chi_{B_2}(\bfb_2') = 1 \neq 0$ for every $\bfb_2' \in B_2$. By Lemma \ref{Lemma: crucial observation for shift inclusion}(a) again, we obtain $O_{B_2}(\chi_{B_2})(\bfb_2) \neq 0$. Take $g = \chi_{B_2}$, then  \eqref{Eq. Goal for the Thm: Shift Inclusion over the entire space is equivalent to the absorption property} holds, and this proves the ``only if'' part of (a).
\end{proof}
\bigskip

In summary, SEs $B_1$ and $B_2$ in $\mathbb{Z}^2$ satisfy the shift inclusion relation if and only if $B_2$ can be expressed as a finite union of translated copies of $B_1$. This equivalence provides an intuitive understanding of the term ``shift inclusion.'' However, it is important to note that these two conditions do not necessarily imply one another, even when considering rectangular image domains $P \subseteq \mathbb{Z}^2$ (see, e.g., Appendix \ref{Section: Shift Inclusion vs. Weak Shift Inclusion}).

\paragraph{Examples}
In Section \ref{Section: Feasible Examples}, we showed that some commonly used sequences of SEs satisfy the shift inclusion relation when the domain $P$ is a rectangle. However, verifying shift inclusion over arbitrary image domains can be challenging, as it requires checking both positive and negative conditions. Fortunately, when the domain is the entire space ($P = \mathbb{Z}^2$), the decomposition theorem offers a systematic approach for constructing shift inclusion sequences. 

\bigskip
\begin{exampleA}
\label{Example: non-symmetirc Shift inclusion first example}
The following examples are shift inclusion sequences in $\mathbb{Z}^2$:
\begin{equation*}
\ytableausetup{centertableaux}
\begin{split}
B_1 &= \begin{ytableau}
\mathbf{0}
\end{ytableau}\quad,\quad
B_2 = \begin{ytableau}
\ & \mathbf{0} \\
\ & \\
\end{ytableau}\quad,\quad
B_3 = \begin{ytableau}
\ & \ & \mathbf{0} \\
\ & \  & \ \\
\end{ytableau}\quad,\quad
B_4 = \begin{ytableau}
\ & \ & \ & \mathbf{0} \\
\ & \ & \ & \ \\
\end{ytableau}\quad. \\
B_1 &= \begin{ytableau}
\mathbf{0}
\end{ytableau}\quad,\quad
B_2 = \begin{ytableau}
\ & \none \\
\mathbf{0} & \\
\end{ytableau}\quad,\quad
B_3 = \begin{ytableau}
\ & \none & \none \\
\ & \  & \none \\
\mathbf{0}  & \ & \ \\
\end{ytableau}\quad,\quad
B_4 = \begin{ytableau}
\ & \none & \none & \none \\
\ & \ & \none  & \none \\
\ & \ & \ & \none \\
\mathbf{0} & \  & \ & \ \\
\end{ytableau}\quad. \\
B_1 &= \begin{ytableau}
\mathbf{0}
\end{ytableau}\quad,\quad
B_2 = \begin{ytableau}
\ & \mathbf{0} \\
\ & \\
\end{ytableau}\quad,\quad
B_3 = \begin{ytableau}
\ & \mathbf{0} & \ \\
\ & \  & \ \\
\end{ytableau}\quad,\quad
B_4 = \begin{ytableau}
\ & \mathbf{0} & \ & \ \\
\ & \ & \ & \ \\
\end{ytableau}\quad.
\end{split}
\end{equation*}
\end{exampleA}
\bigskip
Moreover, beyond ``connected'' SEs, it is also possible to construct other types of SEs that satisfy the shift inclusion relation.
\bigskip
\begin{exampleA}
\label{ExampleA: Shift inclusion non-connected}
The following SEs $B_1$ and $B_2$ satisfy the shift inclusion relation in $\mathbb{Z}^2$:
\begin{equation*}
\ytableausetup{centertableaux}
\begin{split}
B_1 = \begin{ytableau}
 \none & \none & \ \\
 \none & \none & \none \\
 \mathbf{0} & \none & \none \\
\end{ytableau} = \{ (0,0), (2,2) \}\quad \text{and} \quad
B_2 = \begin{ytableau}
 \none & \none & \ \\
 \none & \none & \none \\
 \mathbf{0} & \none & \ \\
 \none & \none & \none \\
 \ & \none & \none \\
\end{ytableau} = B_1 \cup (-2\bfe_2 + B_1),
\end{split}
\end{equation*}
where $\{ \mathbf{e}_1, \mathbf{e}_2 \}$ is the standard $\mathbb{Z}$-basis of $\mathbb{Z}^2$.
\end{exampleA}
\bigskip

The decomposition theorem of shift inclusion offers a systematic method for constructing sequential SEs that satisfy the desired absorption property. However, we will show by examples in Section \ref{Section: Weak Shift Inclusion} that this process cannot be directly applied when $P \neq \mathbb{Z}^2$. Specifically, the decomposition theorem fails in the case where $P \neq \mathbb{Z}^2$, and, in particular, the relation $\subseteq_{S,\mathbb{Z}^2}$ does not generally imply $\subseteq_{S,P}$.

\bigskip
\begin{exampleA}
\label{ExampleA: desired property doesn't imply subseteq_S,P}
Consider the image domain \( P \subseteq \mathbb{Z}^2 \) defined by
\begin{equation*}
P = 
\begin{ytableau}
\mathbf{0} & \ \\
\ & \none \\
\end{ytableau}
= \{ (0,0), (1,0), (0,-1) \},
\end{equation*}
and SEs 
\begin{equation*}
B_1 = \begin{ytableau}
\mathbf{0} \\
\ \\
\end{ytableau} \quad \subseteq \quad
B_2 = \begin{ytableau}
\mathbf{0} & \ \\
\ & \none \\
\end{ytableau} \quad = P.
\end{equation*}
Then \( B_1 \nsubseteq_{S,P} B_2 \). However, the absorption property with respect to \( B_1 \) and \( B_2 \) is satisfied. In other words, for any image $g \in \mathcal{I}_P$,  
\begin{equation*}
O_{B_2}(g) \leq O_{B_1}(g) \quad \text{and} \quad C_{B_1}(g) \leq C_{B_2}(g).
\end{equation*}  
In particular, the shift inclusion relation \( \subseteq_{S,P} \) is not a necessary condition for the absorption property~\eqref{Equation: Absorption property}.
\end{exampleA}
\begin{proof}
We first show that \( B_1 \nsubseteq_{S,P} B_2 \).  Specifically, consider $\mathbf{x} = \mathbf{0} \in P$ and $\mathbf{b}_2 = (1,0) \in B_2(\mathbf{x}; P, +)$. There are exactly two points in $B_1(\mathbf{x}; P, +)$---the points $\mathbf{b}_1 = \mathbf{0} = (0,0)$ and $\mathbf{b}_1' = (0,-1)$. Then,
\begin{equation*}
\begin{split}
B_1 + (\mathbf{b}_2 - \mathbf{b}_1) = B_1 + (1,0) \nsubseteq B_2 \quad \text{and} \quad B_1 + (\mathbf{b}_2 - \mathbf{b}_1') = B_1 + (1,1) \nsubseteq B_2
\end{split}    
\end{equation*}
since $\mathbf{b}_1' + (1,0) = (1,-1) \notin B_2$ and $\mathbf{b}_1 + (1,1) = (1,1) \notin B_2$. This shows that the SEs $B_1$ and $B_2$ do not satisfy the shift inclusion property.

To show $O_{B_2}(g) \leq O_{B_1}(g)$ and $C_{B_1}(g) \leq C_{B_2}(g)$ whenever $g \in \mathcal{I}_P$, it is sufficient to verify they hold for all $g \in \mathcal{BI}_P$ by Proposition \ref{Proposition: Threshold inclusion relation--key prop}. In particular, the set \( \mathcal{BI}_P \) consists of exactly \( 2^3 = 8 \) elements, and we examine all samples in \( \mathcal{BI}_P \).
%  \medskip & \smallskip
\medskip
\begin{enumerate}
\item For $g = \begin{ytableau}
1 & 1 \\
1 & \none \\
\end{ytableau} \ $, $O_{B_2}(g) = \begin{ytableau}
1 & 1 \\
1 & \none \\
\end{ytableau} = O_{B_1}(g) \text{ and } C_{B_1}(g) = \begin{ytableau}
1 & 1 \\
1 & \none \\
\end{ytableau} = C_{B_2}(g).$
\medskip
\item For $g = \begin{ytableau}
1 & 0 \\
1 & \none \\
\end{ytableau} \ $, we have the following order relations:
\begin{equation*}
O_{B_2}(g) = \begin{ytableau}
0 & 0 \\
1 & \none \\
\end{ytableau} < \begin{ytableau}
1 & 0 \\
1 & \none \\
\end{ytableau} = O_{B_1}(g)
\text{ and }
C_{B_1}(g) = \begin{ytableau}
1 & 0 \\
1 & \none \\
\end{ytableau} < \begin{ytableau}
1 & 1 \\
1 & \none \\
\end{ytableau}
= C_{B_2}(g).
\end{equation*}
\medskip
\item For $g = \begin{ytableau}
0 & 1 \\
1 & \none \\
\end{ytableau} \ $, $O_{B_2}(g) = \begin{ytableau}
0 & 1 \\
1 & \none \\
\end{ytableau} = O_{B_1}(g) \text{ and } C_{B_1}(g) = \begin{ytableau}
0 & 1 \\
1 & \none \\
\end{ytableau}
= C_{B_2}(g)$.
\medskip
\item For $g = \begin{ytableau}
1 & 1 \\
0 & \none \\
\end{ytableau} \ $, $O_{B_2}(g) = \begin{ytableau}
0 & 1 \\
0 & \none \\
\end{ytableau} = O_{B_1}(g) \text{ and } C_{B_1}(g) = \begin{ytableau}
1 & 1 \\
1 & \none \\
\end{ytableau} = C_{B_2}(g)$
\medskip
\item For $g = \begin{ytableau}
0 & 0 \\
1 & \none \\
\end{ytableau} \ $, $O_{B_2}(g) = \begin{ytableau}
0 & 0 \\
1 & \none \\
\end{ytableau} = O_{B_1}(g) \text{ and } C_{B_1}(g) = \begin{ytableau}
0 & 0 \\
1 & \none \\
\end{ytableau} = C_{B_2}(g)$
\medskip
\item For $g = \begin{ytableau}
0 & 1 \\
0 & \none \\
\end{ytableau} \ $, $O_{B_2}(g) = \begin{ytableau}
0 & 1 \\
0 & \none \\
\end{ytableau} = O_{B_1}(g) \text{ and } C_{B_1}(g) = \begin{ytableau}
0 & 1 \\
0 & \none \\
\end{ytableau} = C_{B_2}(g)$
\medskip
\item For $g = \begin{ytableau}
1 & 0 \\
0 & \none \\
\end{ytableau} \ $, we have the following order relations:
\begin{equation*}
O_{B_2}(g) = \begin{ytableau}
0 & 0 \\
0 & \none \\
\end{ytableau} = O_{B_1}(g) \text{ and } C_{B_1}(g) = \begin{ytableau}
1 & 0 \\
1 & \none \\
\end{ytableau} < \begin{ytableau}
1 & 1 \\
1 & \none \\
\end{ytableau} = C_{B_2}(g). 
\end{equation*}
\item For $g = \begin{ytableau}
0 & 0 \\
0 & \none \\
\end{ytableau} \ $, $O_{B_2}(g) = \begin{ytableau}
0 & 0 \\
0 & \none \\
\end{ytableau} = O_{B_1}(g) \text{ and } C_{B_1}(g) = \begin{ytableau}
0 & 0 \\
0 & \none \\
\end{ytableau} = C_{B_2}(g)$.
\end{enumerate}
\medskip
By the discussions above, we conclude that $O_{B_2}(g) \leq O_{B_1}(g) \text{ and } C_{B_1}(g) \leq C_{B_2}(g)$ whenever $g \in \mathcal{I}_P$, i.e., the absorption property for the SEs $B_1$ and $B_2$ is satisfied.
\end{proof}
\bigskip 

Example~\ref{ExampleA: desired property doesn't imply subseteq_S,P} demonstrates that the shift inclusion condition is generally not necessary for the absorption property. To find an equivalent condition for SEs, we introduce the concept of \textit{weak shift inclusion} in the next section (Section~\ref{Section: Weak Shift Inclusion}) and prove that it is equivalent to the absorption property (see  \eqref{Equation: Absorption property-X-form}). This result provides a theoretical characterization equivalent to the absorption property for constructing topological filtrations, offering greater potential for investigating topological filtrations associated with images defined on irregular domains.
 
%%%%%%%%%%%%%%%%%%%%%%%%%%%%%%%%%%%%%%%%%%%%%%
%%%%%%%%%%%%%%%%%%%%%%%%%%%%%%%%%%%%%%%%%%%%%%
\subsection{Weak Shift Inclusion}
\label{Section: Weak Shift Inclusion}
In Appendix \ref{Section: Weak Shift Inclusion}, we introduce a generalized shift relation between SEs $B_1$ and $B_2$, termed \textit{weak shift inclusion}, denoted by $B_1 \subseteq_{WS,P} B_2$ with respect to the image domain $P$. This concept extends the notion of $B_1 \subseteq_{S,P} B_2$ and provides an equivalent condition for ensuring the absorption property of opening and closing operations on arbitrary image domains $P \subseteq \mathbb{Z}^2$, as defined in \eqref{Equation: Absorption property-X-form}.
\bigskip
\begin{definitionA}
Let \( P \subseteq \mathbb{Z}^2 \) be an image domain, and let \( B_1, B_2 \) be SEs in \( \mathbb{Z}^2 \). We say that \( B_1 \) is \textbf{weakly shift included} with respect to \( P \) in \( B_2 \), denoted by \( B_1 \subseteq_{WS,P} B_2 \), if \( B_1 \subseteq B_2 \) and the following properties are satisfied:
\begin{itemize}
\item[$({\rm a})$] (\textbf{Weak positive property}, denoted by $B_1 \subseteq_{WS,P,+} B_2$) For any $\bfx \in P$ and $\bfb_2 \in B_2(\bfx;P,+)$, there exists a $\bfb_1 \in B_1(\bfx;P,+)$ such that $B_1(\bfx + \bfb_1;P,-) + (\bfb_2 - \bfb_1) \subseteq B_2$;
\item[$({\rm b})$] (\textbf{Weak negative property}, denoted by $B_1 \subseteq_{WS,P,-} B_2$) For any $\bfx \in P$ and $\bfb_2 \in B_2(\bfx;P,-)$, there exists a $\bfb_1 \in B_1(\bfx;P,-)$ such that $B_1(\bfx - \bfb_1;P,+) + (\bfb_2 - \bfb_1) \subseteq B_2$.
\end{itemize}
The relation \( B_1 \subseteq_{WS,P} B_2 \) between SEs in \( \mathbb{Z}^2 \) is called the \textbf{weak shift inclusion}.
\end{definitionA}
\bigskip
\begin{remarkA}
Analogous to the shift inclusion sequence $B_1 \subseteq_{S,P} B_2 \subseteq_{S,P} \cdots \subseteq_{S,P} B_n$, a sequence $B_1, B_2, ..., B_n$ of SEs such that $B_1 \subseteq_{WS,P} B_2 \subseteq_{WS,P} \cdots \subseteq_{WS,P} B_n$ is called a \textbf{weak shift inclusion sequence}.
\end{remarkA}
\bigskip

In Remark~\ref{Remark: Positive/negative relationship}, we established the relation between the positive and negative properties, namely that \( B_1 \subseteq_{S,P,+} B_2 \) if and only if \( -B_1 \subseteq_{S,P,-} -B_2 \). For the case of weak shift inclusion, a similar result holds: \( B_1 \subseteq_{WS,P,+} B_2 \) if and only if \( -B_1 \subseteq_{WS,P,-} -B_2 \). We provide the formal statement and proof of this fact below:
\bigskip
 
\begin{propositionA}\label{Proposition: relationship between the weak positive and negative properties}
Let $B_1$ and $B_2$ be SEs in $\mathbb{Z}^2$. Then, \( B_1 \subseteq_{WS,P,+} B_2 \) if and only if \( -B_1 \subseteq_{WS,P,-} -B_2 \). 
\end{propositionA}
\begin{proof}
To simplify the notations, we denote $-B_1$ and $-B_2$ by $C_1$ and $C_2$, respectively. Suppose \( B_1 \subseteq_{WS,P,+} B_2 \), we claim \( C_1 \subseteq_{WS,P,-} C_2 \). Suppose $\mathbf{x} \in P$ and $\mathbf{c}_2 \in C_2(\mathbf{x};P,-) \subseteq C_2 = -B_2$, we have $\mathbf{c}_2 = -\mathbf{b}_2$ for some $\mathbf{b}_2 \in B_2$. Furthermore, we have $\mathbf{x} + \mathbf{b}_2 = \mathbf{x} - \mathbf{c}_2 \in $, and this implies $\mathbf{b}_2 \in B_2(\mathbf{x};P,+)$. By the assumption of \( B_1 \subseteq_{WS,P,+} B_2 \), there is a $\mathbf{b}_1 \in B_1(\mathbf{x};P,+)$ such that
\begin{equation}\label{Eq. relationship between the weak positive and negative properties}
B_1(\bfx + \bfb_1;P,-) + (\bfb_2 - \bfb_1) \subseteq B_2.
\end{equation}
Let $\mathbf{c}_1 = -\mathbf{b}_1 \in -B_1 = C_1$. Then, $\mathbf{c}_1 \in C_1(\mathbf{x};P,-)$ since $\mathbf{x} - \mathbf{c}_1 = \mathbf{x} + \mathbf{b}_1 \in P$. To conclude that \( C_1 \subseteq_{WS,P,-} C_2 \), it is sufficient to show that 
\begin{equation*}
C_1(\bfx - \mathbf{c}_1;P,+) + (\mathbf{c}_2 - \mathbf{c}_1) \subseteq C_2.
\end{equation*}
Suppose $\mathbf{d} \in C_1(\bfx - \mathbf{c}_1;P,+)$, we observe that $-\mathbf{d} \in -C_1 = B_1$ and 
\begin{equation*}
\mathbf{x} + \mathbf{b}_1 - (-\mathbf{d}) = \mathbf{x} + \mathbf{b}_1 + \mathbf{d} = \mathbf{x} - \mathbf{c}_1 + \mathbf{d} \in P. 
\end{equation*}
This shows that $-\mathbf{d} \in B_1(\mathbf{x} + \mathbf{b}_1; P,-)$. In particular, we have
\begin{equation*}
\begin{split}
\mathbf{d} + (\mathbf{c}_2 - \mathbf{c}_1) &= -(-\mathbf{d} +(\mathbf{b}_2 - \mathbf{b}_1)).
\end{split}
\end{equation*}
By the inclusion relation in~\eqref{Eq. relationship between the weak positive and negative properties}, as $-\mathbf{d} \in B_1(\mathbf{x} + \mathbf{b}_1; P,-)$, the point $\mathbf{d} + (\mathbf{c}_2 - \mathbf{c}_1)$ is an element in $-B_2 = C_2$. This proves that \( C_1 \subseteq_{WS,P,-} C_2 \).
 
Conversely, suppose \( C_1 \subseteq_{WS,P,-} C_2 \), we claim \( B_1 \subseteq_{WS,P,+} B_2 \). Suppose $\mathbf{x} \in P$ and $\mathbf{b}_2 \in B_2(\mathbf{x};P,+)$, then $-\mathbf{b}_2 \in C_2$ and $\mathbf{x} - (-\mathbf{b}_2) = \mathbf{x} + \mathbf{b}_2 \in P$, i.e., $-\mathbf{b}_2 \in C_2(\mathbf{x};P,-)$. Because $C_1 \subseteq_{WS,P,-} C_2$, there is a $\mathbf{c}_1 \in C_1(\mathbf{x};P,-)$ such that
\begin{equation}\label{Eq. relationship between the weak positive and negative properties-2}
C_1(\mathbf{x}-\mathbf{c}_1; P,+) + (\mathbf{c}_2 - \mathbf{c}_1) \subseteq C_2.    
\end{equation}
Let $\mathbf{b}_1 = -\mathbf{c}_1 \in -C_1 = B_1$. Then, \( \mathbf{b}_1 \in B_1(\mathbf{x}; P, +) \) since \( \mathbf{x} + \mathbf{b}_1 = \mathbf{x} - \mathbf{c}_1 \in P \). Suppose $\mathbf{d} \in B_1(\mathbf{x}+\mathbf{b}_1; P, -)$, we observe that $-\mathbf{d} \in -B_1 = C_1$ and
\begin{equation*}
\mathbf{x} - \mathbf{c}_1 + (-\mathbf{d}) = \mathbf{x} + \mathbf{b}_1 + (-\mathbf{d}) = \mathbf{x} + \mathbf{b}_1 -\mathbf{d} \in P.    
\end{equation*}
This shows that $-\mathbf{d} \in C_1(\mathbf{x}-\mathbf{c}_1;P,+)$. In particular, we have 
\begin{equation*}
\mathbf{d} + (\mathbf{b}_2 - \mathbf{b}_1) = -(-\mathbf{d} + (-\mathbf{b}_2 + \mathbf{b}_1)) = -(-\mathbf{d} + (\mathbf{c}_2 - \mathbf{c}_1)). 
\end{equation*}
By the inclusion relation in~\eqref{Eq. relationship between the weak positive and negative properties-2}, as $-\mathbf{d} \in C_1(\mathbf{x}-\mathbf{c}_1;P,+)$, the point $\mathbf{d} + (\mathbf{b}_2 - \mathbf{b}_1)$ is an element in $-C_2 = B2$. This proves that $B_1 \subseteq_{WS,P,+} B_2$. 
\end{proof}
\bigskip
 
By examining the definitions of the weak positive/negative properties, the following proposition formally establishes that, under these conditions, the shift inclusion relation implies the weak shift inclusion relation. In other words, the shift inclusion relation is a special case of the weak shift inclusion relation, which justifies the use of the term ``weak'' in naming the relations \( \subseteq_{WS,P,+} \), \( \subseteq_{WS,P,-} \), and \( \subseteq_{WS,P} \).

\bigskip
\begin{propositionA}\label{Proposition: S implies WS}
Given an image domain $P \subseteq \mathbb{Z}^2$ and SEs $B_1, B_2 \subseteq \mathbb{Z}^2$, the following statements hold:
\begin{itemize}
\item[$({\rm a})$] If $B_1 \subseteq_{S,P,+} B_2$, then $B_1 \subseteq_{WS,P,+} B_2$.
\item[$({\rm b})$] If $B_1 \subseteq_{S,P,-} B_2$, then $B_1 \subseteq_{WS,P,-} B_2$.
\end{itemize}
\end{propositionA}
\begin{proof}
By a symmetric argument, it is sufficient to prove (a).  Suppose $B_1 \subseteq_{S,P,+} B_2$ and $\mathbf{b}_2 \in B_2(\mathbf{x};P,+)$, then there is a $\mathbf{b}_1 \in B_1(\mathbf{x};P,+)$ such that $B_1 + (\mathbf{b}_2 - \mathbf{b}_1) \subseteq B_2$. Because $B_1(\bfx + \bfb_1;P,-) \subseteq B_1$, the inclusion $B_1(\bfx + \bfb_1;P,-) + (\bfb_2 - \bfb_1) \subseteq B_2$ holds. Therefore, $B_1 \subseteq_{WS,P,+} B_2$.
\end{proof}
%%%%%%%%%%%%%%%%%%%%%%%%%%%%%%%%%%%%%%%%%%%%%%%%%%%%%%%%%%%%%%%%%%%%%%%%%%%%%%%%%%%%%%%%%%%%%%%%%%%%%%%%%%%%
The following theorems, namely Theorems~\ref{Theorem: Equivalence of WS,P and Desired property-form 1} and~\ref{Theorem: Equivalence of WS,P and Desired property-form 2}, constitute a central result in the theory of weak shift inclusion. They establish that, for an arbitrary image domain \( P \subseteq \mathbb{Z}^2 \), the weak shift inclusion relation \( \subseteq_{WS,P} \) is equivalent to the desired absorption property. This demonstrates that weak shift inclusion provides a more general---indeed, strictly weaker---condition for achieving the absorption property than the standard shift inclusion.

%%%%%%%%%%%%%%%%%%%%%%%%%%%%%%%%%%%%%%%%%%%%%%%%%%%%%%%%%%%%%%%%%%%%%%%%%%%%%%%%%%%%%%%%%%%%%%%%%%%%%%%%%%%%
\bigskip
\begin{theoremA}[Main theorem on weak shift inclusion, first form]
\label{Theorem: Equivalence of WS,P and Desired property-form 1}
Let $P \subseteq \mathbb{Z}^2$ be an image domain, and let $B_1 \subseteq B_2 \subseteq \mathbb{Z}^2$ be SEs. Then, the following assertions hold:
\begin{itemize}
\item[$({\rm a})$] $\mathcal{X}_{O_{B_1}(g)} \subseteq \mathcal{X}_{O_{B_2}(g)}$ whenever $g \in \mathcal{BI}_P$ if and only if $B_1 \subseteq_{WS,P,-} B_2$;
\item[$({\rm b})$] $\mathcal{X}_{C_{B_2}(g)} \subseteq \mathcal{X}_{C_{B_1}(g)}$ whenever $g \in \mathcal{BI}_P$ if and only if $B_1 \subseteq_{WS,P,+} B_2$ .
\end{itemize}
\end{theoremA}
\begin{proof}
Instead of proving assertion~(b) by directly imitating the argument for~(a), we adopt an alternative approach that leverages Proposition~\ref{Proposition: relationship between the weak positive and negative properties} to show that it is sufficient to prove assertion~(a). Suppose assertion (a) holds, then $\mathcal{X}_{C_{B_2}(g)} \subseteq \mathcal{X}_{C_{B_1}(g)}$ whenever $g \in \mathcal{BI}_P$ if and only if $C_{B_1}(g) \leq C_{B_2}(g)$ whenever $g \in \mathcal{BI}_P$. By Proposition \ref{Proposition: O2 <= O1 iff C-1 <= C-2}, it is equivalent to the condition: $O_{-B_2}(g) \leq O_{-B_1}(g)$ whenever $g \in \mathcal{BI}_P$. By (a), it is equivalent to $-B_1 \subseteq_{WS,P,-} -B_2$. By Proposition~\ref{Proposition: relationship between the weak positive and negative properties}, it is equivalent to $B_1 \subseteq_{WS,P+} B_2$. This shows that (b) holds if (a) is satisfied.

To prove (a), we first assume $B_1 \subseteq_{WS,P,-} B_2$ and $O_{B_1}(g)(\bfx) = 0$. We prove that the condition of Proposition \ref{Lemma: crucial observation for shift inclusion}(a) holds: for every $\bfx \in P$ and $\mathbf{b}_2 \in B_2(\mathbf{x};P,-)$, there is a $\mathbf{b}_2' \in B_2(\mathbf{x}-\mathbf{b}_2; P, +)$ such that $g(\mathbf{x}-\mathbf{b}_2+\mathbf{b}_2') = 0$.  Specifically, suppose $\bfx \in P$ and $\bfb_2 \in B_2(\bfx;P,-)$. Because $B_1 \subseteq_{WS,P,-} B_2$, there is a $\bfb_1 \in B_1(\bfx;P,-)$ such that
\begin{equation}\label{Eq. the equation of weak shift inclusion}
B_1(\bfx - \bfb_1;P,+) + (\bfb_2 - \bfb_1) \subseteq B_2.  
\end{equation}
On the other hand, by applying Proposition \ref{Lemma: crucial observation for shift inclusion}(a) on the condition $O_{B_1}(g)(\bfx) = 0$, there is a $\bfb_1' \in B_1(\bfx-\bfb_1;P,+)$ such that $g(\bfx - \bfb_1 + \bfb_1') = 0$, as $\mathbf{b}_1 \in B_1(\mathbf{x};P,-)$. By \eqref{Eq. the equation of weak shift inclusion}, there is a  $\bfb_2' \in B_2$ such that  $\mathbf{b}_1' + \mathbf{b}_2 - \mathbf{b}_1 = \mathbf{b}_2'$. In particular, $\mathbf{b}_2' \in B_2(\mathbf{x}-\mathbf{b}_2;P,+)$ since $\mathbf{x}-\mathbf{b}_2+\mathbf{b}_2' = \mathbf{x}-\mathbf{b}_1+\mathbf{b}_1' \in P$. Furthermore, we have 
\begin{equation*}
g(\bfx - \bfb_2 + \bfb_2') = g(\bfx - \bfb_1 + \bfb_1') = 0. 
\end{equation*}
This shows that $O_{B_2}(g)(\mathbf{x}) = 0$. Therefore, we conclude that $\mathcal{X}_{O_{B_1}(g)} \subseteq \mathcal{X}_{O_{B_2}(g)}$ whenever $g \in \mathcal{BI}_P$ if $B_1 \subseteq_{WS,P,-} B_2$.

Conversely, we prove the following assertion: if $B_1 \nsubseteq_{WS,P,-} B_2$, then there is a binary image $g \in \mathcal{BI}_P$ such that $\mathcal{X}_{O_{B_1}(g)} \nsubseteq \mathcal{X}_{O_{B_2}(g)}$. By the definition of the negative property of the weak shift inclusion, there is a point $\bfx_0 \in P$ and a $\bfb_2 \in B_2(\bfx_0;P,-)$ satisfying the following property: for every $\bfb_1 \in B_1(\bfx_0;P,-)$, there is a $\bfb_1' \in B_1(\bfx_0 - \bfb_1;P,+)$ such that $\bfb_1' + \bfb_2 - \bfb_1 \notin B_2$. In particular, $\bfx_0 - \bfb_2 + \bfb_2' \neq \bfx_0 - \bfb_1 + \bfb_1'$ for every $\bfb_2' \in B_2$. Then, consider the binary image $g: P \rightarrow \{ 0,1 \}$ defined by
\begin{equation}
\label{Eq. 1 WS}
 	g(\bfx) = \begin{cases}
 	1 &\null\text{ if } \bfx \in ((\bfx_0 - \bfb_2) + B_2) \cap P, \\
 	0  &\null\text{ otherwise.}
 	\end{cases}
\end{equation}
Note that $((\bfx_0 - \bfb_2) + B_2) \cap P$ is nonempty since it contains $\bfx_0$. This shows that $g(\bfx_0 - \bfb_1 + \bfb_1') = 0$ since $\bfx_0 - \bfb_1 + \bfb_1' \in P$ doesn't belong to the set $(\bfx_0 - \bfb_2) + B_2$.  By Proposition \ref{Lemma: crucial observation for shift inclusion}(a), 
\begin{equation*}
O_{B_1}(g)(\bfx_0) = 0.    
\end{equation*}
On the other hand, we observe that
\begin{equation*}
\epsilon_{B_2}(g)(\mathbf{x}_0 - \mathbf{b}_2) = \min g \bigg( ((\mathbf{x}_0 - \mathbf{b}_2) + B_2) \cap P \bigg) = 1
\end{equation*}
by the \eqref{Eq. 1 WS}. By the definition of opening operation, we obtain
\begin{equation*}
O_{B_2}(g)(\bfx_0) = (\delta_{B_2} \circ \epsilon_{B_2})(g)(\bfx_0) = \max \epsilon_{B_2}(g)\bigg( (\bfx_0 - B_2) \cap P \bigg) = 1 \neq 0.    
\end{equation*}
This establishes that \( \mathcal{X}_{O_{B_1}(g)} \nsubseteq \mathcal{X}_{O_{B_2}(g)} \) whenever \( B_1 \nsubseteq_{WS,P,-} B_2 \). Hence, we have completed the proof of assertion~(a).
\end{proof}
 
%%%%%%%%%%%%%%%%%%%%%%%%%%%%%%%%%%%%%%%%%%%%%%%%%%%%%%%%%%%%%%%%%%%%%%%%%%%%%%%%%%%%%%%%%%%%%%%%%%%%%%%%%%%%
Furthermore, as stated in the following theorem, by leveraging the results of Corollary~\ref{Corollary: Opening and Closing ordering relations}, the desired absorption property of the mathematical opening and closing operations~\eqref{Equation: Absorption property}, defined on an arbitrary image domain \( P \subseteq \mathbb{Z}^2 \), can be interpreted as the following equivalent weak shift inclusion condition.
%%%%%%%%%%%%%%%%%%%%%%%%%%%%%%%%%%%%%%%%%%%%%%%%%%%%%%%%%%%%%%%%%%%%%%%%%%%%%%%%%%%%%%%%%%%%%%%%%%%%%%%%%%%%
\bigskip
\begin{theoremA}[Main theorem on weak shift inclusion, second form]
\label{Theorem: Equivalence of WS,P and Desired property-form 2}
Let $P \subseteq \mathbb{Z}^2$ be an image domain, and let $B_1 \subseteq B_2 \subseteq \mathbb{Z}^2$ be SEs. Then, the following assertions hold:
\begin{itemize}
\item[$({\rm a})$] $O_{B_2} \leq O_{B_1}$ if and only if $B_1 \subseteq_{WS,P,-} B_2$;
\item[$({\rm b})$] $C_{B_1} \leq C_{B_2}$ if and only if $B_1 \subseteq_{WS,P,+} B_2$.
\end{itemize}
\end{theoremA}
\begin{proof}
Corollary~\ref{Corollary: Opening and Closing ordering relations}, $\mathcal{X}_{O_{B_1}(g)} \subseteq \mathcal{X}_{O_{B_2}(g)}$ whenever $g \in \mathcal{BI}_P$ if and only if $\mathcal{X}_{O_{B_1}(g)} \subseteq \mathcal{X}_{O_{B_2}(g)}$ whenever $g \in \mathcal{I}_P$, i.e., $O_{B_2}(g) \leq O_{B_1}(g)$. Similarly, $\mathcal{X}_{C_{B_2}(g)} \subseteq \mathcal{X}_{C_{B_1}(g)}$ whenever $g \in \mathcal{BI}_P$ if and only if $C_{B_1}(g) \leq C_{B_2}(g)$.    
\end{proof}
\bigskip

In summary, Theorem~\ref{Theorem: Equivalence of WS,P and Desired property-form 2} establishes that, for any arbitrary image domain \( P \subseteq \mathbb{Z}^2 \), the weak shift inclusion relation between SEs is equivalent to the absorption property of the corresponding opening and closing operations. In Section~\ref{Section: Shift Inclusion vs. Weak Shift Inclusion}, we further investigate the relationships between shift inclusion and weak shift inclusion, considering both their formulations on \( \mathbb{Z}^2 \) and on general image domains \( P \subseteq \mathbb{Z}^2 \).

%%%%%%%%%%%%%%%%%%%%%%%%%%%%%%%%%%%%%%%%%%%%%%
%%%%%%%%%%%%%%%%%%%%%%%%%%%%%%%%%%%%%%%%%%%%%%
\subsection{Shift Inclusion vs. Weak Shift Inclusion}\label{Section: Shift Inclusion vs. Weak Shift Inclusion}

Section~\ref{Section: Shift Inclusion vs. Weak Shift Inclusion} discusses the relationships between the shift inclusion and weak shift inclusion relations, specifically focusing on the connections among \( \subseteq_{S,P} \), \( \subseteq_{S,\mathbb{Z}^2} \), \( \subseteq_{WS,P} \), and \( \subseteq_{WS,\mathbb{Z}^2} \). These relationships are summarized in the following diagram:

\begin{equation}\label{Equation: Main Scheme}
\xymatrix@+6.0em{
				& \subseteq_{S,\mathbb{Z}^2}
				\ar@{=>}@{<=>}[d]_{\substack{(1) \\ \textbf{Equivalent}, \\ 
				{\rm Theorem}~\ref{Theorem: Equivalence of WS,P and Desired property-form 2},
				\\ {\rm Theorem}~\ref{Decomposition theorem of shift inclusion}}}
				\ar@{=>}@/_1pc/[r]_{(3) \ \textbf{No},~{\rm Example}~\ref{ExampleA: B_1 subseteq_S,Zm B_2 doesn't imply B_1 subseteq_S,P B_2-2}}
                & \subseteq_{S,P}
                \ar@{=>}@/_1pc/[l]_{(2) \ \textbf{No},~{\rm Example}~\ref{ExampleA: S,P do not imply S,M-1}}
                \ar@{=>}@/_1pc/[d]_{\substack{(6) \\ \textbf{Yes}, \\{\rm Proposition}~\ref{Proposition: S implies WS}}}
                \\
        		& \subseteq_{WS,\mathbb{Z}^2}
        		\ar@{=>}@/_1pc/[r]_{(5) \ \textbf{No},~{\rm Example}~\ref{ExampleA: B_1 subseteq_S,Zm B_2 doesn't imply B_1 subseteq_S,P B_2-1}}
				& \subseteq_{WS,P}
				\ar@{=>}@/_1pc/[l]_{(4) \ \textbf{No},~{\rm Example}~\ref{ExampleA: S,P do not imply S,M-1}}
				\ar@{=>}@/_1pc/[u]_{\substack{(7) \\ \textbf{No}, \\ {\rm Example}~\ref{ExampleA: desired property doesn't imply subseteq_S,P}, \\ {\rm Theorem}~\ref{Theorem: Equivalence of WS,P and Desired property-form 2}}}
}
\end{equation}

\subsubsection*{Proof of (1)}
For the equivalence labeled (1) in the diagram, Theorem~\ref{Theorem: Equivalence of WS,P and Desired property-form 2} states that \( B_1 \subseteq_{WS,\mathbb{Z}^2} B_2 \) if and only if \( B_1 \) and \( B_2 \) satisfy the absorption property~\eqref{Equation: Absorption property}. On the other hand, Theorem~\ref{Decomposition theorem of shift inclusion} shows that this condition is also equivalent to \( B_1 \subseteq_{WS,P} B_2 \).

\subsubsection*{Proof of (2) and (3)}

For the arrows (2) and (3) in~\eqref{Equation: Main Scheme}, we provide counterexamples to support the claim that these implications do not hold in general. Specifically, Examples~\ref{ExampleA: B_1 subseteq_S,Zm B_2 doesn't imply B_1 subseteq_S,P B_2-1} and~\ref{ExampleA: B_1 subseteq_S,Zm B_2 doesn't imply B_1 subseteq_S,P B_2-2} support arrow (3), while Example~\ref{ExampleA: S,P do not imply S,M-1} supports arrow (2).

\bigskip
\begin{exampleA}
\label{ExampleA: B_1 subseteq_S,Zm B_2 doesn't imply B_1 subseteq_S,P B_2-1} 
Let \( g : P \rightarrow \{ 0, 1 \} \) be a binary image defined on a subset \( P \subseteq \mathbb{Z}^2 \), given by
\begin{equation*}
\ytableausetup{centertableaux}
g = \begin{ytableau}
0 & 0 & 1 & 0 & 0\\     
0 & 0 & 0 & 0 & 0 \\
1 & 0 & 1 & 0 & 1 \\
\end{ytableau} \ .    
\end{equation*}
$ $\\
Let $B_1 \subseteq B_2$ be SEs defined as in Example \ref{ExampleA: Shift inclusion non-connected}. Then, by the decomposition theorem for shift inclusion (Theorem \ref{Decomposition theorem of shift inclusion}), $B_1 \subseteq_{S,\mathbb{Z}^2} B_2$ since $B_2 = B_1 \cup (-2\bfe_2 + B_1)$. However, 
\begin{equation*}
\ytableausetup{centertableaux}
\begin{split}
 O_{B_1}(g) = \begin{ytableau}
0 & 0 & 1 & 0 & 0\\     
0 & 0 & 0 & 0 & 0 \\
1 & 0 & 0 & 0 & 1 \\
\end{ytableau} \text{ and } O_{B_2}(g) = 
\begin{ytableau}
0 & 0 & 1 & 0 & 0\\     
0 & 0 & 0 & 0 & 0 \\
1 & 0 & 1 & 0 & 1 \\
\end{ytableau} \ \ .
\end{split}
\end{equation*}
This shows that $O_{B_2} \nleq Q_{B_1}$, and we conclude that $B_1 \nsubseteq_{S,P} B_2$ by Theorem \ref{Theorem: Shift Inclusion as a sufficient condition, strong form}.
\end{exampleA}
\bigskip
\begin{exampleA}
\label{ExampleA: B_1 subseteq_S,Zm B_2 doesn't imply B_1 subseteq_S,P B_2-2}
Let \( g : P \rightarrow \{ 0, 1 \} \) be a binary image defined on a subset \( P \subseteq \mathbb{Z}^2 \), given by
\begin{equation*}
\ytableausetup{centertableaux}
g = \begin{ytableau}
0 & 0 & 0 & 0 & 0\\     
0 & 0 & 1 & 0 & 1 \\
0 & 1 & 1 & 1 & 1 \\
\end{ytableau} \ \ .
\end{equation*}
Consider SEs
\begin{equation*}
B_1 = \begin{ytableau}
\none & \ & \none \\
\ & \ & \mathbf{0} \\
\none & \ & \none \\
\end{ytableau} \ \ \subseteq 
B_2 = \begin{ytableau}
\none & \ & \none & \ & \none \\
\ & \ & \mathbf{0} & \ & \ \\
\none & \ & \none & \ & \none \\
\end{ytableau} \ \ .
\end{equation*}
Theorem \ref{Decomposition theorem of shift inclusion} shows that $B_1 \subseteq_{S,\mathbb{Z}^2} B_2$ since $B_2 = B_1 \cup (2 \bfe_1 + B_1)$. On the other hand, we observe that
\begin{equation*}
\ytableausetup{centertableaux}
\begin{split}
O_{B_1}(g) &= \begin{ytableau}
0 & 0 & 0 & 0 & 0\\     
0 & 0 & 1 & 0 & 0 \\
0 & 1 & 1 & 1 & 0 \\
\end{ytableau} \ \ \text{, and } O_{B_2}(g) = 
\begin{ytableau}
0 & 0 & 0 & 0 & 0\\     
0 & 0 & 1 & 0 & 1 \\
0 & 1 & 1 & 1 & 1 \\
\end{ytableau} \ \ .
\end{split}
\end{equation*}
By comparing pixel values, we deduce the set $O_{B_1}(g)^{-1}(0)$ is not a subset of $O_{B_2}(g)^{-1}(0)$. By Theorem \ref{Theorem: Shift Inclusion as a sufficient condition}, $B_1$ and $B_2$ don not satisfy the shift inclusion with respect to $P$, i.e., $B_1 \nsubseteq_{S,P} B_2$.  
\end{exampleA}
\bigskip
Examples~\ref{ExampleA: B_1 subseteq_S,Zm B_2 doesn't imply B_1 subseteq_S,P B_2-1} and~\ref{ExampleA: B_1 subseteq_S,Zm B_2 doesn't imply B_1 subseteq_S,P B_2-2} demonstrate that, although the SEs in these examples are shift included with respect to the entire space \( \mathbb{Z}^2 \), they are not shift included with respect to the domain \( P \). In other words, the relation \( \subseteq_{S,\mathbb{Z}^2} \) does not imply \( \subseteq_{S,P} \). Conversely, the following example shows that \( \subseteq_{S,P} \) does not imply \( \subseteq_{S,\mathbb{Z}^2} \) either; in particular, arrow (2) holds.

\bigskip
\begin{exampleA}
\label{ExampleA: S,P do not imply S,M-1}
Let $P \subseteq \mathbb{Z}^2$ be the rectangular image domain with shape
\begin{equation*}
P =  \begin{ytableau}
\ & \ & \ & \ & \
\end{ytableau} \quad,   
\end{equation*}
and let $B_1$ and $B_2$ be SEs defined by
\begin{equation*}
B_1 = \begin{ytableau}
\ \\
\mathbf{0} 
\end{ytableau} \quad \text{and} \quad B_2 = \begin{ytableau}
\ & \ \\
\mathbf{0} & \none 
\end{ytableau} \quad.   
\end{equation*}
Then \( B_1 \nsubseteq_{S,\mathbb{Z}^2} B_2 \) by Theorem~\ref{Decomposition theorem of shift inclusion}. On the other hand, the shift inclusion \( B_1 \subseteq_{S,P} B_2 \) holds, since \( \mathbf{x} \pm \mathbf{b}_2 \notin P \) for all \( \mathbf{b}_2 \in B_2 \setminus \{ \mathbf{0} \} \).
\end{exampleA}

\subsubsection*{Proof of (4) and (5)}

For arrow (4), we observe that \( \subseteq_{S,P} \) is a stronger condition than \( \subseteq_{WS,P} \); that is, \( \subseteq_{S,P} \) implies \( \subseteq_{WS,P} \). In particular, the SEs \( B_1 \subseteq B_2 \) in Example~\ref{ExampleA: S,P do not imply S,M-1} satisfy the condition \( B_1 \subseteq_{WS,P} B_2 \), while \( B_1 \nsubseteq_{S,\mathbb{Z}^2} B_2 \).

By arrow (1), the relations \( \subseteq_{WS,\mathbb{Z}^2} \) and \( \subseteq_{S,\mathbb{Z}^2} \) are equivalent. Consider the SEs \( B_1 \subseteq B_2 \) satisfying \( B_1 \subseteq_{S,\mathbb{Z}^2} B_2 \) and the image domain \( P \subseteq \mathbb{Z}^2 \) depicted in Example~\ref{ExampleA: B_1 subseteq_S,Zm B_2 doesn't imply B_1 subseteq_S,P B_2-1}. In that example, we showed that there exists an image \( g \in \mathcal{I}_P \) such that \( O_{B_2}(g) \nleq O_{B_1}(g) \). By Theorem~\ref{Theorem: Equivalence of WS,P and Desired property-form 2}, we deduce that \( B_1 \nsubseteq_{WS,P} B_2 \). This shows that \( \subseteq_{WS,\mathbb{Z}^2} \) does not imply \( \subseteq_{WS,P} \), i.e., arrow (5) holds.

\subsubsection*{Proof of (6)}

Arrow (6) corresponds exactly to the result of Proposition~\ref{Proposition: S implies WS}, which shows that \( \subseteq_{S,P} \) is a stronger condition than \( \subseteq_{WS,P} \).

\subsubsection*{Proof of (7)}

For arrow (7), we observe that Example~\ref{ExampleA: desired property doesn't imply subseteq_S,P} presents a case where the image domain \( P \subseteq \mathbb{Z}^2 \) and a tower of SEs \( B_1 \subseteq B_2 \) satisfy \( O_{B_2}(g) \leq O_{B_1}(g) \) and \( C_{B_1}(g) \leq C_{B_2}(g) \) for all \( g \in \mathcal{I}_P \). By Theorem~\ref{Theorem: Equivalence of WS,P and Desired property-form 2}, this implies \( B_1 \subseteq_{WS,P} B_2 \). However, the argument in Example~\ref{ExampleA: desired property doesn't imply subseteq_S,P} also shows that \( B_1 \nsubseteq_{S,P} B_2 \), thereby proving that arrow (7) holds.



%\cite{soille2013}
\end{appendices}

%\bibliographystyle{abbrv}  
%\bibliography{references}
%\bibliography{sn-bibliography,output_refs}% common bib file
\bibliography{output_refs}% common bib file

\end{document}